\documentclass[a4]{aa}
\usepackage[varg]{txfonts}
\usepackage{graphicx}
\usepackage{epstopdf}
\usepackage{units}
\usepackage[utf8]{inputenc}
\usepackage{afterpage}
\DeclareGraphicsExtensions{.pdf,.eps,.png,.jpg,.mps} 
\bibpunct{(}{)}{;}{a}{}{,}

\usepackage{natbib,twoopt}
\usepackage[breaklinks=true]{hyperref} 
\usepackage{color}

\begin{document}
\title{Stellar magnetic activity and variability of oscillation parameters --- An investigation of 24 solar-like stars observed by \textit{Kepler}}

\author{René Kiefer\inst{1}
\and Ariane Schad\inst{1}
\and Guy Davies\inst{2}
\and Markus Roth\inst{1}}

\institute{Kiepenheuer-Institut für Sonnenphysik, Schöneckstraße 6, 79104 Freiburg, Germany 
\and School of Physics and Astronomy, University of Birmingham, Edgbaston, Birmingham B15 2TT, UK}
\date{Received / Accepted }


\titlerunning{Variability of oscillation parameters}
\authorrunning{René Kiefer et al.}

\abstract{The Sun and solar-like stars undergo activity cycles for which the underlying mechanisms are not well understood. The oscillations of the Sun are known to vary with its activity cycle and these changes provide diagnostics on the conditions below the photosphere. \textit{Kepler} has detected oscillations in hundreds of solar-like stars but as of yet, no widespread detection of signatures of magnetic activity cycles in the oscillation parameters of these stars have been reported.}
{We analyse the photometric short cadence \textit{Kepler} time series of a set of 24 solar-like stars, which were observed for at least 960 days each, with the aim to find signatures of stellar magnetic activity in the oscillation parameters.}
{We analyse the temporal evolution of oscillation parameters by measuring mode frequency shifts, changes in the height of the p-mode envelope, as well as granulation time scales.}
{For 23 of the 24 investigated stars, we find significant frequency shifts in time. We present evidence for magnetic activity in six of them. We find that the amplitude of the frequency shifts decreases with stellar age and rotation period. For the most prominent example, KIC 8006161, we find that, similar to the solar case, frequency shifts are smallest for the lowest and largest for the highest p-mode frequencies. }
{These findings show that magnetic activity can be routinely observed in the oscillation parameters for solar-like stars, which opens up the possibility to place the solar activity cycle in the context of other stars by asteroseismology.}
\maketitle

\section{Introduction}
\label{sec:1}
Understanding the solar dynamo, exactly how and where it operates within the Sun, is a major open task in modern astrophysics. Helio- and asteroseismology are outstanding tools to probe the interiors of the Sun and other stars. The vast amount of data which has become available thanks to satellite missions like \textit{CoRoT} \citep{2006ESASP1306...33B, 2009A&A...506..411A} and \textit{Kepler} \citep{Borucki977, 2010ApJ...713L..79K} enables us to investigate stellar cycles for a wide range of stellar types. By putting the Sun and its magnetic activity cycle into context to similar stars, we can widen our perspective on the basic processes at work in dynamos in different physical environments and may be able to decide between opposing dynamo theories. 

The better part of known activity cycles of main-sequence stars was detected from Mount Wilson Observatory \citep{1978ApJ...226..379W, 1995ApJ...438..269B}. Their measurements of fluxes in the line cores of Ca II H and K, which can be used as proxies for stellar activity \citep[][and references therein]{lrsp-2008-2}, suggest a relation between the age, the rotation, and the activity of stars. For a subset of solar-type stars \cite{1995ApJ...438..269B} found that younger stars exhibit higher average levels of activity and rapid rotation rates, while older stars with a slower rotation usually have cycles with lower levels of magnetic activity. A relation of this kind was already proposed by \cite{1972ApJ...171..565S}, who found that stellar Ca II emissions decay as the inverse square root of the age. \cite{1538-4357-498-1-L51} and \cite{0004-637X-524-1-295} made an empirical classification of stars from the Mount Wilson Observatory sample in a rotation-activity scheme. They found that most stars fall into one of three distinct branches when examining the ratio of cycle and rotation periods as a function of chromospheric activity. \cite{Karoff21082013} used ground-based observations in combination with data from the \textit{Kepler} satellite to test age-rotation-activity relations. 

Helioseismology found that the frequencies of solar acoustic oscillations (p modes) are positively correlated with solar magnetic activity \citep[e.g.][]{1985Natur.318..449W,1990Natur.345..779L, 1998A&A...329.1119J}, whereas mode amplitudes are anti-correlated with it \citep[e.g.][]{palle90, 0004-637X-543-1-472, Broomhall20152706}. Activity related frequency shifts are larger for modes of higher frequency \citep{1998A&A...329.1119J}, because these modes have their maximum sensitivity to structural changes in the solar interior in shallower layers than modes of low frequency \citep{2012ApJ...758...43B}. This was recently used by \cite{2015A&A...578A.137S} to study the change of sub-surface solar activity as a function of time and for different depths. 

In an analysis of photometric data of the \textit{CoRoT} satellite, \cite{2010Sci...329.1032G} found evidence for an activity cycle with $P_{\textrm{cyc}}>\unit[120]{d}$ in the F5V star HD 49933. They found shifts in the frequencies of the star's p modes which showed a cycle-like behaviour and, in addition, changes in mode amplitudes which were anti-correlated to the frequency shifts. Later, \cite{2011A&A...530A.127S} found the frequency shifts of HD 49933 to be dependent on mode frequency. The \textit{Kepler} satellite produced a huge data set on solar-like stars with asteroseismic import, which shows great promise for many more detections of the kind \citeauthor{2010Sci...329.1032G} made. \cite{Vida01072014} found hints of cyclic activity in the \textit{Kepler} data of nine fast-rotating late-type stars. Recently, \cite{2016A&A...589A.118S} investigated \textit{Kepler} and spectroscopic data from the young solar analog KIC~10644253. They found variability in the p-mode frequencies of this star with a modulation of about 1.5 years. 

In this paper, we describe our efforts to find signatures of stellar magnetic activity in the oscillation parameters of solar-like stars, focussing on the shifts of p-mode frequencies and the variation of the height of the p-mode envelope. In Sec.~\ref{sec:2} we describe our approach to measure these quantities and to estimate the errors on them. Sec.~\ref{sec:3} is dedicated to a presentation of the results and a more detailed discussion of two stars from our sample. Sec.~\ref{sec:4} gives a conclusion of our findings.

\section{Methodology}
\label{sec:2}
\subsection{Data and computation of periodograms}
We analysed \textit{Kepler} data from 24 stars, which were observed in the satellite's short cadence mode. Only data from the \textit{Kepler} data release 25 was used in this work. In this data release, the scrambling of the short-cadence collateral smear data, which was reported in the Global Erratum for \textit{Kepler} Q0-Q17 \& K2 C0-C5 Short-Cadence Data, and the Dynablack calibration problem, which was reported in the Data Release 24 Notes Q0–Q17 Erratum, are corrected. These 24 stars were selected because they are known to exhibit solar like oscillations \citep[e.g.][]{2012ApJ...749..152M, 2012A&A...543A..54A, 2013ApJ...767..127H} and are known to be similar to the Sun regarding fundamental parameters like mass, radius, and effective temperature (cf. Table~\ref{table:A1} and Table~\ref{table:A2} and references therein). No strict criteria were applied to define this special sample of stars. It is therefore just a first assessment of the variability of stellar p mode parameters in the \textit{Kepler} short cadence sample. The full set of stars, which were observed in \textit{Kepler}’s short cadence mode over extended periods of time, is considerably larger and is a worthwhile subject for future investigation.

The coverage of data for these stars ranges from 10 to 13 \textit{Kepler} quarters of observation (960 to 1147 days). A list of the \textit{Kepler} input catalogue (KIC) numbers of the stars and the available quarters of data is given in the first two columns of Table~\ref{table:1}. Missing quarters are listed in a footnote to Table~\ref{table:1}. The third column gives the number of days the data for each star spans, counting from the first day of observation to the last. The light curves were cleaned of jumps, drifts, and outliers similar to the method described by \cite{2011MNRAS.414L...6G}.

To investigate the temporal variation of p-mode parameters, the time series were divided into consecutive, overlapping segments of equal length. To compromise between frequency resolution and number of independent segments, we chose a typical length of 150 days for the segments. From one segment to the next, the starting point was shifted by 50 days, resulting in a three time overlap for 150 days long segments. If the errors on the eventually measured frequency shifts allowed a decrease in segment length and hence frequency resolution of the periodograms, we reduced it to 100 days. The shift of the starting point remained at 50 days, resulting in a two time overlap. The lengths of the segments are given in the fourth column of Table~\ref{table:1}.

Since the time series are slightly unevenly sampled and are affected by observational gaps, we used the Lomb-Scargle method \citep{1976Ap&SS..39..447L, 1982ApJ...263..835S} in the fast implementation of \cite{press2007numerical} to calculate the periodograms (LS-periodogram). The resulting periodograms are scaled to comply with Parseval's theorem.

The observational gaps lead to spectral leakage of mode power. To study the effect of the gaps in the \textit{Kepler} time series on our analysis, we used the inpainting software of \cite{2015A&A...574A..18P} to fill gaps up to a length of 20 days \citep{2014A&A...568A..10G}. For none of our targets did the high frequency noise level decrease significantly. Also, the results described in the subsequent sections, frequency shifts and mode heights, do not change in a way that would make necessary the inpainting of the time series. This is because the \textit{Kepler} gap structure does not have a large effect on the periodograms of solar-like stars as long as the stars are no fast rotators \citep{2014A&A...568A..10G} and the periodograms are computed from time series segments with a length of only 100 or 150 days.

\begin{figure}
\centering{
\includegraphics[width=0.49\textwidth]{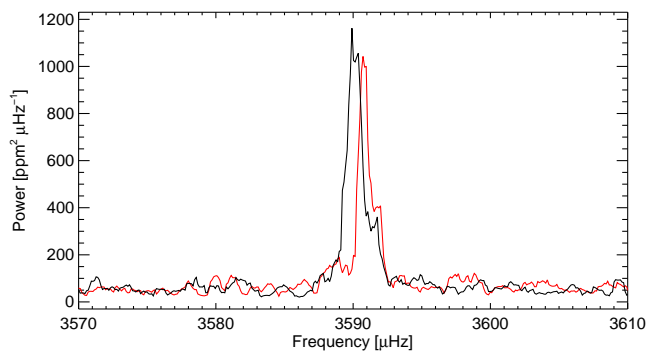}
\caption{Section of the LS-periodograms of the first and nineteenth segment of KIC~8006161 in black and red colour, respectively. The LS-periodograms are boxcar smoothed over $\unit[0.81]{\mu Hz}$. }
\label{fig:1}}
\end{figure}

To get the same frequency axis for the LS-periodogram of each data segment of a given star, the LS-periodograms were interpolated onto the frequency grid of the segment with the lowest fill factor. This way, the LS-periodograms can later be correlated with each other. The minimum fill factor of segments was set to 50\%. Thereby, we ensured a reasonable frequency resolution for the LS-periodograms of all the segments of a star. In Fig.~\ref{fig:1}, the $\unit[0.81]{\mu Hz}$ (7 bins) boxcar smoothed segments between $\unit[3570-3610]{\mu Hz}$ of the LS-periodograms of the first and the nineteenth segment of the time series of KIC~8006161 are shown in black and red colour, respectively. Even by visual inspection of the peak at $\unit[\sim\!\!3590]{\mu Hz}$, a shift towards higher frequency can be spotted for the red peak. 

\subsection{Measuring the frequency shifts}
The first segment of each star was used as the reference for defining zero frequency shift. Due to this, the level of zero frequency shift can be at any level of a stellar activity cycle, e.g. its maximum. We computed the shift relative to the reference segment by calculation of the cross-correlation of that frequency range of the LS-periodograms which contains the p modes and fitting a Lorentzian to the cross-correlation. We selected this frequency range by visually determining the lower and upper boundaries between which p-mode peaks can be spotted in the smoothed LS-periodogram of the entire time series of each star (see last column of Table~\ref{table:1}). Recently, a cross-correlation technique for measuring mode frequency shifts was presented by \cite{Regulo2016}. Their method to measure the frequency shifts is very similar to the approach presented here. However, our approach to estimate the error on the frequency shifts, which is described in the following section, is different in several aspects from that of \cite{Regulo2016}.

\begin{figure}
\centering{
\includegraphics[angle=270,width=0.49\textwidth]{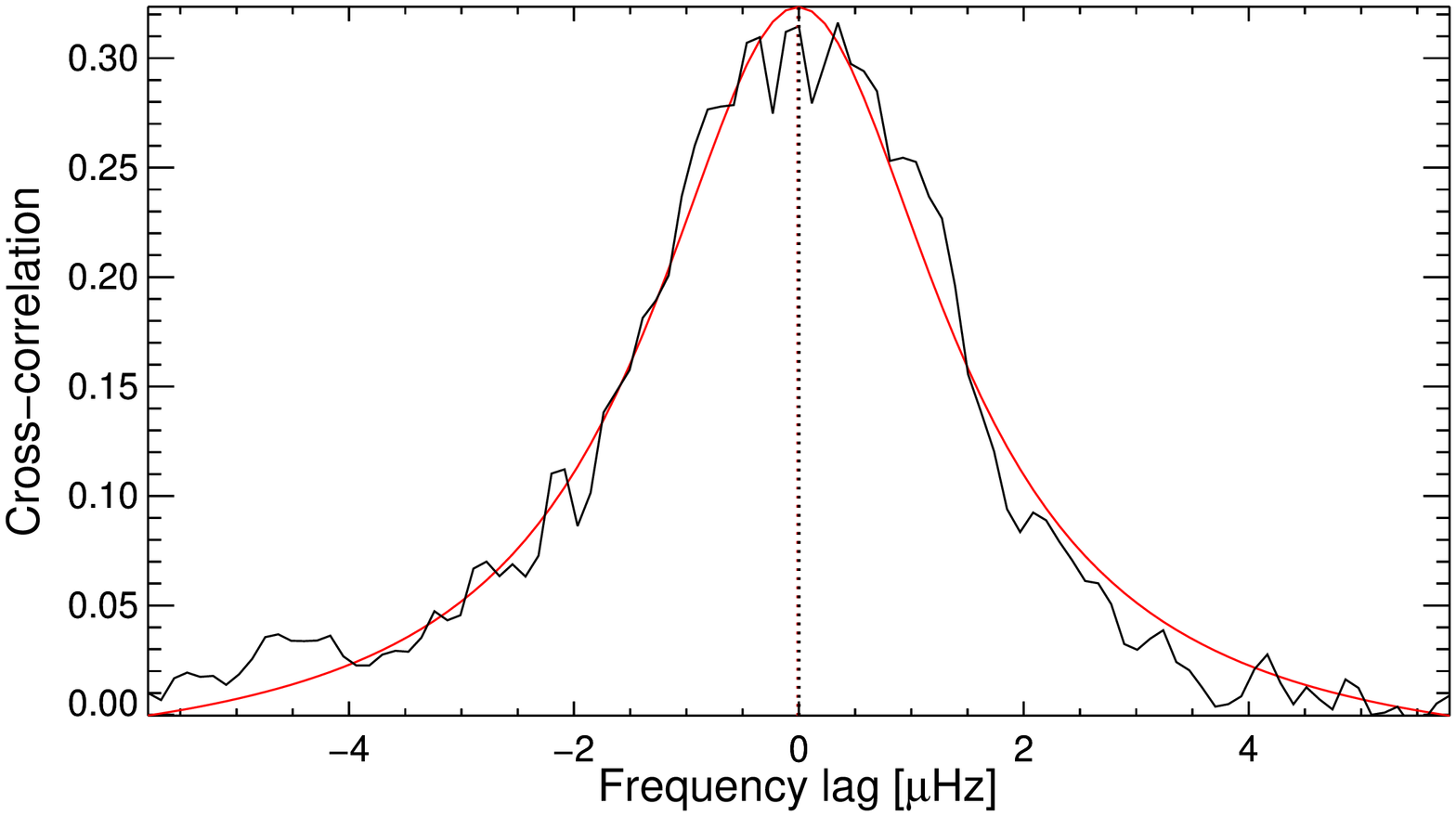}\\
\includegraphics[angle=270,width=0.49\textwidth]{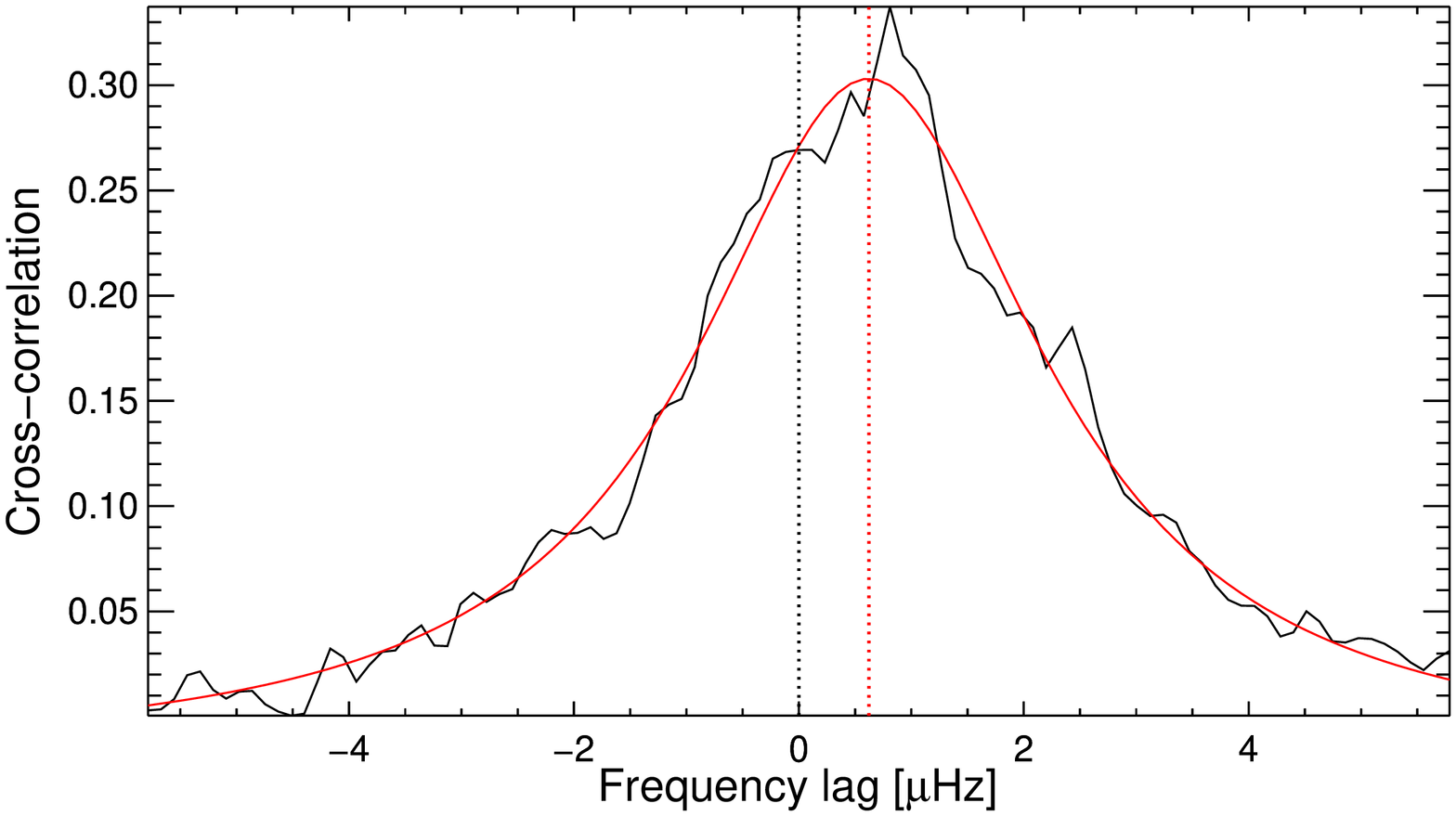}}
\caption{{\it Top panel:} cross-correlation between the part of interest of the LS-periodograms of two realisations of the first segment of KIC~8006161 (black). The Lorentzian fit to the cross-correlation is shown in red. The dotted black vertical line marks zero frequency lag, the dashed red line marks the centre of the fitted Lorentzian. {\it Bottom panel:} same as in the top panel, but for the cross-correlation between a realisation of the first and one of the nineteenth segment.}
\label{fig:2}
\end{figure}

In the top panel of Fig.~\ref{fig:2}, the part between $\pm\unit[5]{\mu Hz}$ of the cross-correlation between the p-mode region of two realisations (see Sect.~\ref{sec:resampling}) of the first segment of KIC~8006161 is shown in black. The Lorentzian fit to the data is shown as a continuous red curve. The dotted black line marks zero frequency lag, the dashed red line marks the center of the fitted Lorentzian. The bottom panel of Fig.~\ref{fig:2} depicts the same as the top panel, but for the cross-correlation between the first and the nineteenth segment of the time series from KIC~8006161. Here, the maximum of the cross-correlation is clearly shifted towards positive values. Peaks of stochastically excited solar and stellar p-mode oscillations appear to be of Lorentzian shape. The Lorentzian is invariant under convolution with a Lorentzian, hence the cross-correlation of two of them can be fitted with a Lorentzian. In our analysis we used a symmetric Lorentzian profile. We found that the extra parameter introduced by a skewness does not decrease the error on the frequency shift or increase the stability of the fits.

\subsubsection{Resampling approach}
\label{sec:resampling}
Since the correlation structure of the estimated cross-correlation function depends on the true unknown cross- and auto-correlation function in a complicated manner, we use a resampling approach to estimate the error on the frequency shift (see \citet{bartlett} and e.g. \citet{2007ApJ...659.1749C}). We generate a sample of new LS-periodograms, which follow the statistical properties of the original periodogram, and from which we compute the sample mean and sample error. In our resampling approach we neglect effects from uneven sampling, since it is small for \textit{Kepler} data. The largest effects on our analysis are assumed to result from gaps in the data. 

For each data segment $i$, we draw $B=200$ complex random series $F^*_i(\nu_k)$ for $\{\pm \nu_k\}_{k=1,\dots,N}$ from a zero mean normal distribution $N(0,1)$ weighted by the estimated spectral power $\hat S_i(\nu_k)$, such that:
\begin{eqnarray}
\Re \{F^*_i(\nu_k)\} &\sim& \sqrt{\hat S_i(\nu_k)} \times N(0,1)\\
\Im \{F^*_i(\nu_k)\} &\sim& \sqrt{\hat S_i(\nu_k)} \times N(0,1)\, ,
\end{eqnarray}
and $F^*(-\nu_k)=\overline{F^*(-\nu_k)}$. For $\nu=0$, we set $F^*_i(\nu=0)=0$, reflecting a zero mean time series. An estimate $\hat S_i$ of the true unknown spectral density $S_i$ is obtained by smoothing the LS-periodogram with a boxcar window. The width of the boxcar varies from star to star, ranging from \unit[0.38--1.52]{$\mu$Hz} (5--12 bins). We chose the width of the boxcar depending on the frequency resolution of the periodograms and the width of the p modes of each individual star. The complex random series $\{F^*_i(\nu_k)\}_i$ may be considered as realisations of Fourier transforms that would be obtained in the ideal case of evenly sampled data without observational gaps but with the same spectral density underlying the respective original data. Note that the real and the imaginary part of $\{F_i^*(\nu_k)\}$ are by construction independent of each other and with respect to $\nu_k$. In order to take into account the effect of gaps in the original data, we generate new time series $\{x_i^*(t_k)\}$ on an evenly sampled grid $\{t_n\}_{n=1,\cdots,N}$ by the inverse Fourier transform of $\{F_i^*(\nu_k)\}$. The grid in time approximates the one for the original data. Data points at the respective grid points with observation gaps are removed. These new time series $\{x_i^*(t_n)\}$ are treated as described above to compute frequency shifts from the cross-correlation function of LS-periodograms. We further note that these resampled time series have no physical meaning, since the phases between real- and imaginary part of the Fourier transforms are random. However, the described procedure conserves the linear spectral properties of the original data.

We emphasise that the LS-periodogram is computed with oversampling factor 1. In case of evenly sampled data without gaps, the LS-periodogram is then equivalent to the classical Fourier based periodogram and follows the same statistical distribution \citep{1982ApJ...263..835S}. An oversampling factor greater than 1 corresponds to an interpolation of the periodogram along frequency bins and gives correlated values at adjacent frequency bins. In this study, we aim to suppress such additional correlations since they lead to an unwanted complex correlation structure in the error analysis and result in a systematic underestimation of the error on the frequency shifts.

Each realisation of the LS-periodogram of a given segment is cross-correlated with a realisation of the LS-periodogram of the reference segment. For the reference segment, two samples of 200 realisations are drawn. The cross-correlation analysis between these two sets yields the error bar on the zero level of the frequency shifts. The standard deviation of the centroid values of the Lorentzian profile of the 200 fits is used as the error for the shift of the p modes, while their mean is used as the value of the shift. The value $B=200$ was chosen, because it is a conservative number of realisations for which the variance of the estimator of the error is small compared to the variance of the distribution of the centroid values of the frequency shifts. Hence, the standard deviation of the 200 centroid values is a reliable estimate for the error on the measured frequency shifts. 

\begin{table*} 
\caption{Overview of the basic data of the investigated time series and their periodograms.}           
\label{table:1}   
\centering                                   
\begin{tabular}{c c c c c}
	\hline\hline
	  KIC    & Data coverage$^*$ & Length of data (d) & Length of segment (d) & Frequency range ($\mu$Hz) \\ \hline
	3632418  & 5-17.2            & 1147               & 150                   & 700-1700                  \\
	3656476  & 7-17.2$^a$        & 960                & 150                   & 1500-2500                 \\
	4914923  & 7-17.2            & 960                & 150                   & 1350-2300                 \\
	5184732  & 7-17.2            & 960                & 100                   & 1400-2700                 \\
	6106415  & 6-16$^b$          & 1018               & 150                   & 1550-3100                 \\
	6116048  & 5-17.2            & 1147               & 150                   & 1550-2600                 \\
	6603624  & 5-17.2            & 1147               & 100                   & 1900-3000                 \\
	6933899  & 5-17.2            & 1147               & 150                   & 1000-1800                 \\
	7680114  & 7-17.2$^c$        & 960                & 150                   & 1350-2100                 \\
	7976303  & 5-17.2            & 1147               & 150                   & 550-1300                  \\
	8006161  & 5-17.2            & 1147               & 100                   & 2800-4400                 \\
	8228742  & 5-17.2            & 1147               & 100                   & 800-1600                  \\
	8379927  & 5-17.2            & 1147               & 150                   & 2100-3700                 \\
	8760414  & 5-17.2            & 1147               & 150                   & 1950-3000                 \\
	9025370  & 5-17.2            & 1147               & 100                   & 2550-3450                 \\
	9955598  & 5-17.2            & 1147               & 150                   & 3000-4100                 \\
	10018963 & 5-17.2            & 1147               & 100                   & 650-1500                  \\
	10516096 & 7-17.2$^d$        & 960                & 100                   & 1200-1900                 \\
	10644253 & 5-17.2            & 1147               & 150                   & 2450-3350                 \\
	10963065 & 5-15$^e$          & 1027               & 150                   & 1700-2700                 \\
	11244118 & 5-17.2            & 1147               & 100                   & 1000-1800                 \\
	11295426 & 5-17.2            & 1147               & 150                   & 1800-2400                 \\
	12009504 & 5-17.2            & 1147               & 100                   & 1400-2300                 \\
	12258514 & 5-17.2            & 1147               & 150                   & 1050-2100                 \\ \hline
\end{tabular}
\tablefoot{$^*$ Data coverage indicated in \textit{Kepler} quarters. Missing quarters: $^a$ Q10, Q14; $^b$ Q9, Q13; $^c$ Q6, Q7.2, Q10; $^d$ Q10.1; $^e$ Q8, Q9, Q12}
\end{table*}

\subsection{Consistency check with BiSON data}
\label{sec:bison}
\begin{figure}
\centering{
\includegraphics[width=0.49\textwidth]{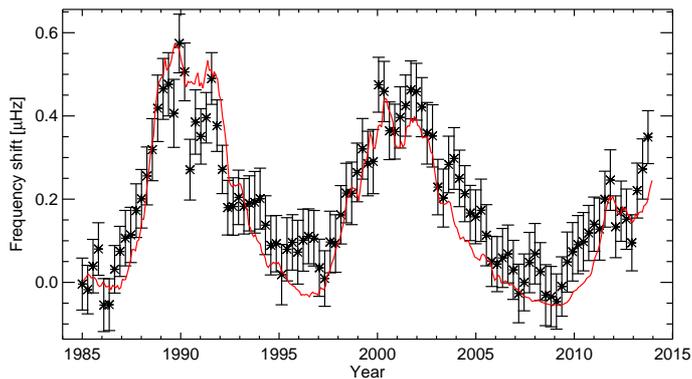}
\caption{Frequency shifts of solar p modes from BiSON data with 1-$\sigma$ error bars (black). The seven month boxcar smoothed monthly sunspot number is shown in red. Source of the sunspot number: WDC-SILSO, Royal Observatory of Belgium, Brussels. }
\label{fig:3}}
\end{figure}

For a consistency check of our method, we analysed data from the BiSON network \citep{2014MNRAS.441.3009D}. The time series encompasses the period from 1985--2014, covering about two and a half solar cycles. Fig.~\ref{fig:3} depicts the frequency shifts and the associated 1-$\sigma$ error bars as black data points. The monthly sunspot number is shown as a continuous red line (boxcar smoothed over seven months). It is scaled to match the amplitude of the variation of the p-mode frequencies in this figure. In this case, we set the length of the segments to 200 days and the step length to 100 days in order to increase the clarity of the plot by reducing the number of data points. We selected the frequency range between \unit[1800--3800]{$\mu$Hz} for the cross-correlation of the periodograms. The frequency shift closely follows the sunspot number, a proxy of solar magnetic activity. 

For the amplitude of the solar p-mode frequency shift we find a value of $\unit[0.62\pm 0.06]{\mu Hz}$. This is in agreement with previous investigations of the p-mode frequency shifts obtained with the cross-correlation method using BiSON data \citep{2007ApJ...659.1749C}. 

\subsection{Height of the mode envelope}
As we are considering the temporal evolution of the frequencies of all p modes in the periodogram, we compute a proxy for the amplitudes of the p modes by measuring their excess over the background in the periodogram. 

For this, we have fitted a background to each of the segments and recorded the summary statistics. We used a variant of "model F" as described in and recommended by \cite{2014A&A...570A..41K}, which consists of two Harvey-like profiles \citep{1985ESASP.235..199H}, a Gaussian p-mode envelope, and an instrumental noise component. Following \citeauthor{2014A&A...570A..41K}, we restricted the exponents of the Harvey-like profiles to a value of 4. 

Here our purpose was to study the impact of stellar activity. So, in order to isolate the activity signal of the p modes from the low frequency activity signal due to surface modulation, we restricted the fitting range to frequencies above $\unit[400]{\mu Hz}$. This lower limit excludes regions of the periodogram that can also contain activity signals, but mitigates problems caused by a \textit{Kepler} noise feature found at $\unit[374]{\mu Hz}$. As the surface activity signal is excluded, we did not include a power-law to account for surface activity in the model.

To fit the model, we used the affine invariant Markov Chain Monte Carlo code \textit{emcee} \citep{2013PASP..125..306F}. From the resulting Markov Chains we quote the summary statistics of the posterior probability distributions, which are normally distributed, as the median value and the $15.9$th and $84.1$th percentile to define the highest posterior density interval, i.e., the credible interval. We used the standard Gaussian form for the mode envelope:
\begin{align*}
N\left(\nu\right) = a\,\exp\left(-\frac{\left(\nu-b\right)^2}{2c^2}\right)
\end{align*}
where $a$ is the height of the Gaussian, $b$ is the frequency of maximum height, and $c$ is the width of the Gaussian. 
In the subsequent sections, we will refer to $a$ as the height of the p-mode envelope, or simply height. 

To exclude contamination of the measured mode heights with a temporally varying instrumental noise component, which might for example arise from the quarterly satellite rolls performed by \textit{Kepler} during its nominal mission phase, we checked its temporal variation for all our targets. We found that there are no considerable jumps in the instrumental noise component at the quarter boundaries of the time series we used for the analyses.

\section{Results}
\label{sec:3}
\subsection{Frequency shifts and background parameters}
In Table \ref{table:2} the peak-to-valley amplitude of the frequency shifts of all 24 stars are listed. The amplitude $A$ was calculated as the difference between the minimum and maximum values of the frequency shifts. With the propagated error $\sigma_A$, we also calculated the significance of the observed shift amplitude $A$/$\sigma_A$, see Table~\ref{table:2}. The observed frequency shifts in our sample range from $\approx$\unit[0.17]{$\mu$Hz} for KIC~4914923 to $\approx$\unit[0.95]{$\mu$Hz} for KIC~8006161. Only one star from our sample, KIC~9955598, does not exhibit a frequency shift amplitude which is larger than the error and is therefore compatible with zero for all the segments. In the top panels of Figures \ref{fig:B1}-\ref{fig:B23} the frequency shifts are plotted as a function of starting time of the segments, while in the bottom panels the temporal variation of the mode heights are depicted.

As described in Sec. \ref{sec:1}, the temporal variation of p-mode frequencies can be a proxy for stellar magnetic activity. For some stars, the frequency shifts follow a cycle-like pattern, e.g. KIC~8379927, KIC~8760414, KIC~10644253. Yet, the available time series are not sufficiently long to draw the conclusion that we found clear evidence for a complete cycle in any of the stars of this sample.  

We calculated the Spearman rank correlation coefficients $\rho$ between the variation of the mode frequency shifts, mode heights, the granulation time scales of the background Harvey profiles, and the noise level, as well as the corresponding p-values $p$ for independent segments of all 24 stars. The values are listed in Tables~\ref{table:B1}-\ref{table:B24}. The correlation coefficients of the frequency shifts and the mode height variations and their p-values are also given in Table~\ref{table:3}.

For the Sun \citep[e.g.][]{palle90, 0004-637X-543-1-472, Broomhall20152706} and one more star, the solar-like CoRoT star HD~49933 \citep{2010Sci...329.1032G}, the shifts in mode frequency and the variation of p-mode amplitudes are anti-correlated over the course of their magnetic cycle. For six stars from our sample, the correlation coefficient between the frequency shifts and the variation of the height of the p-mode envelope is less than $-0.5$ and the amplitude of the frequency shift is significant: KIC~6933899, KIC~8006161, KIC~8760414, KIC~9955598, KIC~10644253, KIC~11244118, and KIC~12258514. For these stars, the evidence for magnetic activity influencing the p-mode parameters is strongest.

There is no clear consensus in the literature, on whether the granulation and super-granulation properties are correlated with the solar activity cycle. \cite{2008A&A...490.1143L} find no correlation between the activity level and granulation velocities or time scales in GOLF data, \cite{2007A&A...475..717M} show that there is an in-phase variation of the contrast of solar granulation with the activity cycle but no such correlation for granulation length scales, \cite{2008A&A...488.1109M} find smaller supergranules at cycle maximum. Since granulation has different properties for different stellar parameters \citep[e.g.][]{2013A&A...558A..49B}, we expect different responses by the granulation to magnetic cycles for different stars.
In the case of HD 49933, \cite{2010Sci...329.1032G} showed that the variation of the granulation time scale, the photon noise, and the mode amplitudes are uncorrelated.

The correlation coefficients we present in Tables~\ref{table:B1}-\ref{table:B24} do not present any systematic correlation between the oscillation and background parameters across the sample of stars. For the two stars, which we will discuss in detail in the subsequent sections, KIC~8006161 and KIC~10644253, both of which show evident signs of magnetic activity, we find very different correlations between the investigated parameters.

For KIC~10644253, which was shown to exhibit magnetic activity by \cite{2016A&A...589A.118S}, we find that the frequency shifts are anti-correlated with the first granulation time scale with $\rho = -0.71$ and $p=0.07$, but correlated with the second time scale with $\rho = 0.86$ and $p=0.01$, cf. Table~\ref{table:B19}. The measured mode heights show the opposite behavior. It is noteworthy that for this evidentially active star \citep{2014A&A...562A.124M,2016A&A...589A.118S} the frequency shifts are correlated with changes in the high frequency noise with $\rho = 0.50$, while the mode heights are anti-correlated to these changes with $\rho = -0.75$.

In contrast to this, KIC~8006161, which exhibits the strongest frequency shifts of the investigated sample, shows virtually no significant correlations or anti-correlations between any of the background parameters, cf. Table~\ref{table:B11}, where the only exception is a correlation between the two granulation time scales with  $\rho = 0.74$ and $p=0.01$.

For four stars, KIC~6106415, KIC~6116048, KIC~6603624 and KIC~7680114, the shifts in p-mode frequency are strongly correlated with the variation in mode height, with correlation coefficients greater than $0.5$. For magnetic activity, which is comparable to that observed on the Sun, this is not expected. As some of these stars show significant temporal variability in their p-mode frequencies as well as their mode heights, e.g. KIC~6116048, this might either be due to a dynamo in these stars, which is working differently than in the solar case, or to another mechanism influencing the oscillation parameters, e.g. temporally varying flows \citep{2003A&A...405..779R}.
Whether there is relationship between basic stellar parameters (e.g. mass, effective temperature, or thickness of the convection zone) and the correlation of seismic activity proxies as well as granulation parameters, will be investigated in a future study of an expanded sample of \textit{Kepler} stars. 

Stellar magnetic activity need not be necessarily cyclic in the solar sense and it might be that the variation of frequency shifts and mode heights is correlated in physical settings which are different from the solar reference case regarding e.g. differential rotation. The frequency shifts and mode heights of KIC~10018963 for example, see Fig~\ref{fig:B17}, are moderately correlated with a correlation coefficient of 0.46. Both, frequencies and mode heights, exhibit significant temporal variations which might be caused by non-cyclic sporadic stellar magnetic activity.

\subsection{Results for the ensemble}
With this systematic investigation of the frequency shifts for a larger sample of stars at hand, we try to find relationships between the measured shift amplitudes and fundamental stellar parameters. We assume here that the observed shifts are solely due to magnetic activity. The available values for stellar radius, mass, and age are listed in Table~\ref{table:A1}. The spectral types, effective temperatures, rotation periods, and special features, e.g. binarity, are presented in Table~\ref{table:A2}. The references for all values we used are given below these two tables.

\begin{figure}[th!]
	\centering{
		\includegraphics[angle=-90, width=0.49\textwidth]{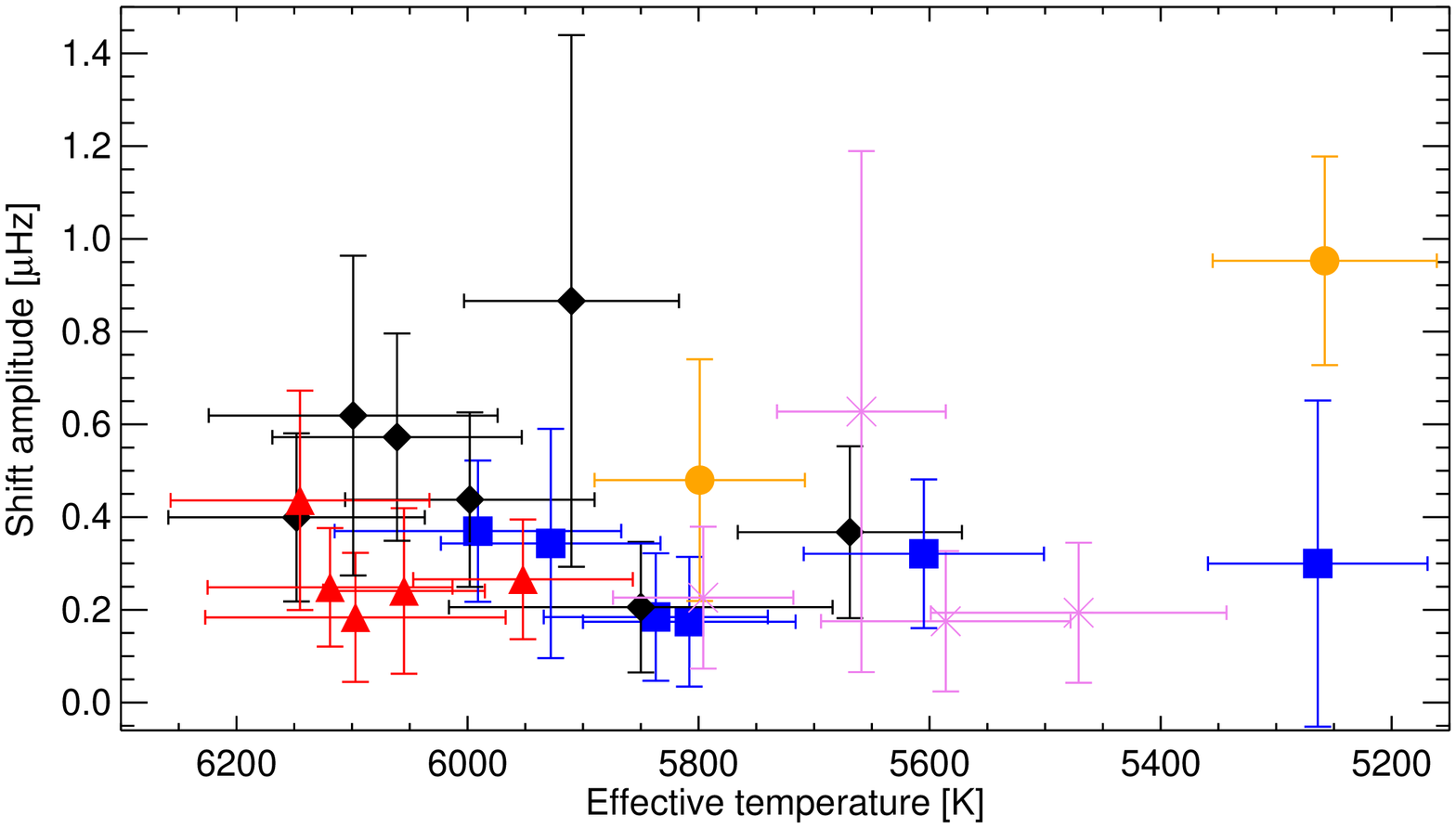}
		\caption{Measured shift amplitudes as a function of effective temperature. Colour and symbol coding: black diamonds represent stars younger than \unit[4]{Gyrs}, red triangles are stars between \unit[4--5]{Gyrs}, orange circles between \unit[5--6]{Gyrs}, blue squares between \unit[6--7]{Gyrs}, and purple asterisks are stars older than \unit[7]{Gyrs}.
			\label{fig:12}}
		\includegraphics[angle=-90, width=0.49\textwidth]{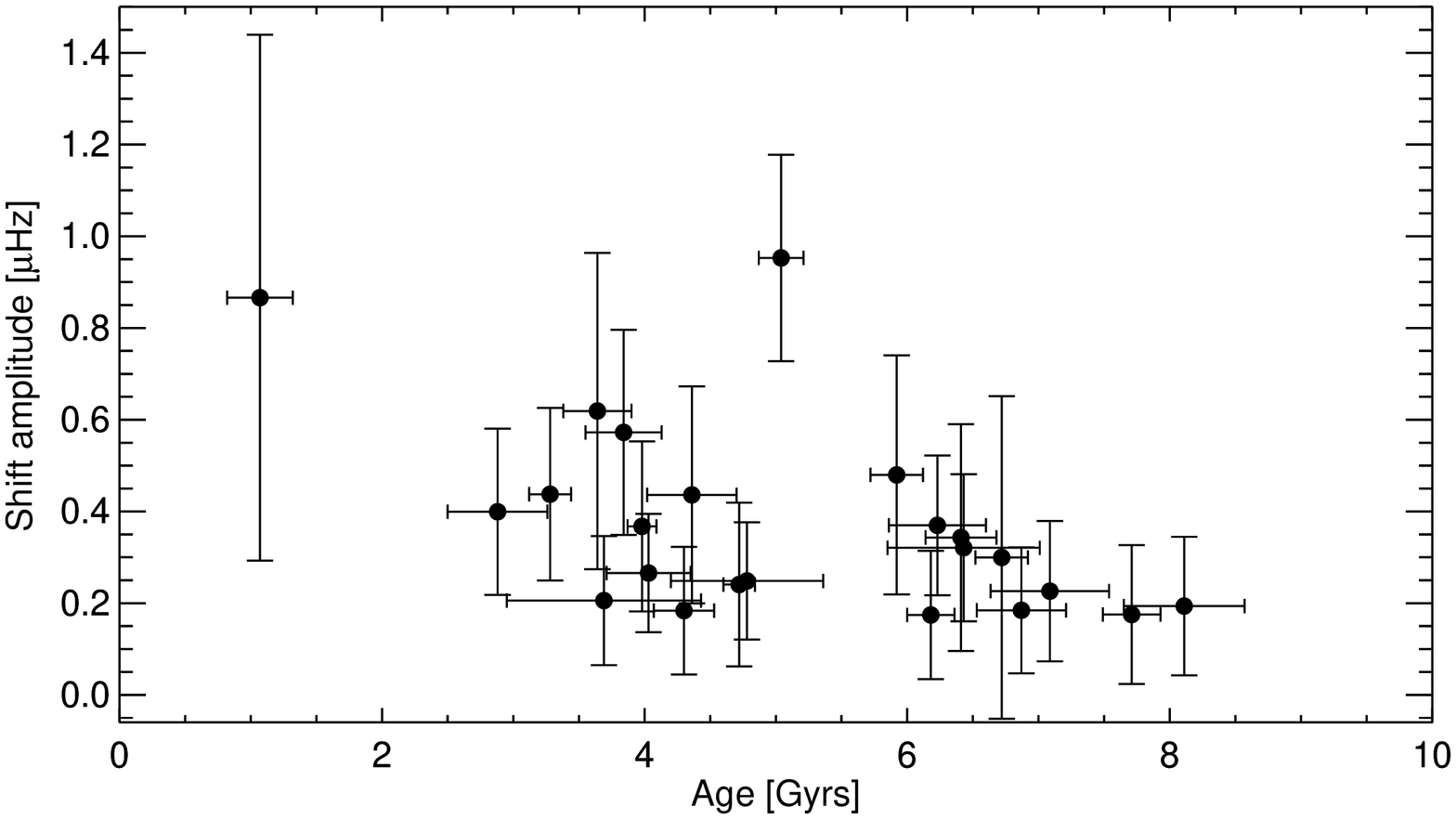}
		\caption{Measured shift amplitudes as a function of stellar age.
			\label{fig:13}}
		\includegraphics[angle=-90, width=0.49\textwidth]{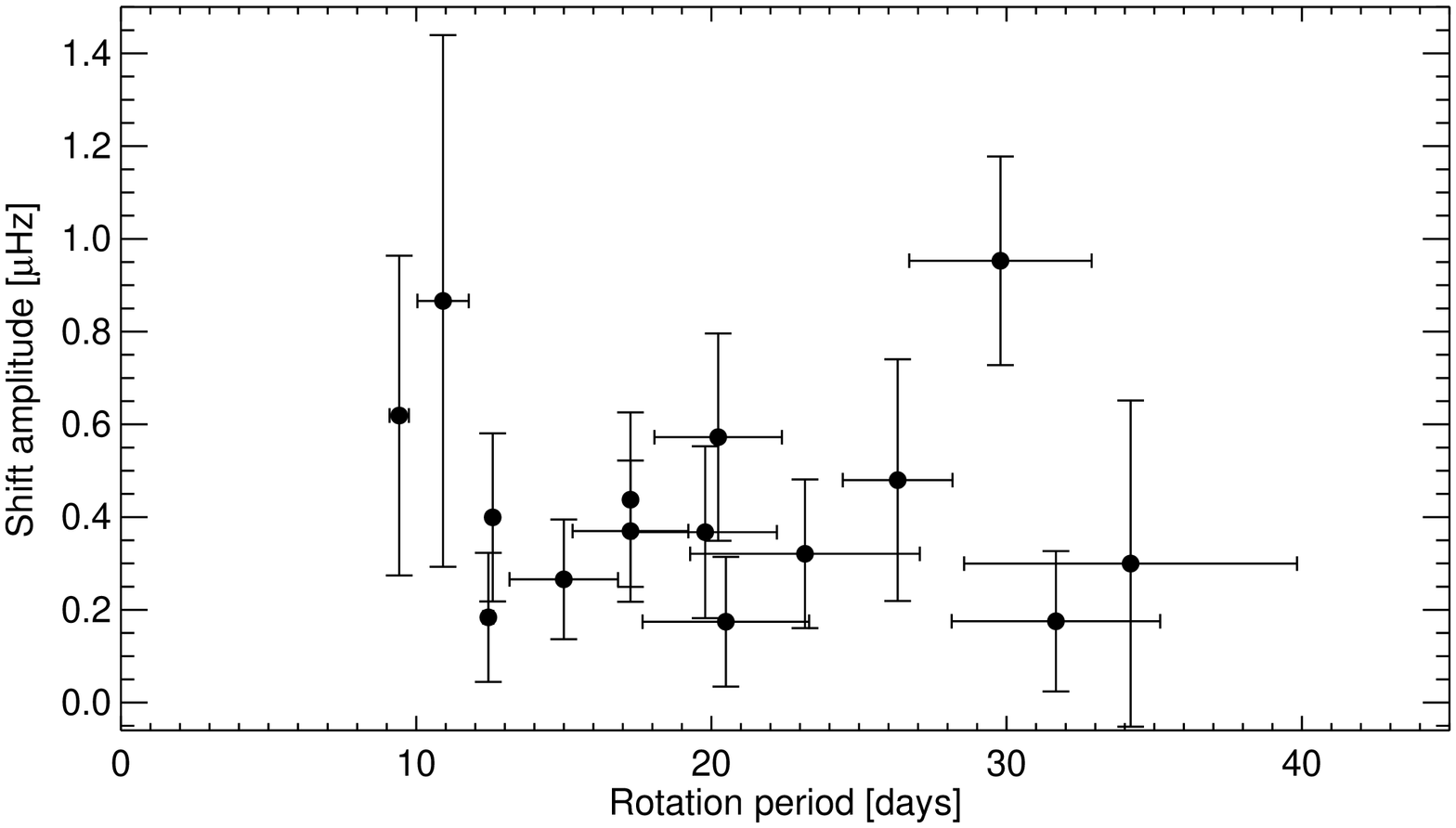}
		\caption{Measured shift amplitudes as a function of rotation period.
			\label{fig:14}}}
\end{figure}

In Fig.~\ref{fig:12} the measured frequency shift amplitudes are plotted as a function of effective temperature. The age of the stars is given by the colour and type of symbol: black diamonds represent stars younger than \unit[4]{Gyrs}, red triangles are stars between \unit[4--5]{Gyrs}, orange circles between \unit[5--6]{Gyrs}, blue squares between \unit[6--7]{Gyrs}, and purple asterisks are stars older than \unit[7]{Gyrs}.

There are two proposed scaling models for the frequency shifts over a stellar activity cycle: \cite{2007MNRAS.377...17C} come to the conclusion that the frequency shifts are directly proportional to the strength of the activity cycle given by $\Delta R_{\text{HK}'}$, which is given by the average fraction of the stellar luminosity that is emitted in the Ca II H and K line cores.
The second model, presented by \cite{2007MNRAS.379L..16M}, assumes that the frequency shifts are proportional to $D/I \cdot\Delta R_{\text{HK}'}$, where $D$ is the depth of the perturbations and $I$ is the mode inertia. In Figure~9 of \cite{2009MNRAS.399..914K}, these two models are compared as a function of effective temperature and for different stellar ages.

The here presented sample of stars does not allow us to decide between the two models. The errors on the frequency shift amplitudes are too large and the sample of 24 stars is too small. However, there appears to be a slight tendency towards greater shift amplitudes for hotter stars. This would support the scaling model by \cite{2007MNRAS.379L..16M}. The outstanding exception from this is KIC~8006161, located at $T_{\text{eff}}=\unit[5258]{K}$ and a shift amplitude of \unit[0.95]{$\mu$Hz}. That KIC~8006161 shows out-of-the-ordinary activity was also noticed by \cite{Karoff21082013}. An expansion of our study to all stars with \textit{Kepler} short cadence data of sufficient length and a reduction of the uncertainty on the frequency shifts, e.g., by peak-bagging of the periodograms of all segments, will help to shed light on the question which of the two models is the correct one.

Figure~\ref{fig:13} shows the frequency shift amplitudes as a function of the stellar ages, which are available in the literature, see Table~\ref{table:A1}. As shown by e.g. \cite{1972ApJ...171..565S} and recently by \cite{2016arXiv160801489S}, the strength of stellar activity is declining over time. Both scaling models for the cycle frequency shifts include the strength of the activity cycle via the quantity $\Delta R_{\text{HK}'}$. It is the youngest star of the investigated sample, KIC~10644253, which shows the second greatest shift amplitude. Stars between \unit[1.07--2.88]{Gyrs} are missing from our sample. We excluded KIC~9025370 from this plot, as its age is only poorly determined. From Fig.~\ref{fig:13}, it can be seen that the amplitude of the frequency shifts is decreasing with stellar age. Again, KIC~8006161, at an age of \unit[5.04]{Gyrs} and a shift amplitude of \unit[0.95]{$\mu$Hz}, is the exception. 

In Fig.~\ref{fig:14} the shift amplitudes are plotted as a function of rotation period. Only 15 stars of the sample have measured rotation periods. As the rotation period is linked to the strength of stellar activity \citep[e.g.][]{1972ApJ...171..565S, Wright2011}, it is straightforward to assume that also the frequency shift amplitudes decrease with increasing rotation period. We note that once more KIC~8006161, found at $P_{\text{rot}}=\unit[29.79]{days}$ and a shift amplitude of $\unit[0.95]{\mu Hz}$, appears to deviate from the behaviour of the remaining sample. Indeed, even though the sample now consists of only 15 stars, there is a trend towards smaller shift amplitudes for longer rotation periods.

\begin{table*} 
\caption{Amplitudes of frequency shifts and their significance.}           
\label{table:2}   
\centering                                   
\begin{tabular}{c c c c ||c c c c} 
\hline\hline
KIC & $A$ (\unit[]{$10^{-7}$Hz}) &$\sigma_{A}$ (\unit[]{$10^{-7}$Hz}) & $A$/$\sigma_{A}$ &  KIC & $A$ (\unit[]{$10^{-7}$Hz})& $\sigma_{A}$ (\unit[]{$10^{-7}$Hz})& $A$/$\sigma_{A}$ \\ 
\hline
3632418 &  3.99 &  1.81 &  2.20 &8228742 &  5.72 &  2.24 &  2.56\\
3656476 &  1.75 &  1.51 &  1.16 &8379927 &  4.38 &  1.88 &  2.32\\ 
4914923 &  1.74 &  1.40 &  1.25 &8760414 &  2.06 &  1.41 &  1.46\\ 
5184732 &  3.67 &  1.85 &  1.98 &9025370 &  6.27 &  5.62 &  1.12\\ 
6106415 &  2.41 &  1.79 &  1.35 &9955598 &  3.00 &  3.52 &  0.85\\ 
6116048 &  3.70 &  1.52 &  2.43 &10018963 &  4.36 &  2.37 &  1.84\\ 
6603624 &  1.94 &  1.51 &  1.28 &10516096 &  3.43 &  2.47 &  1.39\\ 
6933899 &  1.84 &  1.37 &  1.34 &10644253 &  8.66 &  5.73 &  1.51\\ 
7680114 &  4.80 &  2.61 &  1.84 &10963065 &  1.84 &  1.39 &  1.32\\ 
7976303 &  2.49 &  1.28 &  1.95 &11244118 &  3.21 &  1.60 &  2.00\\  
8006161 &  9.53 &  2.25 &  4.23 &11295426 &  2.26 &  1.53 &  1.48\\  
low     &   5.65 &  2.77 &  2.04&12009504 &  6.19 &  3.45 &  1.79\\                         
mid     &  10.81 &  2.96 &  3.65&12258514 &  2.66 &  1.29 &  2.06 \\                        
high    &  12.61 &  4.81 &  2.62& \\
\hline        
\end{tabular}
\end{table*}         

\begin{table*} 
\caption{Correlation coefficients of frequency shifts and p-mode envelope height variations and their p-values}           
\label{table:3}   
\centering                                   
\begin{tabular}{c c c||c c c} 
\hline\hline
KIC & correlation & p &KIC & correlation & p\\ 
\hline                                 
3632418 &  -0.14 &   0.76& 8228742 &  -0.29 &   0.39\\
3656476 &   0.30 &   0.62& 8379927 &   0.43 &   0.34\\
4914923 &   0.49 &   0.33& 8760414 &  -0.54 &   0.22\\
5184732 &  -0.44 &   0.20& 9025370 &  -0.25 &   0.45\\
6106415 &   0.60 &   0.21& 9955598 &  -0.75 &   0.05\\
6116048 &   0.75 &   0.05&10018963 &   0.46 &   0.15\\
6603624 &   0.52 &   0.10&10516096 &  -0.26 &   0.47\\
6933899 &  -0.57 &   0.18&10644253 &  -0.93 &   $3\cdot10^{-3}$\\
7680114 &   0.60 &   0.21&10963065 &  -0.26 &   0.62\\
7976303 &  -0.29 &   0.53&11244118 &  -0.61 &   0.05\\
8006161 &  -0.93 &   $4\cdot10^{-5}$&11295426 &   0.36 &   0.43\\
 low    &  -0.92 &   $7\cdot10^{-5}$&12009504 &  -0.13 &   0.71\\
 mid    &  -0.93 &   $4\cdot10^{-5}$&12258514 &  -0.50 &   0.25\\
 high   &  -0.87 &   $5\cdot10^{-5}$&&&\\
\hline 
\end{tabular}
\end{table*}

\begin{figure}
	\centering{
		\includegraphics[angle=-90,width=0.49\textwidth]{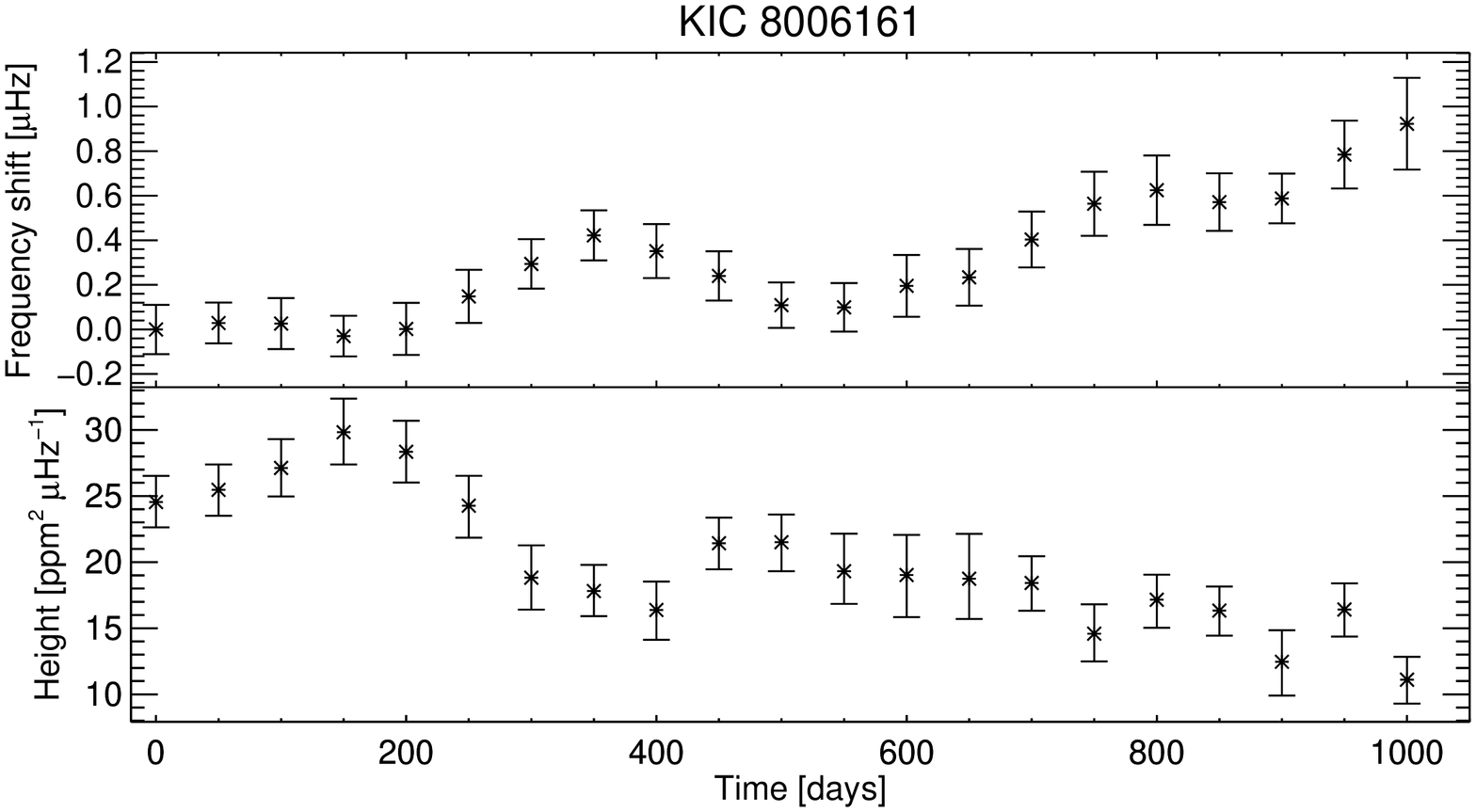}
		\includegraphics[angle=-90,width=0.49\textwidth]{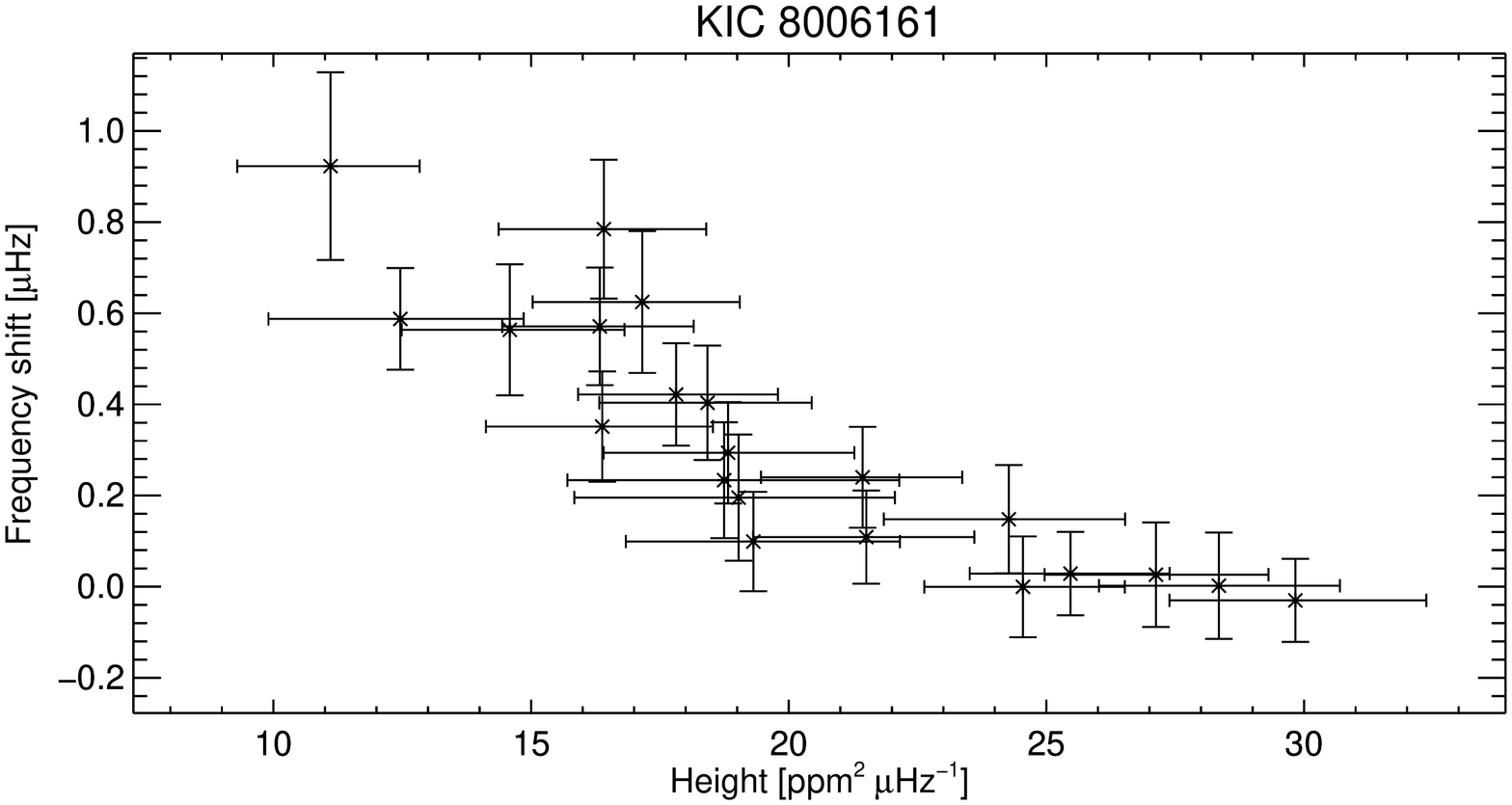}
		\caption{\textit{Top panel:} frequency shifts for KIC~8006161 for the frequency range between \unit[2800--4400]{$\mu$Hz} (top half) and height of the p-mode envelope of KIC~8006161 (bottom half) as a function of time. \textit{Bottom panel:} frequency shift as a function of height of the p-mode envelope. \label{fig:4}}}
\end{figure} 

\subsection{KIC~8006161}
KIC~8006161 is a solar-like star with a mass of $\unit[1.04\pm 0.02]{M_{\odot}}$, a radius of $\unit[0.947\pm 0.007]{R_{\odot}}$, and an age of $\unit[5.04\pm 0.17]{Gyr}$ \citep[model values from][]{2014ApJS..214...27M}. It is a G8V star \citep{2013MNRAS.434.1422M} and its rotation period is $\unit[29.79\pm 3.09]{d}$ \citep{2014A&A...572A..34G}. This star shows the most significant and greatest frequency shifts of our sample with $A$/$\sigma_{A}$=4.23. The variation of the p-mode frequencies in the range between \unit[2800--4400]{$\mu$Hz} is shown in the top half of the top panel of Fig.~\ref{fig:4}. From the zero level at the beginning of the time series, the mode frequencies increase by $\approx$\unit[0.95]{$\mu$Hz} until the end of the time series. The error bars on the frequency shifts, which result from our resampling approach described in Sec.~\ref{sec:resampling}, are only of the order of $\approx$\unit[0.1]{$\mu$Hz}. This small error is due to the narrow width of the p modes in the LS-periodogram of KIC~8006161 and the broad frequency range where modes can be detected. The shifts reach a local maximum at around day 350, followed by a brief decrease. The shifts then increase again towards the end of the time series. This behaviour is reminiscent of the double maximum of the Sun's activity cycle. Unfortunately, the available time series does not cover a full activity cycle of KIC~8006161. Therefore, we can only limit the cycle length to $P_{\mathrm{cyc}}> \unit[1147]{d}$. 

The variation of the mode heights are depicted in the bottom half of the top panel of Fig.~\ref{fig:4}. We find a clear anti-correlation between the shift of mode frequencies and the variation of mode heights with a correlation value of $\rho=-0.93$ with a p value of $4\cdot 10^{-5}$. This linear anti-correlation is clearly seen in the bottom panel of Fig.~\ref{fig:4}, where the frequency shifts are plotted as a function of height of the mode envelope. There is a slight departure from the linear relationship towards higher values of height, which correspond to a state of lower magnetic activity. This behaviour, albeit less pronounced, is also observed for KIC~5184732, see Fig.~\ref{fig:C1}.
As discussed before, the other parameters of the background fit show no correlations or anti-correlations, cf. Table~\ref{table:B11}, where the only exception is a correlation between the two granulation time scales with  $\rho = 0.74$ and $p=0.01$. We therefore conclude that the observed variations in mode frequencies and mode heights are due to magnetic activity on KIC~8006161.

\cite{Karoff21082013} found that the $S$ index of KIC~8006161, a measure for stellar chromospheric activity, is comparable to that of the Sun. They also found that, at the same time, KIC~8006161 exhibits an unusually low excess flux. This flux is associated with magnetic sources on the stellar surface. They argue that due to this odd combination of activity indices, KIC~8006161 might be in a minimum state of activity. Their measurements were taken between 2010 and 2012, which corresponds to \textit{Kepler} quarters 4 through 15, and therefore largely overlap with the time series we used for our analyses, which encompasses Q5-17.2.\footnote{\textit{Kepler} quarter 4 started December 19, 2009; quarter 15 ended January 11, 2013. } In the study of \cite{2014A&A...572A..34G}, KIC~8006161 was found to be more active than the Sun. \citeauthor{2014A&A...572A..34G} used the standard deviation of the time series, a measure of photospheric activity, to determine the activity level.

Since this star is very similar to the Sun regarding its fundamental parameters, our findings for the activity related shift of its p-mode frequencies, and especially the strong anti-correlation between mode frequencies and heights of the p-mode envelope, suggest that the magnetic activity on KIC~8006161 is somewhat stronger than on the Sun, following the results presented by \cite{2014A&A...572A..34G}. 

\begin{figure*}[h!]
	\centering{
		\includegraphics[width=1\textwidth]{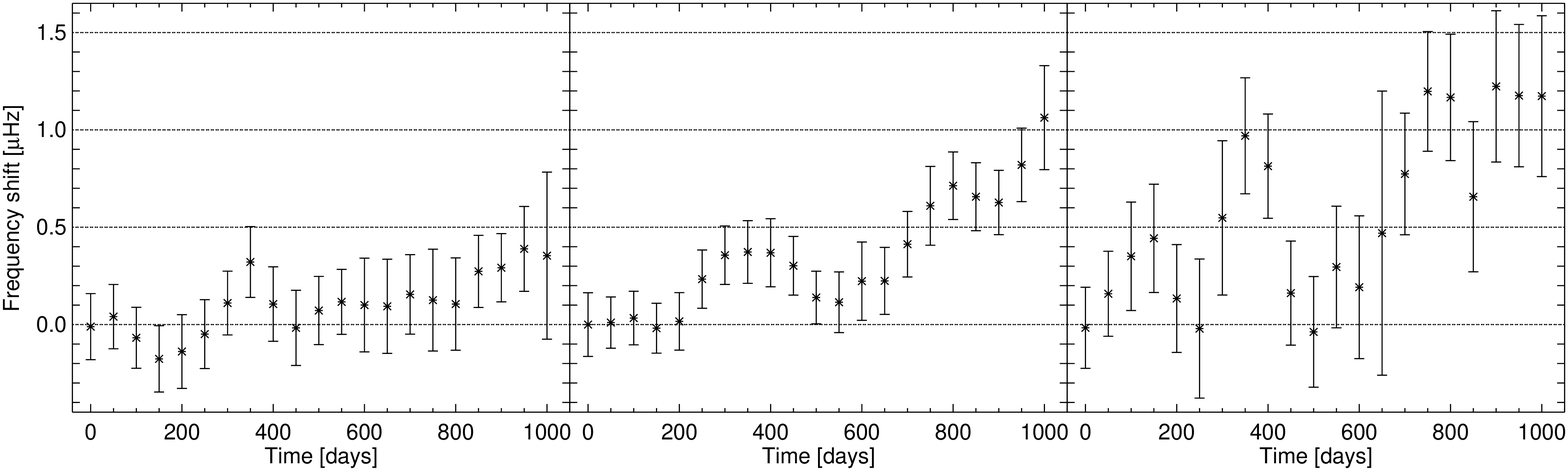}
		\caption{Frequency shifts of KIC~8006161 for three frequency ranges as a function of time: \unit[2800--3300]{$\mu$Hz} ({\it left panel}), \unit[3300--3800]{$\mu$Hz} ({\it middle panel}), and \unit[3800--4400]{$\mu$Hz} ({\it right panel}).}
		\label{fig:6}}
\end{figure*}

For the Sun, it was shown that the amplitude of frequency shifts over the solar cycle are larger for higher frequencies \citep[e.g. ][]{1990Natur.345..779L,  2001A&A...379..622J, 2007ApJ...659.1749C,2015A&A...578A.137S}. We looked for a similar behaviour for KIC~8006161 by dividing the frequency range where p modes are visible in the periodogram into three smaller sections: a low frequencies range between \unit[2800--3300]{$\mu$Hz}, an intermediate range with frequencies \unit[3300--3800]{$\mu$Hz}, and high frequency range between \unit[3800--4400]{$\mu$Hz}. The results are shown in the three panels of Fig.~\ref{fig:6}. As in the solar case, the frequency shifts increase with increasing mode frequency: the modes in the low frequency range vary by only $\approx$\unit[0.6]{$\mu$Hz}, while the modes in the intermediate range shift $\approx$\unit[1.1]{$\mu$Hz}. The modes in the high frequency range even shift by as much as $\approx$\unit[1.2]{$\mu$Hz}. The measuring of the frequency shifts via the cross-correlation and the estimation of the error bars on them are sensitive to the mode widths and amplitudes. Due to the fact that mode widths increase with frequency and mode amplitudes follow a Gaussian envelope centred around the frequency of maximum power, the estimated error bars on the frequency shifts are smallest in the intermediate range and largest in the high frequency range (cf. Fig.~\ref{fig:6}, Table~\ref{table:2}, and \cite{Regulo2016})  The overall course of the variation of the frequencies is similar for all three frequency ranges. The characteristics of the variation for the entire frequency range are dominated by the variation of the modes in the intermediate frequency range, since these modes have the largest amplitudes and thus contribute strongest to the cross-correlation. As for the entire frequency range, the frequency shifts of these sub-ranges are strongly anti-correlated to the variation of the mode heights with correlation coefficients of -0.92 (low frequency range), -0.93 (intermediate frequency range), and -0.87 (high frequency range). This is only the fourth star for which a frequency dependence of the activity related shift of p-mode frequencies has been detected after the Sun, HD~49933 \citep{2011A&A...530A.127S}, and KIC~10644253 \citep{2016A&A...589A.118S}.

\subsection{KIC~10644253}
This star has a mass of $\unit[1.13\pm 0.05]{M_{\odot}}$, a radius of $\unit[1.108\pm 0.016]{R_{\odot}}$ \citep{2014A&A...562A.124M}, and an age of $\unit[1.07\pm 0.25]{Gyr}$ \citep[model values from][]{2014ApJS..214...27M}. It is a G0V star \citep{2013MNRAS.434.1422M} with a surface rotation rate of $\unit[10.91\pm 0.87]{d}$ \citep{2014A&A...572A..34G}. 

\begin{figure}
	\centering{
		\includegraphics[angle=-90,width=0.49\textwidth]{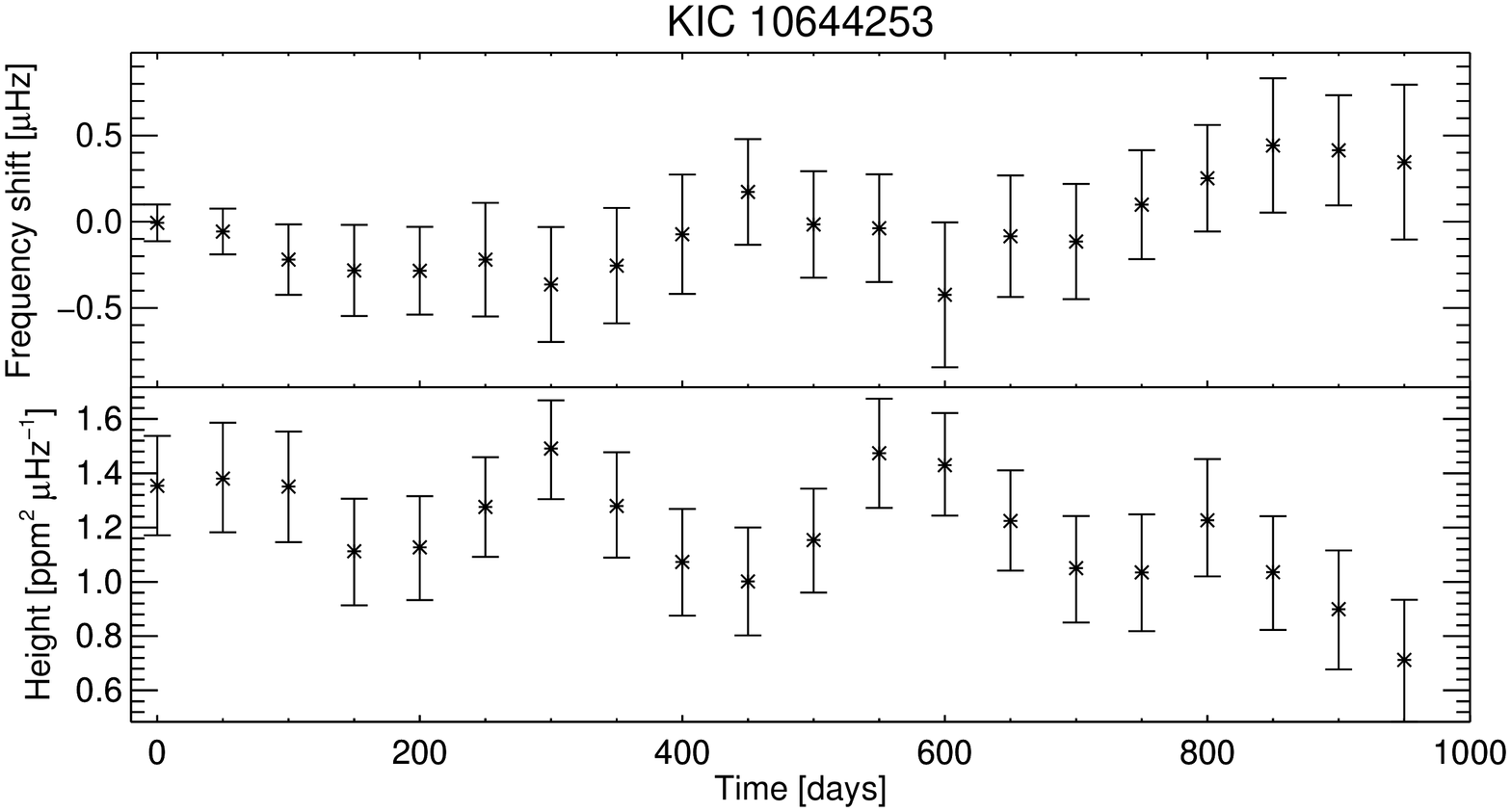}
		\includegraphics[angle=-90,width=0.49\textwidth]{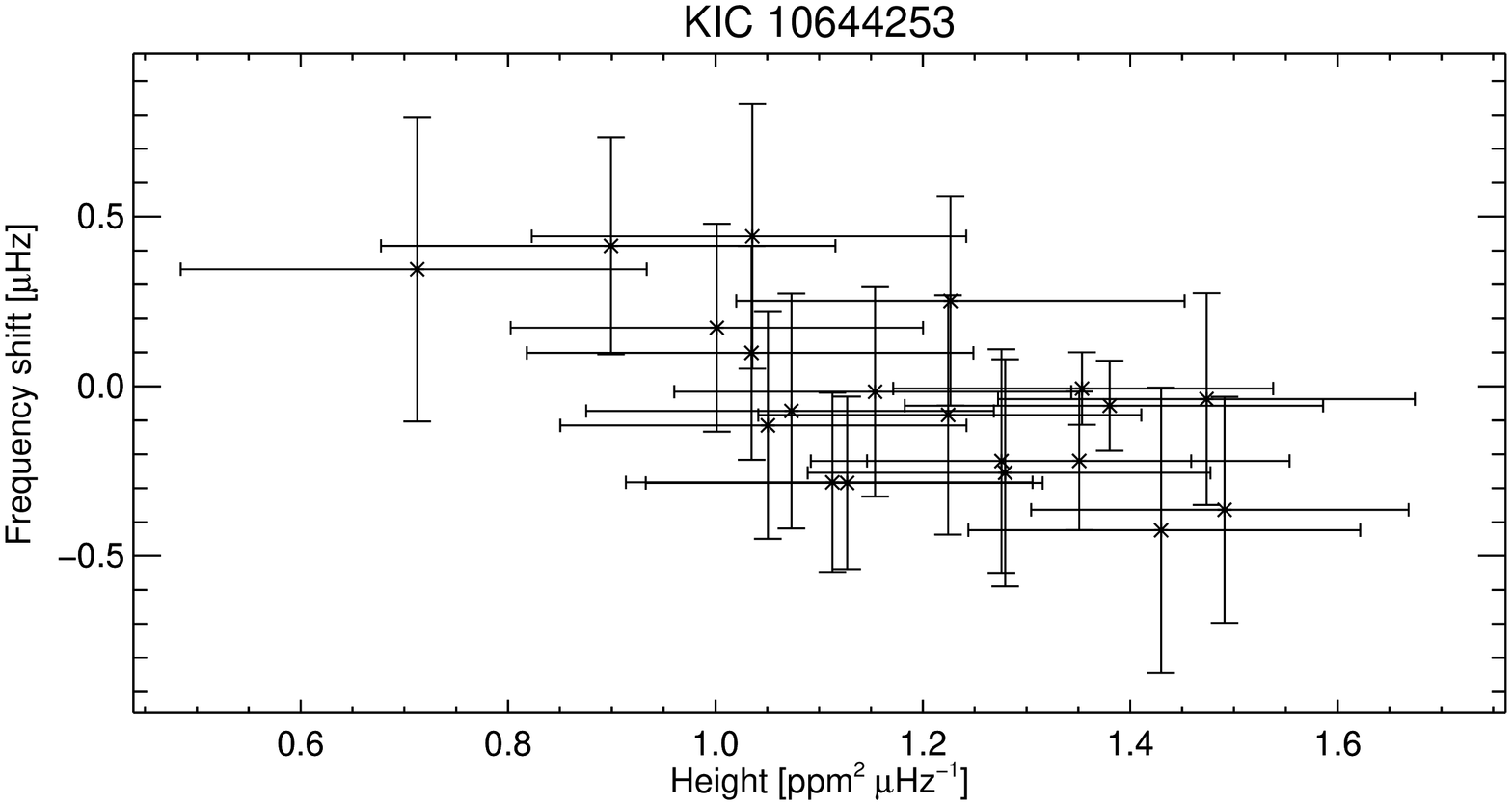}
		\caption{\textit{Top panel:} frequency shifts for KIC~10644253 for the frequency range between \unit[2450--3350]{$\mu$Hz} (top half) and height of the p-mode envelope of KIC~10644253 as a function of time (bottom half). \textit{Bottom panel:} frequency shift as a function of height of the p-mode envelope.}
		\label{fig:7}}
\end{figure}

The frequency shifts we found for this star follow a cycle-like pattern, see top half of the top panel of Fig.~\ref{fig:7}, and have the second greatest peak-to-valley amplitude in our sample of stars with a value of \unit[$\approx$0.87]{$\mu$Hz}. The heights of the mode envelope, see bottom half of the top panel of Fig.~\ref{fig:7}, are strongly anti-correlated to the frequency shifts with a correlation coefficient of $\rho=-0.93$ and a p value of $3\cdot 10^{-3}$. This linear anti-correlation is seen in the bottom panel of Fig.~\ref{fig:7}, where the frequency shifts of KIC~10644253 are plotted as a function of height of the mode envelope.  

In their study of 22 F stars \citeauthor{2014A&A...562A.124M} investigated the long cadence \textit{Kepler} time series of KIC~10644253 regarding signs of magnetic activity and activity cycles. In their magnetic proxy for this star, they found indications of two activity cycles during the first 16 quarters of \textit{Kepler} data. They use the scale average variance, which is a projection of the wavelet power spectrum in a small region around the stellar rotation period onto the time axis, as a magnetic proxy. The overall magnetic activity of this star was among the highest from their sample of stars. \citeauthor{2014A&A...562A.124M} spotted the strongest peak in their cycle proxy of KIC~10644253 during Quarter 10, which corresponds to a time during day $\approx$400-450 for our data set. This was recently confirmed by \cite{2016A&A...589A.118S} who find a positive shift in p-mode frequencies and a large stellar magnetic variability in this time period. Indeed, in our analysis the frequency shifts are at a local maximum and the mode heights are at a local minimum during this period. The local minima found in the frequency shifts before and after this maximum are also reflected in the magnetic proxy of \cite{2014A&A...562A.124M} and the frequency shifts and stellar magnetic variability of \cite{2016A&A...589A.118S}. 

Interestingly, we find strong positive frequency shifts and small mode heights, which indicates strong magnetic activity, during the last 250 days of the time series. This is not reflected in the proxy of \cite{2014A&A...562A.124M} or in the stellar magnetic variability of \cite{2016A&A...589A.118S}. During the same time period, there is a discrepancy between the mode frequency shifts and the photometric activity proxy presented by \cite{2016A&A...589A.118S}.  

There are different values for the inclination angle of KIC~10644253 in the literature. \cite{2014A&A...562A.124M} state $i=\unit[43.44\pm 14.48]{^{\circ}}$, whereas \cite{2016A&A...589A.118S} find values of  $i\approx\unit[11]{^{\circ}}$ derived from spectroscopic measurements\footnote{For the weighted average of the inclination angle, we estimated the error on this value to be $\unit[\pm 11]{^{\circ}}$}, $i=\unit[7\substack{+9\\-9}]{^{\circ}}$ derived from the stellar radius, rotation period and the asteroseismic $v\sin i$, and $i=\unit[48\substack{+11\\-9}]{^{\circ}}$ derived with another value of the spectroscopic $v\sin i$. Therefore, the inclination of the rotation axis of KIC~10644253 to the line of sight seems to be rather low, with a weighted average value of $i=\unit[23\pm 6]{^{\circ}}$. 

Due to this low inclination, it is conceivable that the regions of high activity are largely confined to the nearly out-of-sight hemisphere of the star during the last 250 days of the time series, where the discrepancy between the activity proxies are found. The global, low degree modes we use for our analyses are susceptible to magnetic activity throughout the star. Therefore, the frequency shifts and variation of the mode heights might reveal signatures of magnetic activity which are hidden to photometry based techniques like that of \cite{2014A&A...562A.124M}.

\section{Summary and conclusion}
\label{sec:4}
We investigated the \textit{Kepler} data of 24 solar-like stars for the temporal variation of p-mode frequencies and height of the p-mode envelope. For this, we split the stars' time series into shorter segments.  The shifts of p-mode frequencies were measured by using a cross-correlation analysis of the segments' periodograms. The errors on the frequency shifts were estimated with a resampling approach. This approach will be further investigated and compared to similar approaches used in helio- and asteroseismology in a future study. The temporal variation of the heights of the p-mode envelope, the high frequency noise, and two granulation time scales were measured by fitting the segments' periodograms. 

We found significant variation of p-mode frequencies (above 1-$\sigma$), which can be signatures of stellar magnetic activity, during the observed period for 23 of the 24 stars of our sample. A cycle-like behaviour could be spotted in the frequency shifts of several stars (e.g. KIC~3632418, KIC~6933899, KIC~8760414, KIC~10644253). The correlation between the frequency shifts and the variation of mode heights was found to be strongly negative (below -0.5) for six stars (KIC~6933899, KIC~8006161, KIC~8760414, KIC~9955598, KIC~10644253, KIC~11244118, and KIC~12258514). Together, the significant shift in p-mode frequencies and this anti-correlation strongly supports asteroseismic detection of solar-like magnetic activity on these stars. Due to the limited length of the time series, we refrain from assigning cycle periods to any of these stars. However, we can limit the cycle period of KIC~8006161 to $P_{\text{cyc}}>\unit[1147]{d}$ and state that more detailed and longer observations of the respective stars with evidence for cyclic activity could shed light on the period lengths of their magnetic activity cycles. Moreover, there are hints of sporadic activity on some of the investigated stars. The frequency shifts and mode heights of KIC~10018963 and KIC~11295426 exhibit significant temporal variations which might be caused by such non-cyclic sporadic stellar magnetic activity. 
It is also noteworthy that the correlation of frequency shifts and mode heights is strongly negative, following the solar reference case, for only 7 out of 24 stars, and even strongly positive for a couple of stars. Whether there is something to be learned from this about the connection of stellar magnetic activity or the dynamos at work in stars slightly different than the Sun to the temporal evolution of the mode frequency shifts and the mode heights is currently subject of further investigations.

The background parameters -- specifically mode heights, granulation timescales, and high frequency noise level -- and the frequency shifts do not show any systematic correlation across the sample of stars. We found very different correlations between the investigated parameters for the two stars, which we discussed in detail, KIC~8006161 and KIC~10644253, both of which show clear signs of magnetic activity. It will need an expanded sample of stars with a broader range of fundamental parameters, optimally all available \textit{Kepler} stars with long enough short cadence coverage, to find the cause for this varying behaviour of the oscillation and background parameters.

The star KIC~10644253 exhibited shifts in p-mode frequencies and changes of the mode heights which might be caused by two stellar cycles during the available data set, as was also recently discovered by \cite{2016A&A...589A.118S}. When compared to the activity indices of \cite{2014A&A...562A.124M} and \cite{2016A&A...589A.118S}, we conclude that we found indications of stellar activity on the nearly out-of-sight hemisphere of this star during the last 250 days of \textit{Kepler} observations. 

For the solar-like star KIC~8006161 we were able to investigate the frequency dependence of the mode frequency shifts. We found that the shifts are larger for p modes of higher frequencies, just like in the solar case \citep{1998A&A...329.1119J}, for HD 49933 \citep{2011A&A...530A.127S}, and for KIC~10644253 \citep{2016A&A...589A.118S}. It is only the fourth star for which this frequency dependence of the activity related frequency shifts could be measured. KIC~8006161 is very similar to the Sun regarding its mass, age, and rotation period. It is therefore noteworthy that the frequency shifts are larger than for the Sun, even though the time series does clearly not cover a full cycle. This raises the question what sets this star's magnetic cycle apart from the Sun's cycle. Further investigation of KIC~8006161 and its magnetic activity cycle is therefore worthwhile. When examining the frequency shifts as a function of mode height there is a departure from the negative linear relationship between these two quantities for higher values of mode height. This is also observed for KIC~5184732 and raises the question whether this observed pattern is part of a hysteresis curve of the frequency shifts over a stellar cycle, similar as observed for the Sun \citep{1998A&A...329.1119J}.

We found that the frequency shift amplitude decreases with stellar age and rotation period. The results for our sample did not enable us to reject one of the two opposing scaling models for cycle frequency shift amplitudes by \cite{2007MNRAS.377...17C} and \cite{2007MNRAS.379L..16M}, while the data seems to show a slight tendency towards the scaling of \cite{2007MNRAS.379L..16M}.

\begin{acknowledgements}
The research leading to these results received funding from the European Research Council under the European Union’s Seventh Framework Program (FP/2007-2013)/ERC Grant Agreement no. 307117. GRD acknowledges the support of the UK Science and Technology Facilities Council (STFC). This research has made use of the SIMBAD database, operated at CDS, Strasbourg, France.
\end{acknowledgements}
\bibliographystyle{aa}
\bibliography{references} 

\begin{thebibliography}{67}
\expandafter\ifx\csname natexlab\endcsname\relax\def\natexlab#1{#1}\fi

\bibitem[{{Appourchaux} {et~al.}(2012){Appourchaux}, {Chaplin},
  {Garc{\'{\i}}a}, {Gruberbauer}, {Verner}, {Antia}, {Benomar}, {Campante},
  {Davies}, {Deheuvels}, {Handberg}, {Hekker}, {Howe}, {R{\'e}gulo},
  {Salabert}, {Bedding}, {White}, {Ballot}, {Mathur}, {Silva Aguirre},
  {Elsworth}, {Basu}, {Gilliland}, {Christensen-Dalsgaard}, {Kjeldsen},
  {Uddin}, {Stumpe}, \& {Barclay}}]{2012A&A...543A..54A}
{Appourchaux}, T., {Chaplin}, W.~J., {Garc{\'{\i}}a}, R.~A., {et~al.} 2012,
  \aap, 543, A54

\bibitem[{{Auvergne} {et~al.}(2009){Auvergne}, {Bodin}, {Boisnard}, {Buey},
  {Chaintreuil}, {Epstein}, {Jouret}, {Lam-Trong}, {Levacher}, {Magnan},
  {Perez}, {Plasson}, {Plesseria}, {Peter}, {Steller}, {Tiph{\`e}ne}, {Baglin},
  {Agogu{\'e}}, {Appourchaux}, {Barbet}, {Beaufort}, {Bellenger}, {Berlin},
  {Bernardi}, {Blouin}, {Boumier}, {Bonneau}, {Briet}, {Butler}, {Cautain},
  {Chiavassa}, {Costes}, {Cuvilho}, {Cunha-Parro}, {de Oliveira Fialho},
  {Decaudin}, {Defise}, {Djalal}, {Docclo}, {Drummond}, {Dupuis}, {Exil},
  {Faur{\'e}}, {Gaboriaud}, {Gamet}, {Gavalda}, {Grolleau}, {Gueguen},
  {Guivarc'h}, {Guterman}, {Hasiba}, {Huntzinger}, {Hustaix}, {Imbert},
  {Jeanville}, {Johlander}, {Jorda}, {Journoud}, {Karioty}, {Kerjean},
  {Lafond}, {Lapeyrere}, {Landiech}, {Larqu{\'e}}, {Laudet}, {Le Merrer},
  {Leporati}, {Leruyet}, {Levieuge}, {Llebaria}, {Martin}, {Mazy}, {Mesnager},
  {Michel}, {Moalic}, {Monjoin}, {Naudet}, {Neukirchner}, {Nguyen-Kim},
  {Ollivier}, {Orcesi}, {Ottacher}, {Oulali}, {Parisot}, {Perruchot},
  {Piacentino}, {Pinheiro da Silva}, {Platzer}, {Pontet}, {Pradines},
  {Quentin}, {Rohbeck}, {Rolland}, {Rollenhagen}, {Romagnan}, {Russ}, {Samadi},
  {Schmidt}, {Schwartz}, {Sebbag}, {Smit}, {Sunter}, {Tello}, {Toulouse},
  {Ulmer}, {Vandermarcq}, {Vergnault}, {Wallner}, {Waultier}, \&
  {Zanatta}}]{2009A&A...506..411A}
{Auvergne}, M., {Bodin}, P., {Boisnard}, L., {et~al.} 2009, \aap, 506, 411

\bibitem[{{Baglin} {et~al.}(2006){Baglin}, {Auvergne}, {Barge}, {Deleuil},
  {Catala}, {Michel}, {Weiss}, \& {COROT Team}}]{2006ESASP1306...33B}
{Baglin}, A., {Auvergne}, M., {Barge}, P., {et~al.} 2006, in ESA Special
  Publication, Vol. 1306, The CoRoT Mission Pre-Launch Status - Stellar
  Seismology and Planet Finding, ed. M.~{Fridlund}, A.~{Baglin}, J.~{Lochard},
  \& L.~{Conroy}, 33

\bibitem[{{Baliunas} {et~al.}(1995){Baliunas}, {Donahue}, {Soon}, {Horne},
  {Frazer}, {Woodard-Eklund}, {Bradford}, {Rao}, {Wilson}, {Zhang}, {Bennett},
  {Briggs}, {Carroll}, {Duncan}, {Figueroa}, {Lanning}, {Misch}, {Mueller},
  {Noyes}, {Poppe}, {Porter}, {Robinson}, {Russell}, {Shelton}, {Soyumer},
  {Vaughan}, \& {Whitney}}]{1995ApJ...438..269B}
{Baliunas}, S.~L., {Donahue}, R.~A., {Soon}, W.~H., {et~al.} 1995, \apj, 438,
  269

\bibitem[{{Bartlett}(1978)}]{bartlett}
{Bartlett}, M.~S. 1978, An Introduction to Stochastic Processes: With Special
  Reference to Methods and Applications, 3rd edn. (Cambridge University Press)

\bibitem[{{Basu} {et~al.}(2012){Basu}, {Broomhall}, {Chaplin}, \&
  {Elsworth}}]{2012ApJ...758...43B}
{Basu}, S., {Broomhall}, A.-M., {Chaplin}, W.~J., \& {Elsworth}, Y. 2012, \apj,
  758, 43

\bibitem[{{Beeck} {et~al.}(2013){Beeck}, {Cameron}, {Reiners}, \&
  {Sch{\"u}ssler}}]{2013A&A...558A..49B}
{Beeck}, B., {Cameron}, R.~H., {Reiners}, A., \& {Sch{\"u}ssler}, M. 2013,
  \aap, 558, A49

\bibitem[{Borucki {et~al.}(2010)Borucki, Koch, Basri, Batalha, Brown, Caldwell,
  Caldwell, Christensen-Dalsgaard, Cochran, DeVore, Dunham, Dupree, Gautier,
  Geary, Gilliland, Gould, Howell, Jenkins, Kondo, Latham, Marcy, Meibom,
  Kjeldsen, Lissauer, Monet, Morrison, Sasselov, Tarter, Boss, Brownlee, Owen,
  Buzasi, Charbonneau, Doyle, Fortney, Ford, Holman, Seager, Steffen, Welsh,
  Rowe, Anderson, Buchhave, Ciardi, Walkowicz, Sherry, Horch, Isaacson,
  Everett, Fischer, Torres, Johnson, Endl, MacQueen, Bryson, Dotson, Haas,
  Kolodziejczak, Van~Cleve, Chandrasekaran, Twicken, Quintana, Clarke, Allen,
  Li, Wu, Tenenbaum, Verner, Bruhweiler, Barnes, \& Prsa}]{Borucki977}
Borucki, W.~J., Koch, D., Basri, G., {et~al.} 2010, Science, 327, 977

\bibitem[{Brandenburg {et~al.}(1998)Brandenburg, Saar, \&
  Turpin}]{1538-4357-498-1-L51}
Brandenburg, A., Saar, S.~H., \& Turpin, C.~R. 1998, \apjl, 498, L51

\bibitem[{Broomhall {et~al.}(2015)Broomhall, Pugh, \&
  Nakariakov}]{Broomhall20152706}
Broomhall, A.-M., Pugh, C., \& Nakariakov, V. 2015, Advances in Space Research,
  56, 2706 , advances in Solar Physics

\bibitem[{{Bruntt} {et~al.}(2012){Bruntt}, {Basu}, {Smalley}, {Chaplin},
  {Verner}, {Bedding}, {Catala}, {Gazzano}, {Molenda-{\.Z}akowicz}, {Thygesen},
  {Uytterhoeven}, {Hekker}, {Huber}, {Karoff}, {Mathur}, {Mosser},
  {Appourchaux}, {Campante}, {Elsworth}, {Garc{\'{\i}}a}, {Handberg},
  {Metcalfe}, {Quirion}, {R{\'e}gulo}, {Roxburgh}, {Stello},
  {Christensen-Dalsgaard}, {Kawaler}, {Kjeldsen}, {Morris}, {Quintana}, \&
  {Sanderfer}}]{2012MNRAS.423..122B}
{Bruntt}, H., {Basu}, S., {Smalley}, B., {et~al.} 2012, \mnras, 423, 122

\bibitem[{{Chaplin} {et~al.}(2014){Chaplin}, {Basu}, {Huber}, {Serenelli},
  {Casagrande}, {Silva Aguirre}, {Ball}, {Creevey}, {Gizon}, {Handberg},
  {Karoff}, {Lutz}, {Marques}, {Miglio}, {Stello}, {Suran}, {Pricopi},
  {Metcalfe}, {Monteiro}, {Molenda-{\.Z}akowicz}, {Appourchaux},
  {Christensen-Dalsgaard}, {Elsworth}, {Garc{\'{\i}}a}, {Houdek}, {Kjeldsen},
  {Bonanno}, {Campante}, {Corsaro}, {Gaulme}, {Hekker}, {Mathur}, {Mosser},
  {R{\'e}gulo}, \& {Salabert}}]{2014ApJS..210....1C}
{Chaplin}, W.~J., {Basu}, S., {Huber}, D., {et~al.} 2014, \apjs, 210, 1

\bibitem[{{Chaplin} {et~al.}(2007{\natexlab{a}}){Chaplin}, {Elsworth},
  {Houdek}, \& {New}}]{2007MNRAS.377...17C}
{Chaplin}, W.~J., {Elsworth}, Y., {Houdek}, G., \& {New}, R.
  2007{\natexlab{a}}, \mnras, 377, 17

\bibitem[{{Chaplin} {et~al.}(2007{\natexlab{b}}){Chaplin}, {Elsworth},
  {Miller}, {Verner}, \& {New}}]{2007ApJ...659.1749C}
{Chaplin}, W.~J., {Elsworth}, Y., {Miller}, B.~A., {Verner}, G.~A., \& {New},
  R. 2007{\natexlab{b}}, \apj, 659, 1749

\bibitem[{{Davies} {et~al.}(2014){Davies}, {Chaplin}, {Elsworth}, \&
  {Hale}}]{2014MNRAS.441.3009D}
{Davies}, G.~R., {Chaplin}, W.~J., {Elsworth}, Y., \& {Hale}, S.~J. 2014,
  \mnras, 441, 3009

\bibitem[{{Foreman-Mackey} {et~al.}(2013){Foreman-Mackey}, {Hogg}, {Lang}, \&
  {Goodman}}]{2013PASP..125..306F}
{Foreman-Mackey}, D., {Hogg}, D.~W., {Lang}, D., \& {Goodman}, J. 2013, \pasp,
  125, 306

\bibitem[{{Garc{\'{\i}}a} {et~al.}(2014{\natexlab{a}}){Garc{\'{\i}}a},
  {Ceillier}, {Salabert}, {Mathur}, {van Saders}, {Pinsonneault}, {Ballot},
  {Beck}, {Bloemen}, {Campante}, {Davies}, {do Nascimento}, {Mathis},
  {Metcalfe}, {Nielsen}, {Su{\'a}rez}, {Chaplin}, {Jim{\'e}nez}, \&
  {Karoff}}]{2014A&A...572A..34G}
{Garc{\'{\i}}a}, R.~A., {Ceillier}, T., {Salabert}, D., {et~al.}
  2014{\natexlab{a}}, \aap, 572, A34

\bibitem[{{Garc{\'{\i}}a} {et~al.}(2011){Garc{\'{\i}}a}, {Hekker}, {Stello},
  {Guti{\'e}rrez-Soto}, {Handberg}, {Huber}, {Karoff}, {Uytterhoeven},
  {Appourchaux}, {Chaplin}, {Elsworth}, {Mathur}, {Ballot},
  {Christensen-Dalsgaard}, {Gilliland}, {Houdek}, {Jenkins}, {Kjeldsen},
  {McCauliff}, {Metcalfe}, {Middour}, {Molenda-Zakowicz}, {Monteiro}, {Smith},
  \& {Thompson}}]{2011MNRAS.414L...6G}
{Garc{\'{\i}}a}, R.~A., {Hekker}, S., {Stello}, D., {et~al.} 2011, \mnras, 414,
  L6

\bibitem[{{Garc{\'{\i}}a} {et~al.}(2014{\natexlab{b}}){Garc{\'{\i}}a},
  {Mathur}, {Pires}, {R{\'e}gulo}, {Bellamy}, {Pall{\'e}}, {Ballot},
  {Barcel{\'o} Forteza}, {Beck}, {Bedding}, {Ceillier}, {Roca Cort{\'e}s},
  {Salabert}, \& {Stello}}]{2014A&A...568A..10G}
{Garc{\'{\i}}a}, R.~A., {Mathur}, S., {Pires}, S., {et~al.} 2014{\natexlab{b}},
  \aap, 568, A10

\bibitem[{{Garc{\'{\i}}a} {et~al.}(2010){Garc{\'{\i}}a}, {Mathur}, {Salabert},
  {Ballot}, {R{\'e}gulo}, {Metcalfe}, \& {Baglin}}]{2010Sci...329.1032G}
{Garc{\'{\i}}a}, R.~A., {Mathur}, S., {Salabert}, D., {et~al.} 2010, Science,
  329, 1032

\bibitem[{{Gilliland} {et~al.}(2013){Gilliland}, {Marcy}, {Rowe}, {Rogers},
  {Torres}, {Fressin}, {Lopez}, {Buchhave}, {Christensen-Dalsgaard},
  {D{\'e}sert}, {Henze}, {Isaacson}, {Jenkins}, {Lissauer}, {Chaplin}, {Basu},
  {Metcalfe}, {Elsworth}, {Handberg}, {Hekker}, {Huber}, {Karoff}, {Kjeldsen},
  {Lund}, {Lundkvist}, {Miglio}, {Charbonneau}, {Ford}, {Fortney}, {Haas},
  {Howard}, {Howell}, {Ragozzine}, \& {Thompson}}]{2013ApJ...766...40G}
{Gilliland}, R.~L., {Marcy}, G.~W., {Rowe}, J.~F., {et~al.} 2013, \apj, 766, 40

\bibitem[{{Griffin}(2007)}]{2007Obs...127..313G}
{Griffin}, R.~F. 2007, The Observatory, 127, 313

\bibitem[{Hall(2008)}]{lrsp-2008-2}
Hall, J.~C. 2008, Living Reviews in Solar Physics, 5

\bibitem[{{Harvey}(1985)}]{1985ESASP.235..199H}
{Harvey}, J. 1985, in ESA Special Publication, Vol. 235, Future Missions in
  Solar, Heliospheric \& Space Plasma Physics, ed. E.~{Rolfe} \& B.~{Battrick}

\bibitem[{{Howell} {et~al.}(2012){Howell}, {Rowe}, {Bryson}, {Quinn}, {Marcy},
  {Isaacson}, {Ciardi}, {Chaplin}, {Metcalfe}, {Monteiro}, {Appourchaux},
  {Basu}, {Creevey}, {Gilliland}, {Quirion}, {Stello}, {Kjeldsen},
  {Christensen-Dalsgaard}, {Elsworth}, {Garc{\'{\i}}a}, {Houdek}, {Karoff},
  {Molenda-{\.Z}akowicz}, {Thompson}, {Verner}, {Torres}, {Fressin}, {Crepp},
  {Adams}, {Dupree}, {Sasselov}, {Dressing}, {Borucki}, {Koch}, {Lissauer},
  {Latham}, {Buchhave}, {Gautier}, {Everett}, {Horch}, {Batalha}, {Dunham},
  {Szkody}, {Silva}, {Mighell}, {Holberg}, {Ballot}, {Bedding}, {Bruntt},
  {Campante}, {Handberg}, {Hekker}, {Huber}, {Mathur}, {Mosser}, {R{\'e}gulo},
  {White}, {Christiansen}, {Middour}, {Haas}, {Hall}, {Jenkins}, {McCaulif},
  {Fanelli}, {Kulesa}, {McCarthy}, \& {Henze}}]{2012ApJ...746..123H}
{Howell}, S.~B., {Rowe}, J.~F., {Bryson}, S.~T., {et~al.} 2012, \apj, 746, 123

\bibitem[{{Huber} {et~al.}(2013){Huber}, {Chaplin}, {Christensen-Dalsgaard},
  {Gilliland}, {Kjeldsen}, {Buchhave}, {Fischer}, {Lissauer}, {Rowe},
  {Sanchis-Ojeda}, {Basu}, {Handberg}, {Hekker}, {Howard}, {Isaacson},
  {Karoff}, {Latham}, {Lund}, {Lundkvist}, {Marcy}, {Miglio}, {Silva Aguirre},
  {Stello}, {Arentoft}, {Barclay}, {Bedding}, {Burke}, {Christiansen},
  {Elsworth}, {Haas}, {Kawaler}, {Metcalfe}, {Mullally}, \&
  {Thompson}}]{2013ApJ...767..127H}
{Huber}, D., {Chaplin}, W.~J., {Christensen-Dalsgaard}, J., {et~al.} 2013,
  \apj, 767, 127

\bibitem[{{Huber} {et~al.}(2014){Huber}, {Silva Aguirre}, {Matthews},
  {Pinsonneault}, {Gaidos}, {Garc{\'{\i}}a}, {Hekker}, {Mathur}, {Mosser},
  {Torres}, {Bastien}, {Basu}, {Bedding}, {Chaplin}, {Demory}, {Fleming},
  {Guo}, {Mann}, {Rowe}, {Serenelli}, {Smith}, \&
  {Stello}}]{2014ApJS..211....2H}
{Huber}, D., {Silva Aguirre}, V., {Matthews}, J.~M., {et~al.} 2014, \apjs, 211,
  2

\bibitem[{{Jim{\'e}nez-Reyes} {et~al.}(2001){Jim{\'e}nez-Reyes}, {Corbard},
  {Pall{\'e}}, {Roca Cort{\'e}s}, \& {Tomczyk}}]{2001A&A...379..622J}
{Jim{\'e}nez-Reyes}, S.~J., {Corbard}, T., {Pall{\'e}}, P.~L., {Roca
  Cort{\'e}s}, T., \& {Tomczyk}, S. 2001, \aap, 379, 622

\bibitem[{{Jimenez-Reyes} {et~al.}(1998){Jimenez-Reyes}, {Regulo}, {Palle}, \&
  {Roca Cortes}}]{1998A&A...329.1119J}
{Jimenez-Reyes}, S.~J., {Regulo}, C., {Palle}, P.~L., \& {Roca Cortes}, T.
  1998, \aap, 329, 1119

\bibitem[{{Kallinger} {et~al.}(2014){Kallinger}, {De Ridder}, {Hekker},
  {Mathur}, {Mosser}, {Gruberbauer}, {Garc{\'{\i}}a}, {Karoff}, \&
  {Ballot}}]{2014A&A...570A..41K}
{Kallinger}, T., {De Ridder}, J., {Hekker}, S., {et~al.} 2014, \aap, 570, A41

\bibitem[{{Karoff} {et~al.}(2009){Karoff}, {Metcalfe}, {Chaplin}, {Elsworth},
  {Kjeldsen}, {Arentoft}, \& {Buzasi}}]{2009MNRAS.399..914K}
{Karoff}, C., {Metcalfe}, T.~S., {Chaplin}, W.~J., {et~al.} 2009, \mnras, 399,
  914

\bibitem[{Karoff {et~al.}(2013)Karoff, Metcalfe, Chaplin, Frandsen, Grundahl,
  Kjeldsen, Christensen-Dalsgaard, Nielsen, Frimann, Thygesen, Arentoft, Amby,
  Sousa, \& Buzasi}]{Karoff21082013}
Karoff, C., Metcalfe, T.~S., Chaplin, W.~J., {et~al.} 2013, \mnras, 433, 3227

\bibitem[{{Koch} {et~al.}(2010){Koch}, {Borucki}, {Basri}, {Batalha}, {Brown},
  {Caldwell}, {Christensen-Dalsgaard}, {Cochran}, {DeVore}, {Dunham},
  {Gautier}, {Geary}, {Gilliland}, {Gould}, {Jenkins}, {Kondo}, {Latham},
  {Lissauer}, {Marcy}, {Monet}, {Sasselov}, {Boss}, {Brownlee}, {Caldwell},
  {Dupree}, {Howell}, {Kjeldsen}, {Meibom}, {Morrison}, {Owen}, {Reitsema},
  {Tarter}, {Bryson}, {Dotson}, {Gazis}, {Haas}, {Kolodziejczak}, {Rowe}, {Van
  Cleve}, {Allen}, {Chandrasekaran}, {Clarke}, {Li}, {Quintana}, {Tenenbaum},
  {Twicken}, \& {Wu}}]{2010ApJ...713L..79K}
{Koch}, D.~G., {Borucki}, W.~J., {Basri}, G., {et~al.} 2010, \apjl, 713, L79

\bibitem[{Komm {et~al.}(2000)Komm, Howe, \& Hill}]{0004-637X-543-1-472}
Komm, R.~W., Howe, R., \& Hill, F. 2000, \apj, 543, 472

\bibitem[{{Lefebvre} {et~al.}(2008){Lefebvre}, {Garc{\'{\i}}a},
  {Jim{\'e}nez-Reyes}, {Turck-Chi{\`e}ze}, \& {Mathur}}]{2008A&A...490.1143L}
{Lefebvre}, S., {Garc{\'{\i}}a}, R.~A., {Jim{\'e}nez-Reyes}, S.~J.,
  {Turck-Chi{\`e}ze}, S., \& {Mathur}, S. 2008, \aap, 490, 1143

\bibitem[{{L{\'e}pine} \& {Shara}(2005)}]{2005AJ....129.1483L}
{L{\'e}pine}, S. \& {Shara}, M.~M. 2005, \aj, 129, 1483

\bibitem[{{Libbrecht} \& {Woodard}(1990)}]{1990Natur.345..779L}
{Libbrecht}, K.~G. \& {Woodard}, M.~F. 1990, \nat, 345, 779

\bibitem[{{Lomb}(1976)}]{1976Ap&SS..39..447L}
{Lomb}, N.~R. 1976, \apss, 39, 447

\bibitem[{{Marcy} {et~al.}(2014){Marcy}, {Isaacson}, {Howard}, {Rowe},
  {Jenkins}, {Bryson}, {Latham}, {Howell}, {Gautier}, {Batalha}, {Rogers},
  {Ciardi}, {Fischer}, {Gilliland}, {Kjeldsen}, {Christensen-Dalsgaard},
  {Huber}, {Chaplin}, {Basu}, {Buchhave}, {Quinn}, {Borucki}, {Koch}, {Hunter},
  {Caldwell}, {Van Cleve}, {Kolbl}, {Weiss}, {Petigura}, {Seager}, {Morton},
  {Johnson}, {Ballard}, {Burke}, {Cochran}, {Endl}, {MacQueen}, {Everett},
  {Lissauer}, {Ford}, {Torres}, {Fressin}, {Brown}, {Steffen}, {Charbonneau},
  {Basri}, {Sasselov}, {Winn}, {Sanchis-Ojeda}, {Christiansen}, {Adams},
  {Henze}, {Dupree}, {Fabrycky}, {Fortney}, {Tarter}, {Holman}, {Tenenbaum},
  {Shporer}, {Lucas}, {Welsh}, {Orosz}, {Bedding}, {Campante}, {Davies},
  {Elsworth}, {Handberg}, {Hekker}, {Karoff}, {Kawaler}, {Lund}, {Lundkvist},
  {Metcalfe}, {Miglio}, {Silva Aguirre}, {Stello}, {White}, {Boss}, {Devore},
  {Gould}, {Prsa}, {Agol}, {Barclay}, {Coughlin}, {Brugamyer}, {Mullally},
  {Quintana}, {Still}, {Thompson}, {Morrison}, {Twicken}, {D{\'e}sert},
  {Carter}, {Crepp}, {H{\'e}brard}, {Santerne}, {Moutou}, {Sobeck}, {Hudgins},
  {Haas}, {Robertson}, {Lillo-Box}, \& {Barrado}}]{2014ApJS..210...20M}
{Marcy}, G.~W., {Isaacson}, H., {Howard}, A.~W., {et~al.} 2014, \apjs, 210, 20

\bibitem[{{Mathur} {et~al.}(2014){Mathur}, {Garc{\'{\i}}a}, {Ballot},
  {Ceillier}, {Salabert}, {Metcalfe}, {R{\'e}gulo}, {Jim{\'e}nez}, \&
  {Bloemen}}]{2014A&A...562A.124M}
{Mathur}, S., {Garc{\'{\i}}a}, R.~A., {Ballot}, J., {et~al.} 2014, \aap, 562,
  A124

\bibitem[{{Mathur} {et~al.}(2012){Mathur}, {Metcalfe}, {Woitaszek}, {Bruntt},
  {Verner}, {Christensen-Dalsgaard}, {Creevey}, {Do{\v g}an}, {Basu}, {Karoff},
  {Stello}, {Appourchaux}, {Campante}, {Chaplin}, {Garc{\'{\i}}a}, {Bedding},
  {Benomar}, {Bonanno}, {Deheuvels}, {Elsworth}, {Gaulme}, {Guzik}, {Handberg},
  {Hekker}, {Herzberg}, {Monteiro}, {Piau}, {Quirion}, {R{\'e}gulo}, {Roth},
  {Salabert}, {Serenelli}, {Thompson}, {Trampedach}, {White}, {Ballot},
  {Brand{\~a}o}, {Molenda-{\.Z}akowicz}, {Kjeldsen}, {Twicken}, {Uddin}, \&
  {Wohler}}]{2012ApJ...749..152M}
{Mathur}, S., {Metcalfe}, T.~S., {Woitaszek}, M., {et~al.} 2012, \apj, 749, 152

\bibitem[{{McQuillan} {et~al.}(2013){McQuillan}, {Mazeh}, \&
  {Aigrain}}]{2013ApJ...775L..11M}
{McQuillan}, A., {Mazeh}, T., \& {Aigrain}, S. 2013, \apjl, 775, L11

\bibitem[{{McQuillan} {et~al.}(2014){McQuillan}, {Mazeh}, \&
  {Aigrain}}]{2014ApJS..211...24M}
{McQuillan}, A., {Mazeh}, T., \& {Aigrain}, S. 2014, \apjs, 211, 24

\bibitem[{{Metcalfe} {et~al.}(2014){Metcalfe}, {Creevey}, {Do{\u g}an},
  {Mathur}, {Xu}, {Bedding}, {Chaplin}, {Christensen-Dalsgaard}, {Karoff},
  {Trampedach}, {Benomar}, {Brown}, {Buzasi}, {Campante}, {{\c C}elik},
  {Cunha}, {Davies}, {Deheuvels}, {Derekas}, {Di Mauro}, {Garc{\'{\i}}a},
  {Guzik}, {Howe}, {MacGregor}, {Mazumdar}, {Montalb{\'a}n}, {Monteiro},
  {Salabert}, {Serenelli}, {Stello}, {St{\c e}{\' s}licki}, {Suran},
  {Y{\i}ld{\i}z}, {Aksoy}, {Elsworth}, {Gruberbauer}, {Guenther}, {Lebreton},
  {Molaverdikhani}, {Pricopi}, {Simoniello}, \& {White}}]{2014ApJS..214...27M}
{Metcalfe}, T.~S., {Creevey}, O.~L., {Do{\u g}an}, G., {et~al.} 2014, \apjs,
  214, 27

\bibitem[{{Metcalfe} {et~al.}(2007){Metcalfe}, {Dziembowski}, {Judge}, \&
  {Snow}}]{2007MNRAS.379L..16M}
{Metcalfe}, T.~S., {Dziembowski}, W.~A., {Judge}, P.~G., \& {Snow}, M. 2007,
  \mnras, 379, L16

\bibitem[{{Meunier} {et~al.}(2008){Meunier}, {Roudier}, \&
  {Rieutord}}]{2008A&A...488.1109M}
{Meunier}, N., {Roudier}, T., \& {Rieutord}, M. 2008, \aap, 488, 1109

\bibitem[{{Molenda-{\.Z}akowicz} {et~al.}(2013){Molenda-{\.Z}akowicz}, {Sousa},
  {Frasca}, {Uytterhoeven}, {Briquet}, {Van Winckel}, {Drobek}, {Niemczura},
  {Lampens}, {Lykke}, {Bloemen}, {Gameiro}, {Jean}, {Volpi}, {Gorlova},
  {Mortier}, {Tsantaki}, \& {Raskin}}]{2013MNRAS.434.1422M}
{Molenda-{\.Z}akowicz}, J., {Sousa}, S.~G., {Frasca}, A., {et~al.} 2013,
  \mnras, 434, 1422

\bibitem[{{Muller} {et~al.}(2007){Muller}, {Hanslmeier}, \&
  {Salda{\~n}a-Mu{\~n}oz}}]{2007A&A...475..717M}
{Muller}, R., {Hanslmeier}, A., \& {Salda{\~n}a-Mu{\~n}oz}, M. 2007, \aap, 475,
  717

\bibitem[{Pallé {et~al.}(1990)Pallé, Régulo, \& Roca~Cortés}]{palle90}
Pallé, P., Régulo, C., \& Roca~Cortés, T. 1990, in Lecture Notes in Physics,
  Vol. 367, Progress of Seismology of the Sun and Stars, ed. Y.~Osaki \&
  H.~Shibahashi (Springer Berlin Heidelberg), 129--134

\bibitem[{{Pinsonneault} {et~al.}(2012){Pinsonneault}, {An},
  {Molenda-{\.Z}akowicz}, {Chaplin}, {Metcalfe}, \&
  {Bruntt}}]{2012ApJS..199...30P}
{Pinsonneault}, M.~H., {An}, D., {Molenda-{\.Z}akowicz}, J., {et~al.} 2012,
  \apjs, 199, 30

\bibitem[{{Pires} {et~al.}(2015){Pires}, {Mathur}, {Garc{\'{\i}}a}, {Ballot},
  {Stello}, \& {Sato}}]{2015A&A...574A..18P}
{Pires}, S., {Mathur}, S., {Garc{\'{\i}}a}, R.~A., {et~al.} 2015, \aap, 574,
  A18

\bibitem[{Press(2007)}]{press2007numerical}
Press, W.~H. 2007, Numerical recipes 3rd edition: The art of scientific
  computing (Cambridge University Press)

\bibitem[{{R{\'e}gulo} {et~al.}(2016){R{\'e}gulo}, {Garc{\'{\i}}a}, \&
  {Ballot}}]{Regulo2016}
{R{\'e}gulo}, C., {Garc{\'{\i}}a}, R.~A., \& {Ballot}, J. 2016, \aap, 589, A103

\bibitem[{{Roth} \& {Stix}(2003)}]{2003A&A...405..779R}
{Roth}, M. \& {Stix}, M. 2003, \aap, 405, 779

\bibitem[{Saar \& Brandenburg(1999)}]{0004-637X-524-1-295}
Saar, S.~H. \& Brandenburg, A. 1999, \apj, 524, 295

\bibitem[{{Salabert} {et~al.}(2016{\natexlab{a}}){Salabert}, {Garcia}, {Beck},
  {Egeland}, {Palle}, {Mathur}, {Metcalfe}, {do Nascimento}, {Ceillier},
  {Andersen}, \& {Trivino Hage}}]{2016arXiv160801489S}
{Salabert}, D., {Garcia}, R.~A., {Beck}, P.~G., {et~al.} 2016{\natexlab{a}},
  ArXiv e-prints [\eprint[arXiv]{1608.01489}]

\bibitem[{{Salabert} {et~al.}(2015){Salabert}, {Garc{\'{\i}}a}, \&
  {Turck-Chi{\`e}ze}}]{2015A&A...578A.137S}
{Salabert}, D., {Garc{\'{\i}}a}, R.~A., \& {Turck-Chi{\`e}ze}, S. 2015, \aap,
  578, A137

\bibitem[{{Salabert} {et~al.}(2011){Salabert}, {R{\'e}gulo}, {Ballot},
  {Garc{\'{\i}}a}, \& {Mathur}}]{2011A&A...530A.127S}
{Salabert}, D., {R{\'e}gulo}, C., {Ballot}, J., {Garc{\'{\i}}a}, R.~A., \&
  {Mathur}, S. 2011, \aap, 530, A127

\bibitem[{{Salabert} {et~al.}(2016{\natexlab{b}}){Salabert}, {R{\'e}gulo},
  {Garc{\'{\i}}a}, {Beck}, {Ballot}, {Creevey}, {P{\'e}rez Hern{\'a}ndez}, {do
  Nascimento}, {Corsaro}, {Egeland}, {Mathur}, {Metcalfe}, {Bigot}, {Ceillier},
  \& {Pall{\'e}}}]{2016A&A...589A.118S}
{Salabert}, D., {R{\'e}gulo}, C., {Garc{\'{\i}}a}, R.~A., {et~al.}
  2016{\natexlab{b}}, \aap, 589, A118

\bibitem[{{Scargle}(1982)}]{1982ApJ...263..835S}
{Scargle}, J.~D. 1982, \apj, 263, 835

\bibitem[{{Silva Aguirre} {et~al.}(2015){Silva Aguirre}, {Davies}, {Basu},
  {Christensen-Dalsgaard}, {Creevey}, {Metcalfe}, {Bedding}, {Casagrande},
  {Handberg}, {Lund}, {Nissen}, {Chaplin}, {Huber}, {Serenelli}, {Stello}, {Van
  Eylen}, {Campante}, {Elsworth}, {Gilliland}, {Hekker}, {Karoff}, {Kawaler},
  {Kjeldsen}, \& {Lundkvist}}]{2015MNRAS.452.2127S}
{Silva Aguirre}, V., {Davies}, G.~R., {Basu}, S., {et~al.} 2015, \mnras, 452,
  2127

\bibitem[{{Skumanich}(1972)}]{1972ApJ...171..565S}
{Skumanich}, A. 1972, \apj, 171, 565

\bibitem[{{van Leeuwen}(2007)}]{2007A&A...474..653V}
{van Leeuwen}, F. 2007, \aap, 474, 653

\bibitem[{Vida {et~al.}(2014)Vida, Oláh, \& Szabó}]{Vida01072014}
Vida, K., Oláh, K., \& Szabó, R. 2014, \mnras, 441, 2744

\bibitem[{{Wilson}(1978)}]{1978ApJ...226..379W}
{Wilson}, O.~C. 1978, \apj, 226, 379

\bibitem[{{Woodard} \& {Noyes}(1985)}]{1985Natur.318..449W}
{Woodard}, M.~F. \& {Noyes}, R.~W. 1985, \nat, 318, 449

\bibitem[{{Wright} {et~al.}(2011){Wright}, {Drake}, {Mamajek}, \&
  {Henry}}]{Wright2011}
{Wright}, N.~J., {Drake}, J.~J., {Mamajek}, E.~E., \& {Henry}, G.~W. 2011,
  \apj, 743, 48

\end{thebibliography}
\newpage

\onecolumn
\begin{appendix}
\section{Tables of stellar parameters}
 \begin{table*}[h!]
	\caption{Radius, mass, and age of the investigated stars}           
	\label{table:A1}   
	\centering                                   
	\begin{tabular}[b!]{c c c c c c c c}
		\hline\hline
		KIC      & $R$ [$R_{\odot}$] & $\sigma_{R}$ [$R_{\odot}$] & $M$ [$M_{\odot}$] & $\sigma_{M}$ [$M_{\odot}$] & Age [Gyr] & $\sigma_{\text{Age}}$ [Gyr]  & Reference$^*$ \\ \hline
		3632418  & 1.835             & 0.034                      & 1.27              & 0.03                       & 2.88      & 0.38                         & (2)           \\
		3656476  & 1.32              & 0.03                       & 1.09              & 0.01                       & 7.71      & 0.22                         & (1)           \\
		4914923  & 1.37              & 0.05                       & 1.10              & 0.01                       & 6.18      & 0.18                         & (1)           \\
		5184732  & 1.36              & 0.01                       & 1.25              & 0.01                       & 3.98      & 0.11                         & (1)           \\
		6106415  & 1.24              & 0.01                       & 1.12              & 0.02                       & 4.72      & 0.12                         & (1)           \\
		6116048  & 1.219             & 0.09                       & 1.01              & 0.03                       & 6.23      & 0.37                         & (2)           \\
		6603624  & 1.181             & 0.015                      & 1.09              & 0.03                       & 8.11      & 0.46                         & (2)           \\
		6933899  & 1.599             & 0.018                      & 1.14              & 0.03                       & 6.87      & 0.34                         & (2)           \\
		7680114  & 1.45              & 0.03                       & 1.19              & 0.01                       & 5.92      & 0.20                         & (1)           \\
		7976303  & 1.961             & 0.041                      & 1.10              & 0.05                       & 4.78      & 0.58                         & (2)           \\
		8006161  & 0.947             & 0.007                      & 1.04              & 0.02                       & 5.04      & 0.17                         & (2)           \\
		8228742  & 1.809             & 0.014                      & 1.27              & 0.02                       & 3.84      & 0.29                         & (2)           \\
		8379927  & 1.11              & 0.02                       & 1.09              & 0.03                       & 3.28      & 0.16                         & (1)           \\
		8760414  & 1.010             & 0.004                      & 0.78              & 0.01                       & 3.69      & 0.74                         & (2)           \\
		9025370  & 0.960             & $\substack{+0.04 \\-0.03}$ & 0.83              & $\substack{+0.12 \\-0.06}$ & 11.8      & {$\substack{+3.8   \\-5.6}$} & (3), (4)      \\
		9955598  & 0.883             & 0.008                      & 0.89              & 0.02                       & 6.72      & 0.20                         & (2)           \\
		10018963 & 1.915             & 0.020                      & 1.18              & 0.03                       & 4.36      & 0.34                         & (2)           \\
		10516096 & 1.42              & 0.03                       & 1.12              & 0.03                       & 6.41      & 0.27                         & (1)           \\
		10644253 & 1.108             & 0.016                      & 1.13              & 0.05                       & 1.07      & 0.25                         & (2)           \\
		10963065 & 1.213             & 0.008                      & 1.05              & 0.02                       & 4.30      & 0.23                         & (2)           \\
		11244118 & 1.589             & 0.026                      & 1.10              & 0.05                       & 6.43      & 0.58                         & (2)           \\
		11295426 & 1.243             & 0.019                      & 1.079             & 0.051                      & 7.087     & 0.451                        & (3), (5)      \\
		12009504 & 1.375             & 0.015                      & 1.12              & 0.03                       & 3.64      & 0.26                         & (2)           \\
		12258514 & 1.573             & 0.010                      & 1.19              & 0.03                       & 4.03      & 0.32                         & (2)           \\ \hline
	\end{tabular}
	\tablefoot{$^*$ If one reference is given, it applies to R, M, and age. If two references are given, the first one applies to R and M and the second to the age. References: (1) \cite{2012ApJ...749..152M}, (2) \cite{2014ApJS..214...27M}, (3) \cite{2014ApJS..211....2H}, {(4) \cite{2014ApJS..210....1C}, (5) \cite{2015MNRAS.452.2127S}}}
\end{table*}
\clearpage
\begin{table*} 
	\caption{Spectral type, effective temperature, and rotation period of the investigated stars}           
	\label{table:A2}   
	\centering                                   
	\begin{tabular}{c c c c c c c c c c }
		\hline\hline
		  KIC    & Spectral type & $T_{\text{eff}}$ [K] & $\sigma_{T_{\text{eff}}}$ [K] & Reference$^*$ & $P_{\text{rot}}$ [d] & $\sigma_{P_{\text{rot}}}$ [d] & Reference & Notes\\ 
		  \hline
		3632418  & F6IV          & 6148                 & 111        & (1)           & 12.591               & 0.036          & (5)      & Planet host (8)\\
		3656476  & G5IV          & 5586                 & 108        & (1)           & 31.67                & 3.53           & (6)      & High proper motion (9)\\
		4914923  & G1.5V         & 5808                 & 92         & (1)           & 20.49                & 2.82           & (6)      & ---\\
		5184732  & G4V           & 5669                 & 97         & (1)           & 19.79                & 2.43           & (6)      & ---\\
		6106415  & G0            & 6055                 & 70         & (2), (3)      & ---                  & ---            & ---      & ---\\
		6116048  & F9IV-V        & 5991                 & 124        & (1)           & 17.26                & 1.96           & (6)      & ---\\
		6603624  & G8IV-V        & 5471                 & 128        & (1)           & ---                  & ---            & ---      & ---\\
		6933899  & G0.5IV        & 5837                 & 97         & (1)           & ---                  & ---            & ---      & ---\\
		7680114  & G0V           & 5799                 & 91         & (1)           & 26.31                & 1.86           & (6)      & ---\\
		7976303  & F8V           & 6119                 & 106        & (1)           & ---                  & ---            & ---      & ---\\
		8006161  & G8V           & 5258                 & 97         & (1)           & 29.79                & 3.09           & (6)      & High proper motion (9)\\
		8228742  & F9IV-V        & 6061                 & 108        & (1)           & 20.23                & 2.16           & (6)      & ---\\
		8379927  & F9IV-V        & 5998                 & 108        & (1)           & 17.259               & 0.026          & (7)      & Spectroscopic binary (10)\\
		8760414  & G0IV          & 5850                 & 166        & (1)           & ---                  & ---            & ---      & High proper motion (11)\\
		9025370  & F8            & 5659                 & 73         & (2), (4)      & ---                  & ---            & ---      & ---\\
		9955598  & K0V           & 5264                 & 95         & (1)           & 34.20                & 5.64           & (6)      & Planet host (12)\\
		10018963 & F6IV          & 6145                 & 112        & (1)           & ---                  & ---            & ---      & ---\\
		10516096 & F9IV-V        & 5928                 & 95         & (1)           & ---                  & ---            & ---      & ---\\
		10644253 & G0V           & 5910                 & 93         & (1)           & 10.91                & 0.87           & (6)      & ---\\
		10963065 & F8V           & 6097                 & 130        & (1)           & 12.444               & 0.172          & (5)      & Planet host (12)\\
		11244118 & G5IV          & 5605                 & 104        & (1)           & 23.17                & 3.89           & (6)      & ---\\
		11295426 & ---           & 5796                 & 78         & (4)           & ---                  & ---            & ---      & Planet host (13)\\
		12009504 & F9IV-V        & 6099                 & 125        & (1)           & 9.426                & 0.327          & (7)      & ---\\
		12258514 & G0.5IV        & 5952                 & 95         & (1)           & 15.00                & 1.84           & (6)      & ---\\ 
		\hline
	\end{tabular}
	\tablefoot{$^*$ If one reference is given, it applies to the spectral type and $T_{\text{eff}}$. If two references are given, the first one applies to the spectral type and the second to $T_{\text{eff}}$. References: (1) \cite{2013MNRAS.434.1422M}, (2) SIMBAD\footnote{http://simbad.u-strasbg.fr/simbad/} entry without reference, (3) \cite{2012MNRAS.423..122B}, (4) \cite{2012ApJS..199...30P}, (5) \cite{2013ApJ...775L..11M}, (6) \cite{2014A&A...572A..34G}, (7) \cite{2014ApJS..211...24M}, (8) \cite{2012ApJ...746..123H}, (9) \cite{2007A&A...474..653V}, (10) \cite{2007Obs...127..313G}, (11) \cite{2005AJ....129.1483L}, (12) \cite{2014ApJS..210...20M}, (13) \cite{2013ApJ...766...40G}}
\end{table*}
\clearpage
\pagebreak

\section{Plots of frequency shifts and variations of the height of the p-mode envelope, correlation coefficients of the background model parameters and the measured frequency shifts}
\vspace{0.1em}

\begin{table*}[h!]
 	\caption{Correlation coefficients between parameters of the background model and the frequency shifts for KIC 3632418}           
 	\label{table:B1}
 	\centering
 	\begin{tabular}{c|c c c c c c c c }
 		\hline\hline
 		3632418 &shifts       &p  &height       &p   &noise       &p &$\tau_1$      &p        \\
 		\hline
 		height &   -0.14 &   0.76   &        &       &         &      &        &             \\
 		noise &   -0.18 &   0.70   & 0.68   & 0.09  &         &      &        &             \\
 		$\tau_1$&   -0.21 &   0.64   & 0.25   & 0.59  & -0.21   & 0.64 &        &             \\
 		$\tau_2$&   -0.68 &   0.09   & 0.68   & 0.09  &  0.75   & 0.05 &  -0.07 &   0.88      \\
 	\end{tabular}
\end{table*}
\begin{figure*}[h!]
	\centering{
		\includegraphics[angle=-90,width=0.70\textwidth]{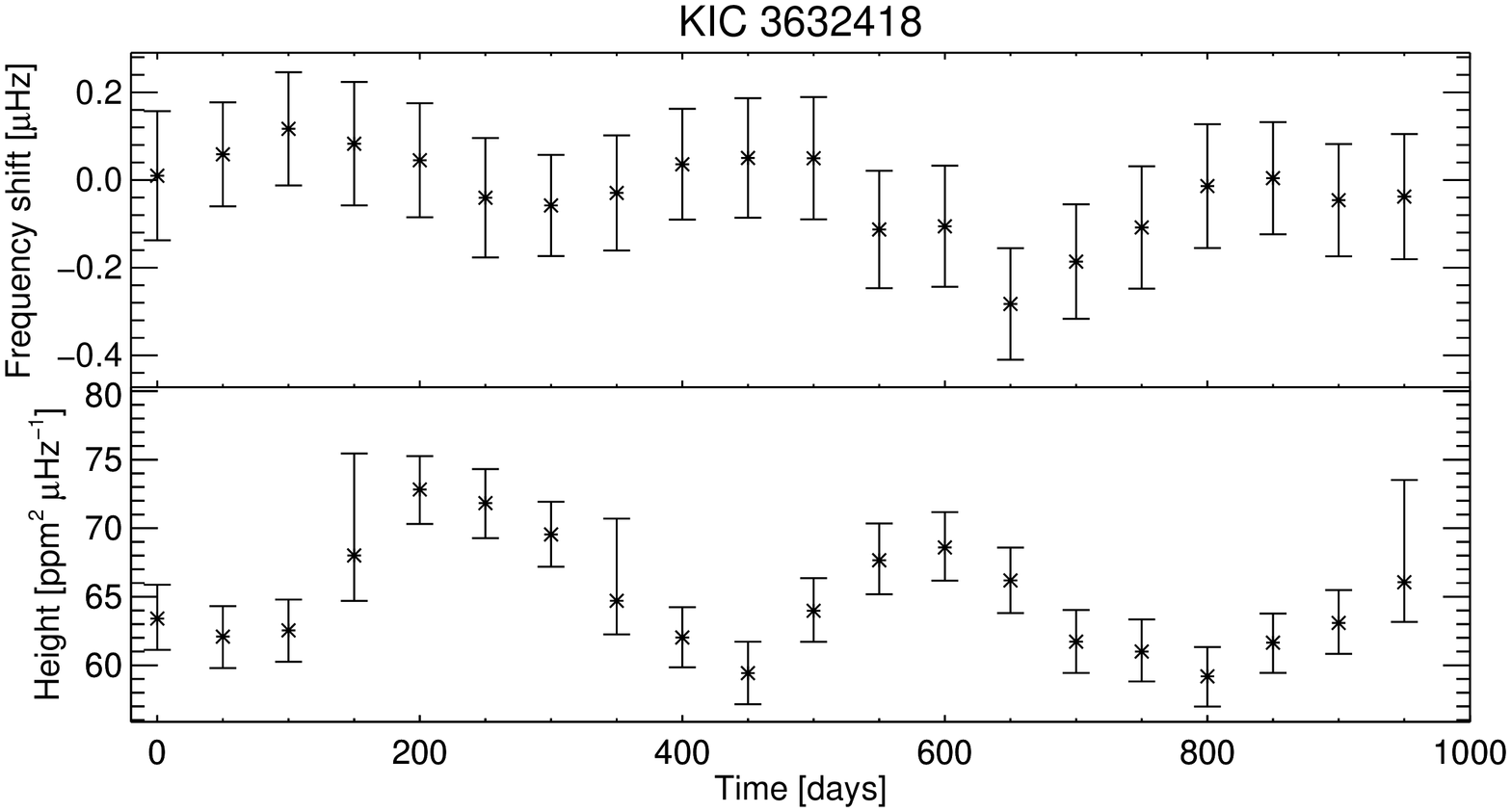}
		\vspace{-1.em}
		\caption{Frequency shifts (top half) and height of the p-mode envelope (bottom half) for KIC 3632418 as a function of time.}
		\label{fig:B1}}
\end{figure*}

\begin{table*}[h!]
 	\caption{Correlation coefficients between parameters of the background model and the frequency shifts for KIC 3656476}           
 	\label{table:B2}
 	\centering
\begin{tabular}{c|c c c c c c c c }
	\hline\hline
	3656476 &shifts       &p  &height       &p   &noise       &p &$\tau_1$      &p        \\
	\hline
	height&    0.30  &  0.62  &         &       &         &       &        &            \\
	noise&    0.70  &  0.19  &  0.70   & 0.19  &         &       &        &            \\
	$\tau_1$ &   0.80  &  0.10  &  0.80   & 0.10  &  0.80   & 0.10  &        &            \\
	$\tau_2$ &  -0.80  &  0.10  & -0.80   & 0.10  & -0.80   & 0.10  & -1.00  &  0.00      \\
\end{tabular}
 \end{table*}
 \begin{figure*}[h!]
 	\centering{
 		\includegraphics[angle=-90,width=0.70\textwidth]{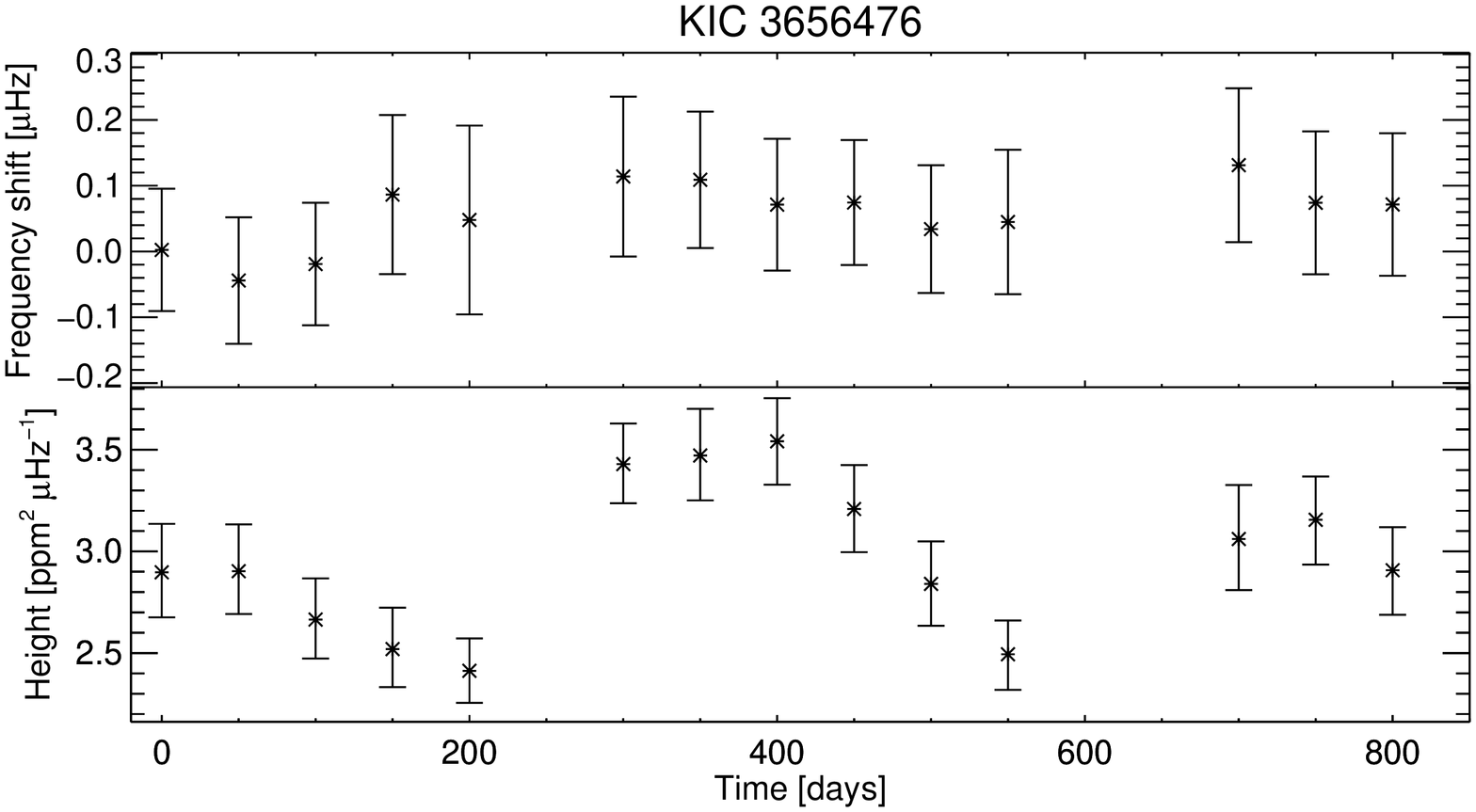}
 		\vspace{-1.em}
 		\caption{Frequency shifts (top half) and height of the p-mode envelope (bottom half) for KIC 3656476 as a function of time.}
 		\label{fig:B2}}
 \end{figure*} 

\clearpage
\pagebreak
 \begin{table*}
 	\caption{Correlation coefficients between parameters of the background model and the frequency shifts for KIC 4914923}           
 	\label{table:B3}
 	\centering
\begin{tabular}{c|c c c c c c c c }
	\hline\hline
	4914923 &shifts       &p  &height       &p   &noise       &p &$\tau_1$      &p        \\
	\hline
	height&    0.49  &  0.33  &        &        &        &        &       &             \\
	noise&    0.20  &  0.70  & -0.54  &  0.27  &        &        &       &             \\
	$\tau_1$ &  -0.20  &  0.70  & -0.26  &  0.62  &  0.09  &  0.87  &       &             \\
	$\tau_2$ &  -0.60  &  0.21  & -0.77  &  0.07  & -0.03  &  0.96  &  0.03 &   0.96      \\
\end{tabular}
\end{table*}
\begin{figure*}
	\centering{
		\includegraphics[angle=-90,width=0.70\textwidth]{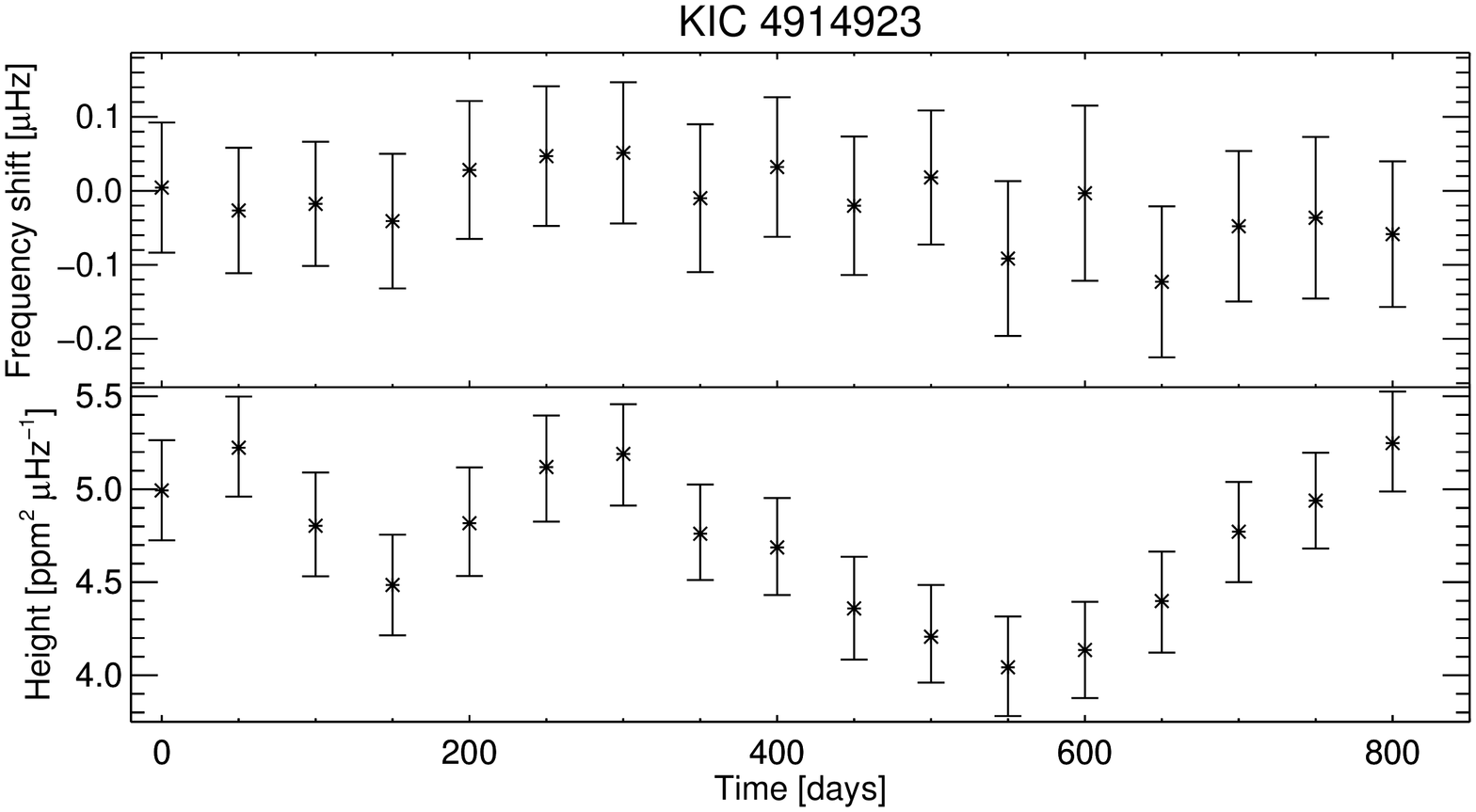}
		\vspace{-1.em}
		\caption{Frequency shifts (top half) and height of the p-mode envelope (bottom half) for KIC 4914923 as a function of time.}
		\label{fig:B3}}
\end{figure*} 

\begin{table*}
	\caption{Correlation coefficients between parameters of the background model and the frequency shifts for KIC 5184732}           
	\label{table:B4}
 	\centering
\begin{tabular}{c|c c c c c c c c }
	\hline\hline
	5184732 &shifts       &p  &height       &p   &noise       &p &$\tau_1$      &p        \\
	\hline
	height&   -0.44  &  0.20  &         &       &       &        &        &             \\
	noise&   -0.43  &  0.21  &  0.25   & 0.49  &       &        &        &             \\
	$\tau_1$ &   0.32  &  0.37  & -0.77   & 0.01  & -0.35 &   0.33 &        &             \\
	$\tau_2$ &  -0.13  &  0.73  & -0.08   & 0.83  & -0.30 &   0.40 &   0.26 &   0.47      \\
\end{tabular}
 \end{table*}
 \begin{figure*}
 	\centering{
 		\includegraphics[angle=-90,width=0.70\textwidth]{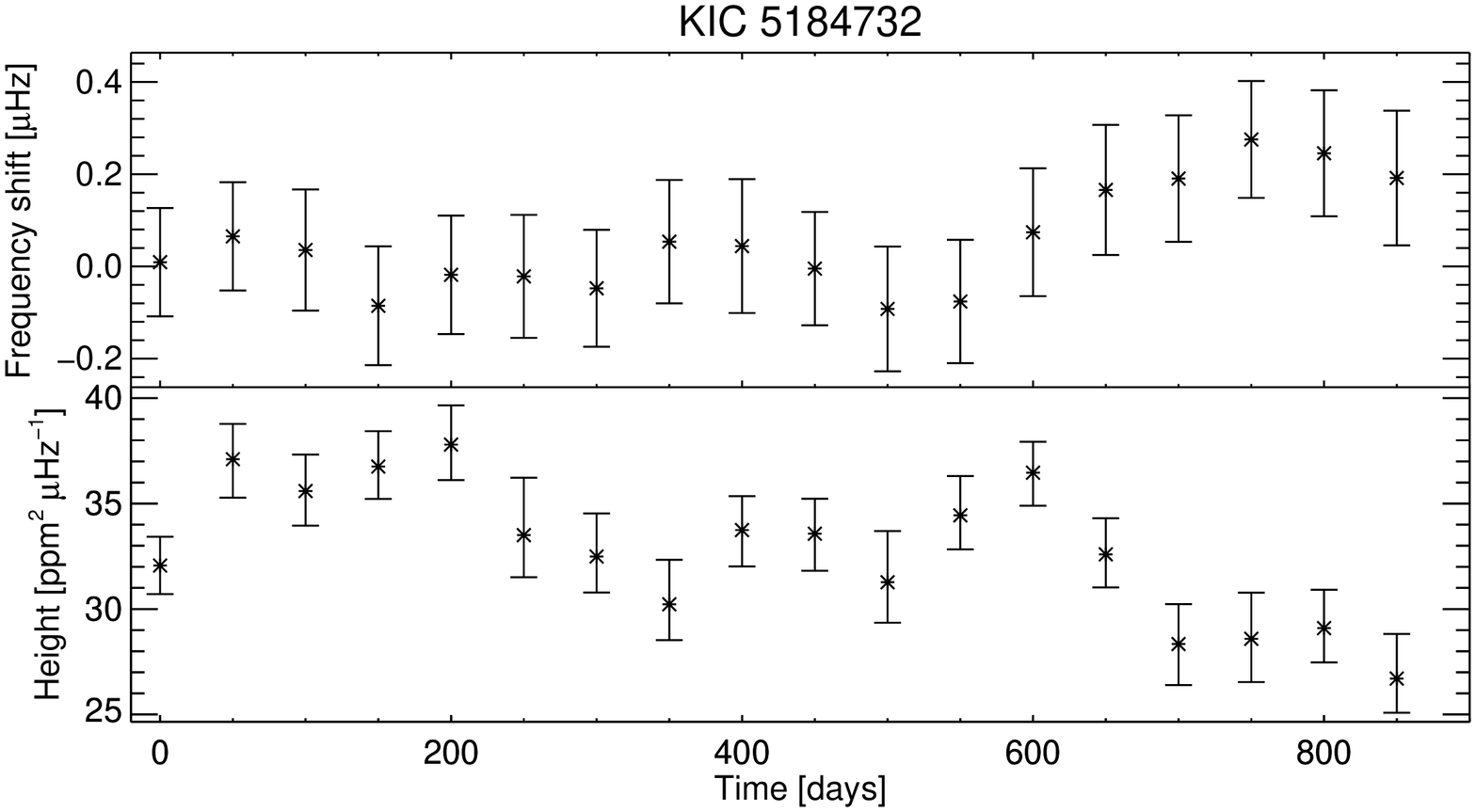}
 		\vspace{-1.em}
 		\caption{Frequency shifts (top half) and height of the p-mode envelope (bottom half) for KIC 5184732 as a function of time.}
 		\label{fig:B4}}
 \end{figure*} 
\clearpage
\pagebreak

 \begin{table*}
 	\caption{Correlation coefficients between parameters of the background model and the frequency shifts for KIC 6106415}           
 	\label{table:B5}
 	\centering
\begin{tabular}{c|c c c c c c c c }
	\hline\hline
	6106415 &shifts       &p  &height       &p   &noise       &p &$\tau_1$      &p        \\
	\hline
	height&    0.60   & 0.21  &       &          &      &        &         &            \\
	noise&    0.14   & 0.79  & -0.26 &   0.62   &      &        &         &            \\
	$\tau_1$ &  -0.31   & 0.54  & -0.83 &   0.04   & 0.14 &   0.79 &         &            \\
	$\tau_2$ &  -0.43   & 0.40  & -0.60 &   0.21   & 0.31 &   0.54 &   0.09  &  0.87      \\
\end{tabular}
 \end{table*}
 \begin{figure*}
 	\centering{
 		\includegraphics[angle=-90,width=0.70\textwidth]{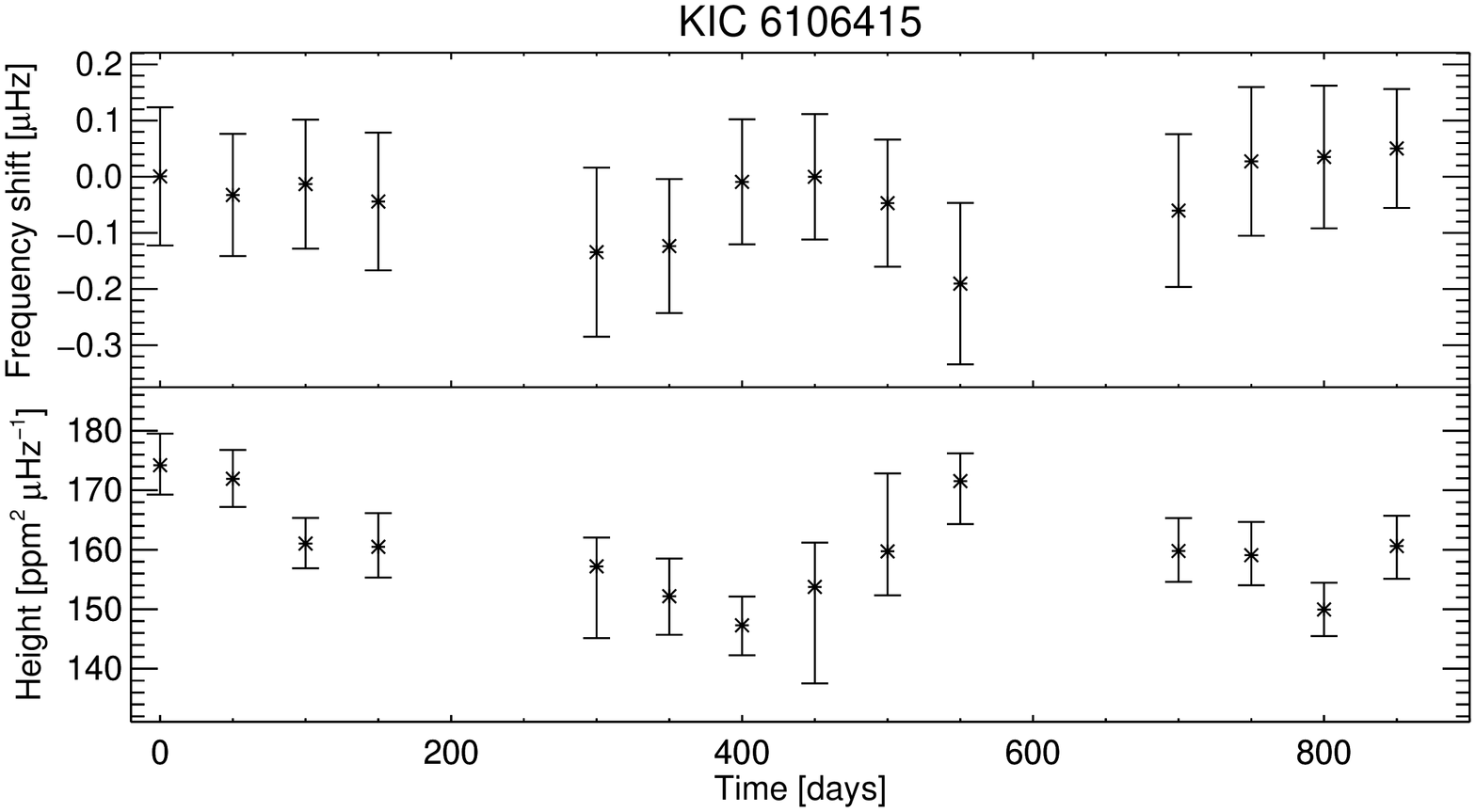}
 		\vspace{-1.em}
 		\caption{Frequency shifts (top half) and height of the p-mode envelope (bottom half) for KIC 6106415 as a function of time.}
 		\label{fig:B5}}
 \end{figure*} 
 
 \begin{table*}
 	\caption{Correlation coefficients between parameters of the background model and the frequency shifts for KIC 6116048}           
 	\label{table:B6}
 	\centering
\begin{tabular}{c|c c c c c c c c }
	\hline\hline
	6116048 &shifts       &p  &height       &p   &noise       &p &$\tau_1$      &p        \\
	\hline
	height&    0.75  &  0.05  &        &        &      &         &        &             \\
	noise&   -0.25  &  0.59  &  0.18  &  0.70  &      &         &        &             \\
	$\tau_1$ &  -0.61  &  0.15  & -0.36  &  0.43  &  0.57&    0.18 &        &             \\
	$\tau_2$ &  -0.32  &  0.48  & -0.71  &  0.07  & -0.14&    0.76 &   0.32 &   0.48      \\
\end{tabular}
 \end{table*}
 \begin{figure*}
 	\centering{
 		\includegraphics[angle=-90,width=0.70\textwidth]{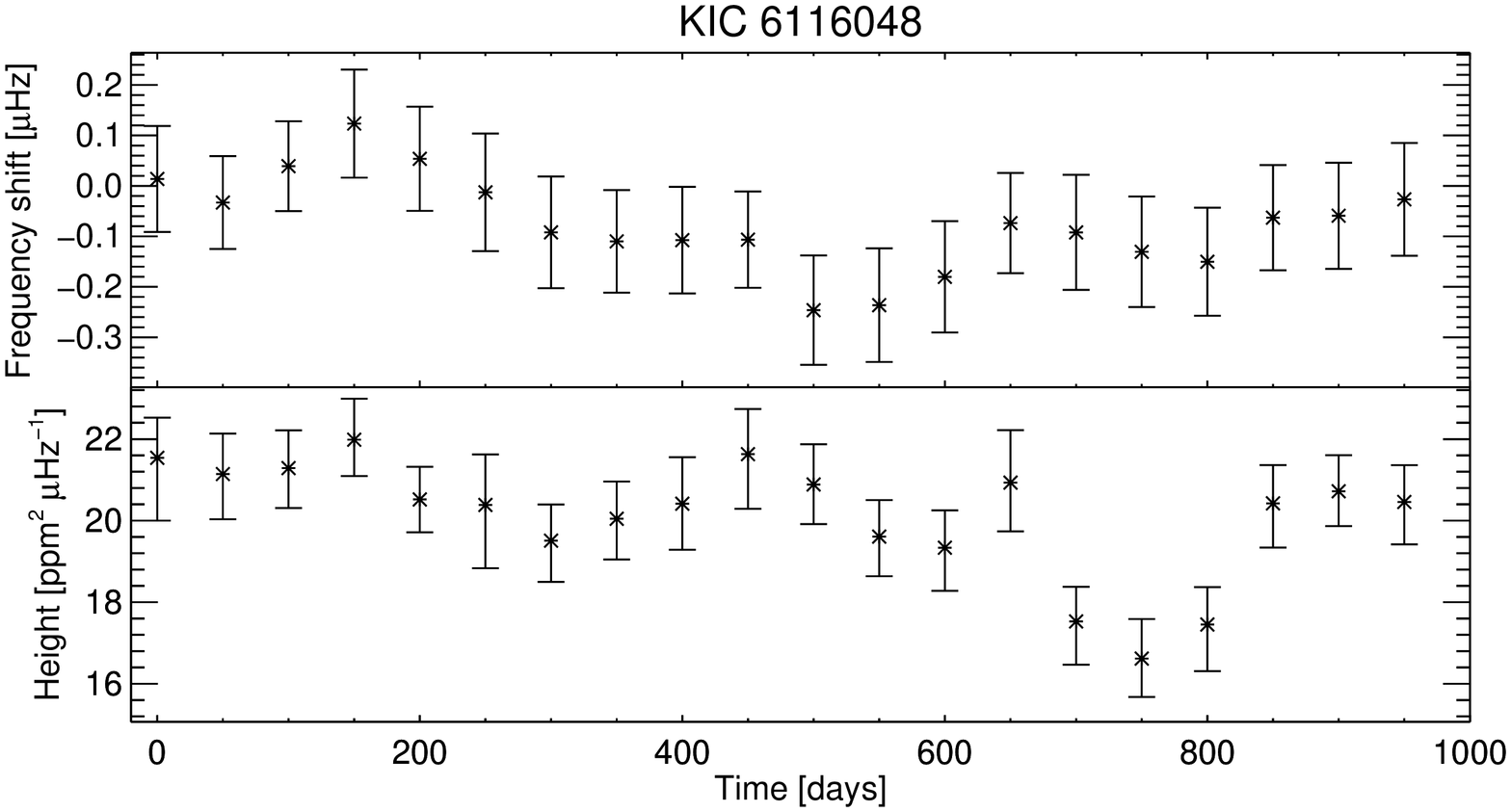}
 		\vspace{-1.em}
 		\caption{Frequency shifts (top half) and height of the p-mode envelope (bottom half) for KIC 6116048 as a function of time.}
 		\label{fig:B6}}
 \end{figure*} 
\clearpage
\pagebreak

 \begin{table*}
 	\caption{Correlation coefficients between parameters of the background model and the frequency shifts for KIC 6603624}           
 	\label{table:B7}
 	\centering
\begin{tabular}{c|c c c c c c c c }
	\hline\hline
	6603624 &shifts       &p  &height       &p   &noise       &p &$\tau_1$      &p        \\
	\hline
	height&    0.52  &  0.10  &       &         &        &       &         &            \\
	noise&    0.28  &  0.40  &  0.15 &   0.67  &        &       &         &            \\
	$\tau_1$  & -0.01  &  0.98  & -0.51 &   0.11  &  0.31  &  0.36 &         &            \\
	$\tau_2$  & -0.45  &  0.17  & -0.25 &   0.45  &  0.16  &  0.63 &   0.52  &  0.10      \\
\end{tabular}
 \end{table*}
 \begin{figure*}
 	\centering{
 		\includegraphics[angle=-90,width=0.70\textwidth]{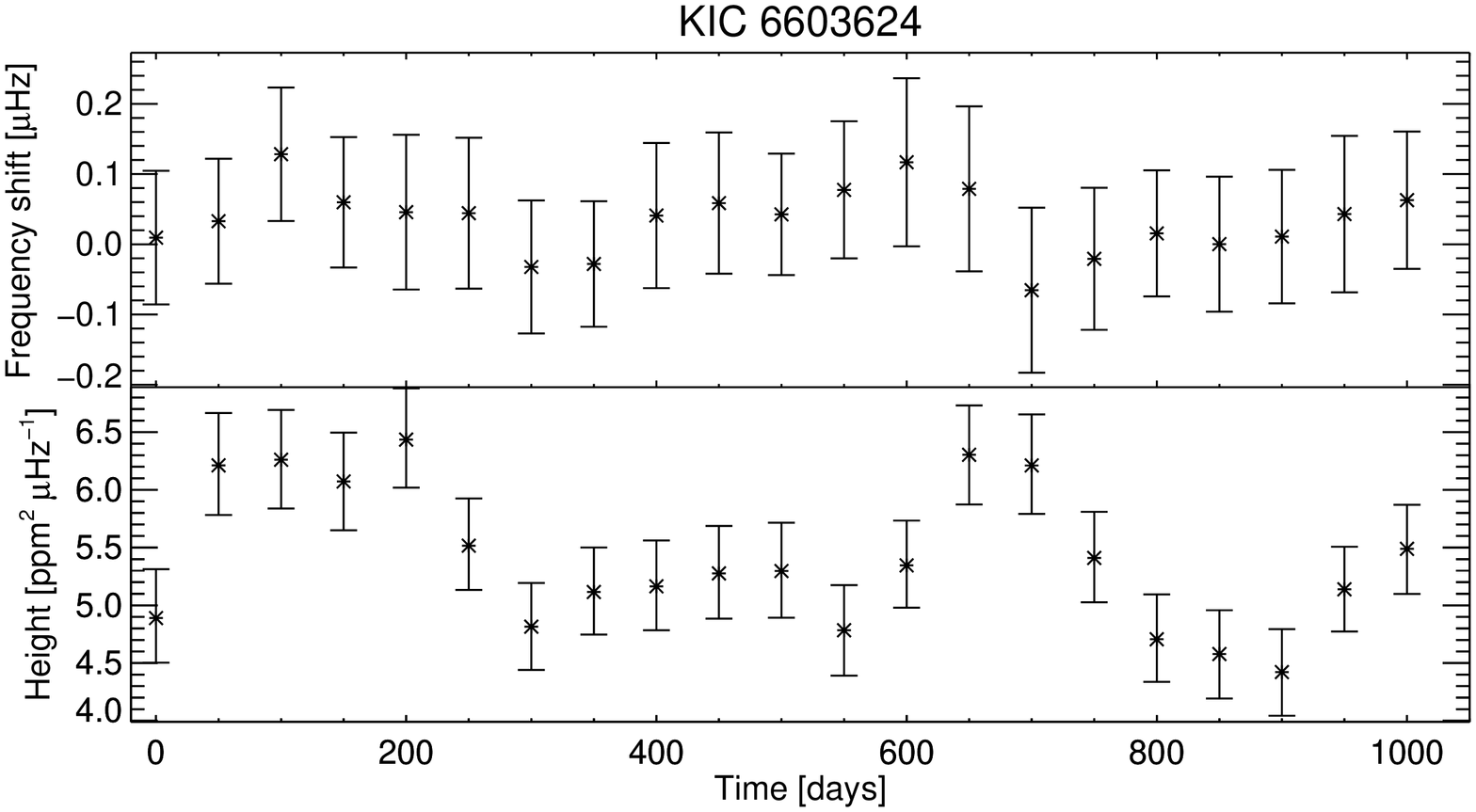}
 		\vspace{-1.em}
 		\caption{Frequency shifts (top half) and height of the p-mode envelope (bottom half) for KIC 6603624 as a function of time.}
 		\label{fig:B7}}
 \end{figure*} 

 \begin{table*}
 	\caption{Correlation coefficients between parameters of the background model and the frequency shifts for KIC 6933899}           
 	\label{table:B8}
 	\centering
\begin{tabular}{c|c c c c c c c c }
	\hline\hline
	6933899 &shifts       &p  &height       &p   &noise       &p &$\tau_1$      &p        \\
	\hline
	height&   -0.57 &   0.18  &      &          &       &         &         &           \\
	noise&   -0.39 &   0.38  &  0.68&    0.09  &       &         &         &           \\
	$\tau_1$ &   0.11 &   0.82  &  0.07&    0.88  & -0.36 &   0.43  &         &           \\
	$\tau_2$ &   0.50 &   0.25  & -0.04&    0.94  &  0.07 &   0.88  & -0.57   & 0.18      \\
\end{tabular}
 \end{table*}
 \begin{figure*}
 	\centering{
 		\includegraphics[angle=-90,width=0.70\textwidth]{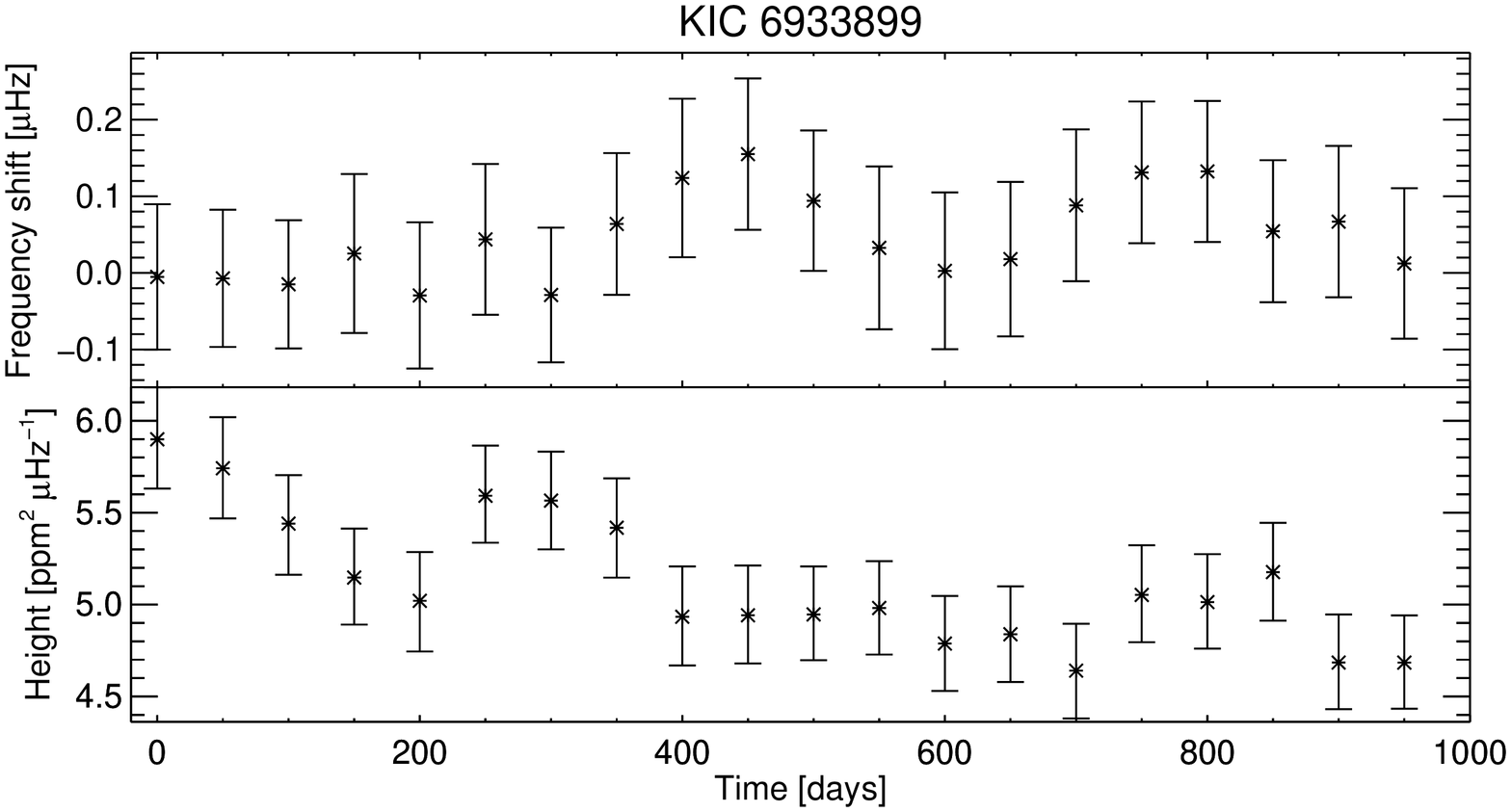}
 		\vspace{-1.em}
 		\caption{Frequency shifts (top half) and height of the p-mode envelope (bottom half) for KIC 6933899 as a function of time.}
 		\label{fig:B8}}
 \end{figure*} 
\clearpage
\pagebreak
 \begin{table*}
 	\caption{Correlation coefficients between parameters of the background model and the frequency shifts for KIC 7680114}           
 	\label{table:B9}
 	\centering
\begin{tabular}{c|c c c c c c c c }
	\hline\hline
	7680114 &shifts       &p  &height       &p   &noise       &p &$\tau_1$      &p        \\
	\hline
	height&    0.60   & 0.21 &          &         &      &        &       &            \\
	noise&    0.89   & 0.02 &   0.54   & 0.27    &      &        &       &             \\
	$\tau_1$ &   0.31   & 0.54 &  -0.20   & 0.70    &0.43  &  0.40  &       &             \\
	$\tau_2$ &   1.00   & 0.00 &   0.60   & 0.21    &0.89  &  0.02  &  0.31 &   0.54      \\
\end{tabular}
 \end{table*}
 \begin{figure*}
 	\centering{
 		\includegraphics[angle=-90,width=0.70\textwidth]{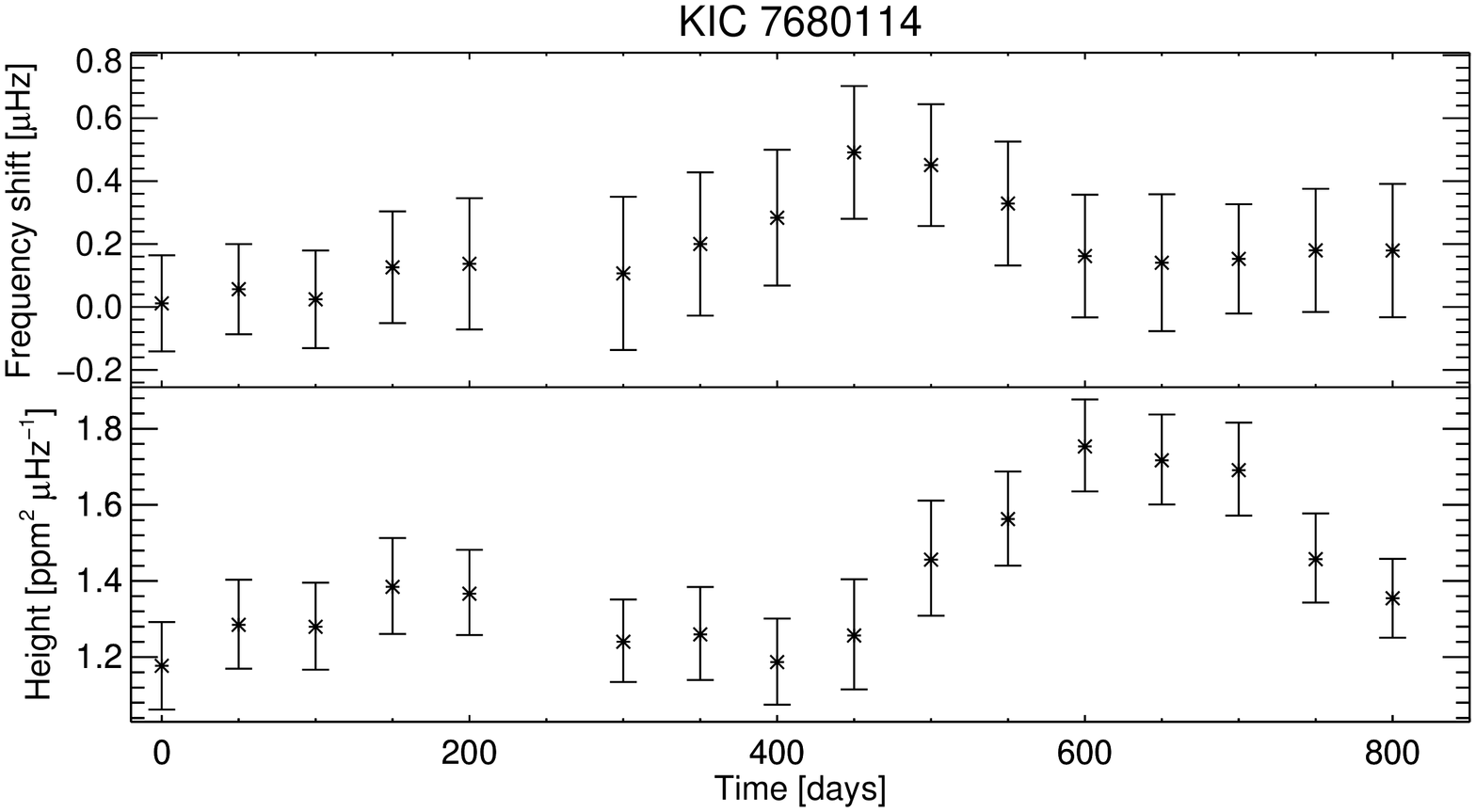}
 		\vspace{-1.em}
 		\caption{Frequency shifts (top half) and height of the p-mode envelope (bottom half) for KIC 7680114 as a function of time.}
 		\label{fig:B9}}
 \end{figure*} 

 \begin{table*}
	 \caption{Correlation coefficients between parameters of the background model and the frequency shifts for KIC 7976303}           
	 \label{table:B10}
 	\centering
\begin{tabular}{c|c c c c c c c c }
	\hline\hline
	7976303 &shifts       &p  &height       &p   &noise       &p &$\tau_1$      &p        \\
	\hline
	height&   -0.29   & 0.53 &        &         &        &       &         &            \\
	noise&   -0.14   & 0.76 &  -0.50 &   0.25  &        &       &         &            \\
	$\tau_1$ &   0.39   & 0.38 &  -0.57 &   0.18  &  0.79  &  0.04 &         &            \\
	$\tau_2$ &  -0.36   & 0.43 &   0.36 &   0.43  & -0.57  &  0.18 &  -0.82  &  0.02      \\
\end{tabular}
 \end{table*}
 \begin{figure*}
 	\centering{
 		\includegraphics[angle=-90,width=0.70\textwidth]{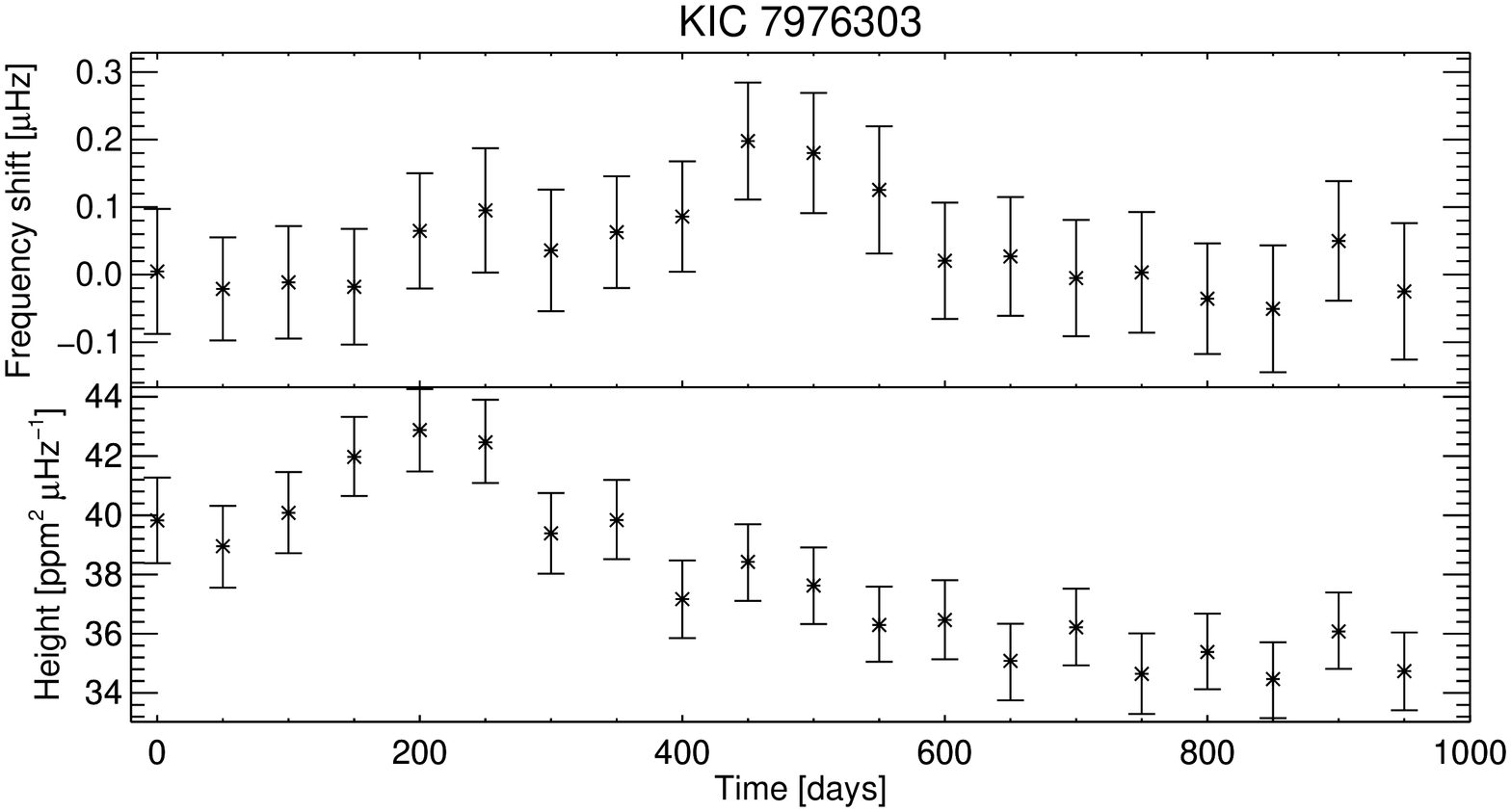}
 		\vspace{-1.em}
 		\caption{Frequency shifts (top half) and height of the p-mode envelope (bottom half) for KIC 7976303 as a function of time.}
 		\label{fig:B10}}
 \end{figure*} 
\clearpage
\pagebreak
 \begin{table*}
 	\caption{Correlation coefficients between parameters of the background model and the frequency shifts for KIC 8006161}           
 	\label{table:B11}
 	\centering
\begin{tabular}{c|c c c c c c c c }
	\hline\hline
	8006161 &shifts       &p  &height       &p   &noise       &p &$\tau_1$      &p        \\
	\hline
	height&   -0.93   & $4\cdot 10^{-5}$ &        &         &        &       &         &            \\
	noise&    0.23   & 0.50 &  -0.09 &   0.79  &        &       &         &            \\
	$\tau_1$ &  -0.27   & 0.42 &   0.37 &   0.26  & -0.24  &  0.48 &         &            \\
	$\tau_2$ &  -0.01   & 0.98 &   0.14 &   0.69  &  0.04  &  0.92 &   0.74  &  0.01      \\
\end{tabular}
 \end{table*}
 \begin{figure*}
 	\centering{
 		\includegraphics[angle=-90,width=0.70\textwidth]{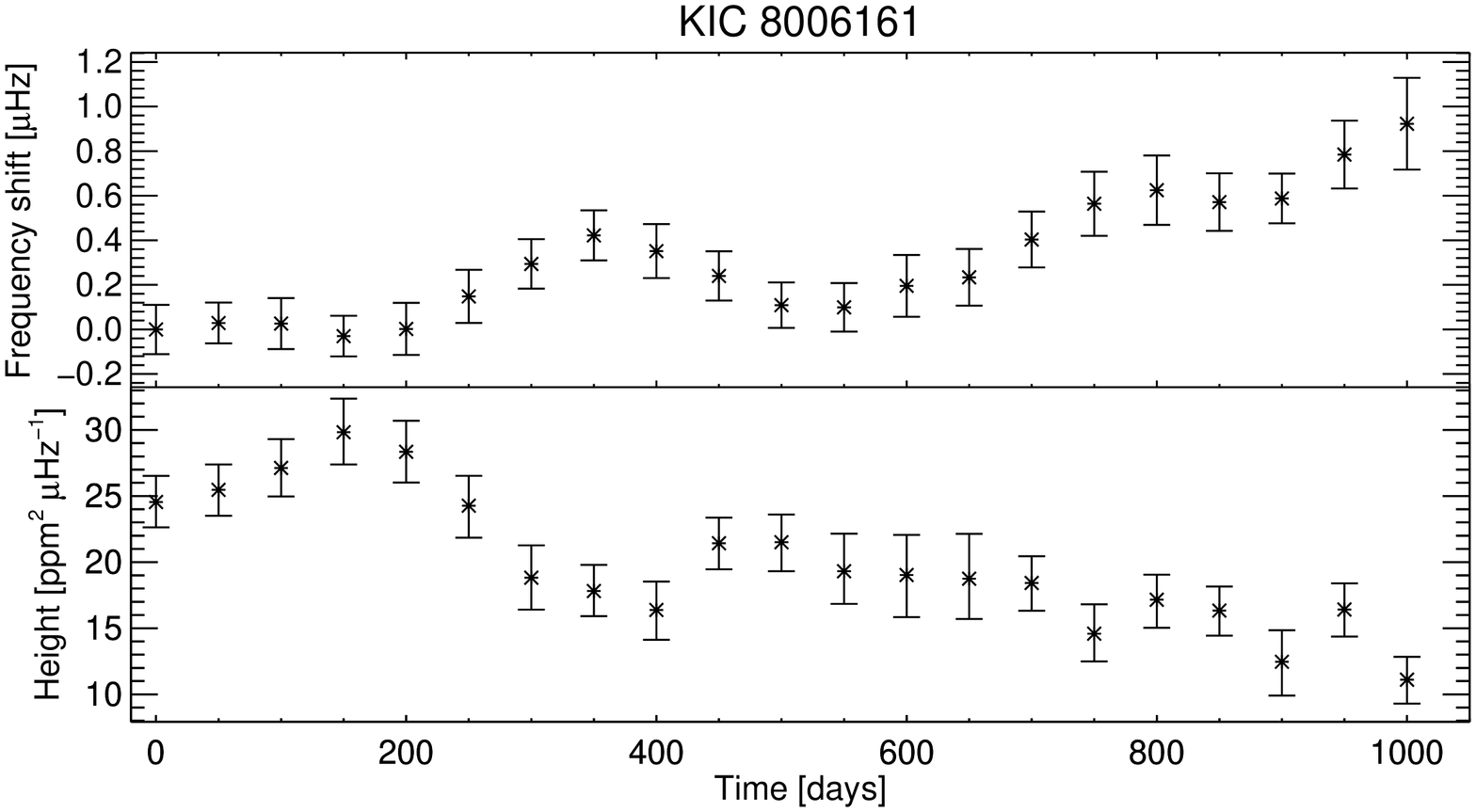}
 		\vspace{-1.em}
 		\caption{Frequency shifts (top half) and height of the p-mode envelope (bottom half) for KIC 8006161 as a function of time.}
 		\label{fig:B11}}
 \end{figure*} 

 \begin{table*}
 	\caption{Correlation coefficients between parameters of the background model and the frequency shifts for KIC 8228742}           
 	\label{table:B12}
 	\centering
\begin{tabular}{c|c c c c c c c c }
	\hline\hline
	8228742 &shifts       &p  &height       &p   &noise       &p &$\tau_1$      &p        \\
	\hline
	height&   -0.29  &  0.39  &        &        &        &        &        &            \\
	noise&   -0.41  &  0.21  &  0.61  &  0.05  &        &        &        &            \\
	$\tau_1$  & -0.07  &  0.83  &  0.04  &  0.92  & -0.21  &  0.54  &        &            \\
	$\tau_2$  & -0.02  &  0.96  & -0.14  &  0.69  & -0.14  &  0.69  & -0.43  &  0.19      \\
\end{tabular}
 \end{table*}
 \begin{figure*}
 	\centering{
 		\includegraphics[angle=-90,width=0.70\textwidth]{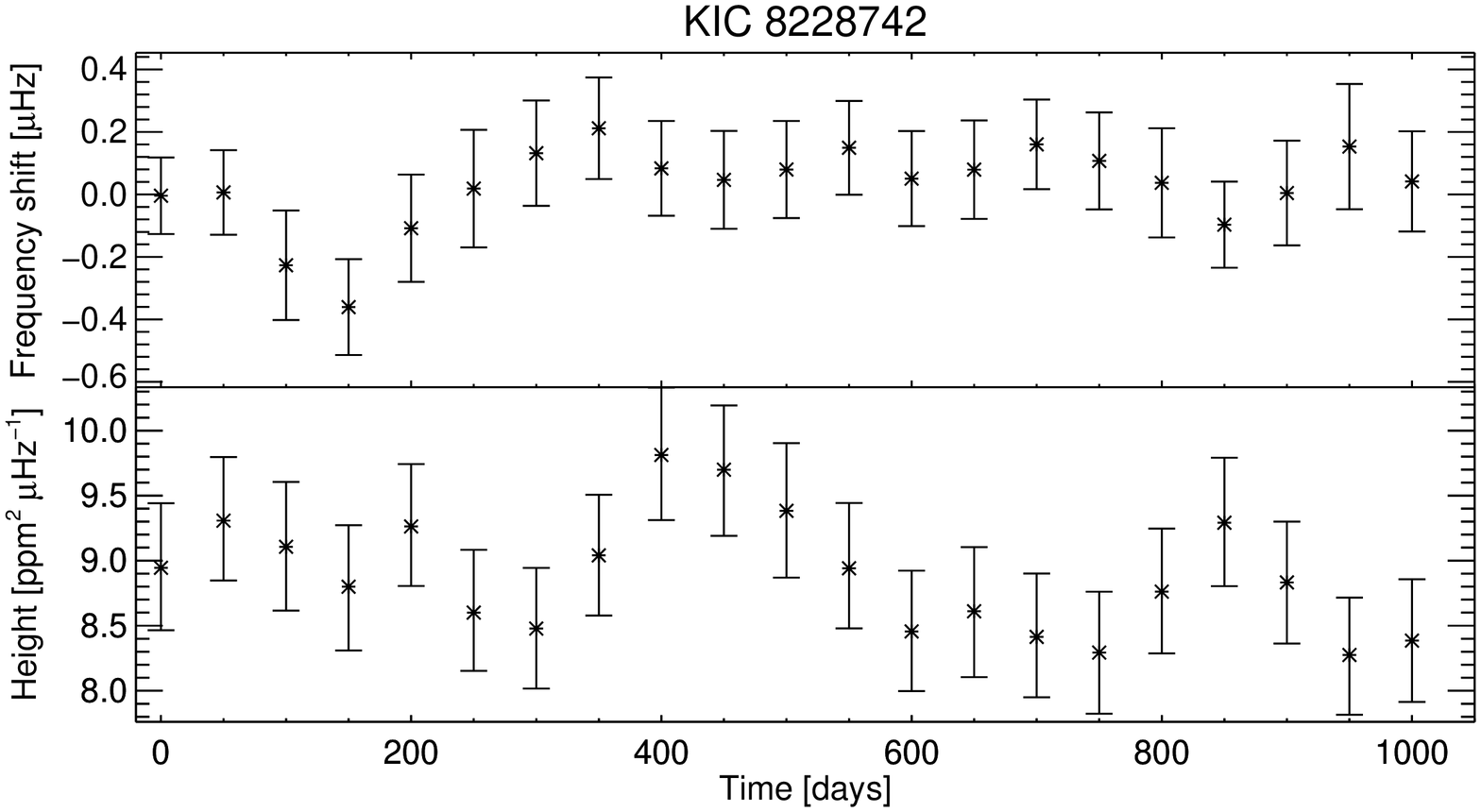}
 		\vspace{-1.em}
 		\caption{Frequency shifts (top half) and height of the p-mode envelope (bottom half) for KIC 8228742 as a function of time.}
 		\label{fig:B12}}
 \end{figure*} 
\clearpage
\pagebreak
 \begin{table*}
 	\caption{Correlation coefficients between parameters of the background model and the frequency shifts for KIC 8379927}           
 	\label{table:B13}
 	\centering
\begin{tabular}{c|c c c c c c c c }
	\hline\hline
	8379927 &shifts       &p  &height       &p   &noise       &p &$\tau_1$      &p        \\
	\hline
	height&    0.43 &   0.34  &       &         &        &        &       &             \\
	noise&   -0.21 &   0.64  &  0.25 &   0.59  &        &        &       &             \\
	$\tau_1$  & -0.11 &   0.82  & -0.04 &   0.94  & -0.04  &  0.94  &       &             \\
	$\tau_2$  &  0.00 &   1.00  & -0.64 &   0.12  & -0.57  &  0.18  &  0.39 &   0.38      \\
\end{tabular}
 \end{table*}
 \begin{figure*}
 	\centering{
 		\includegraphics[angle=-90,width=0.70\textwidth]{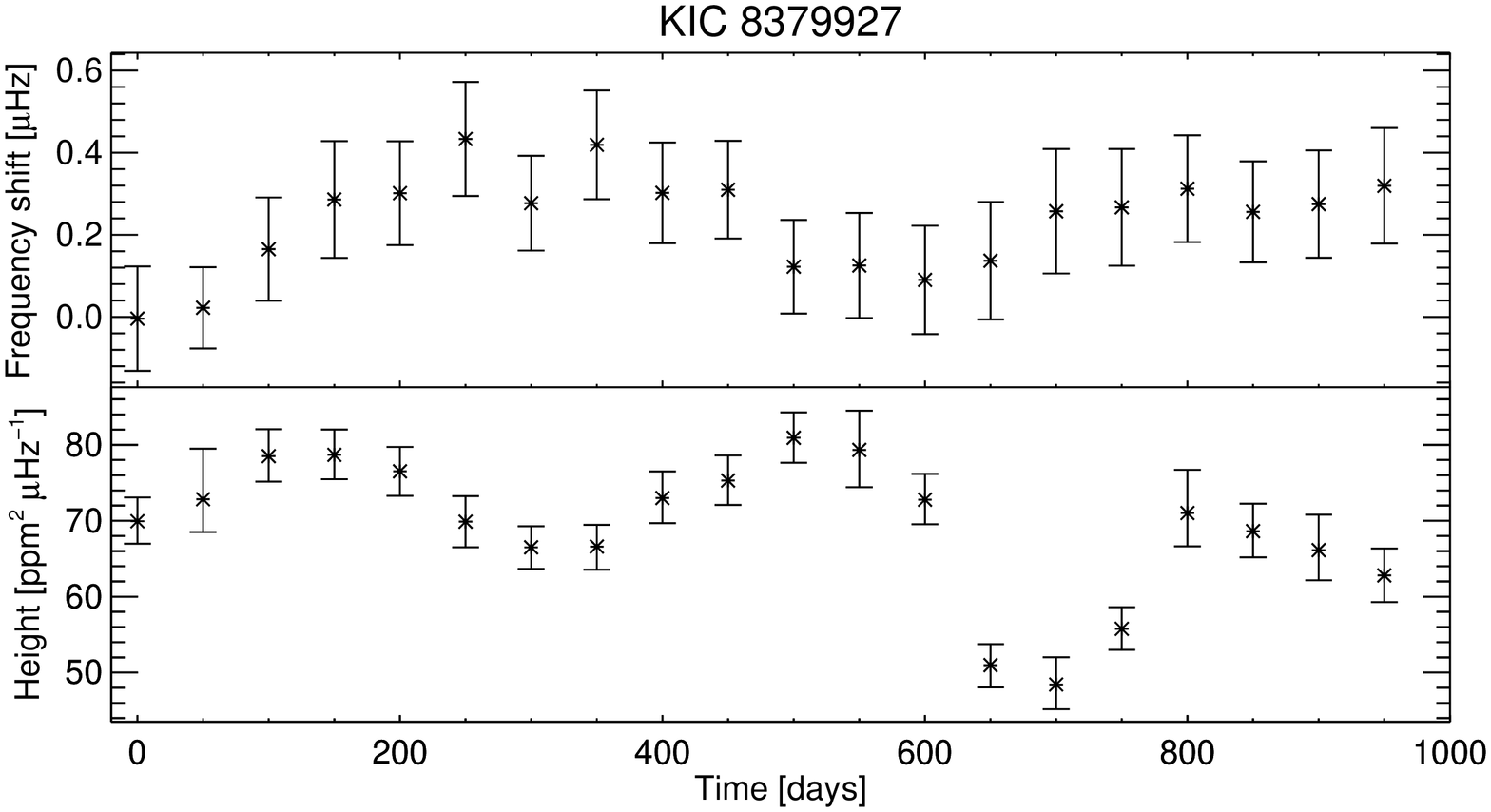}
 		\vspace{-1.em}
 		\caption{Frequency shifts (top half) and height of the p-mode envelope (bottom half) for KIC 8379927 as a function of time.}
 		\label{fig:B13}}
 \end{figure*} 

 \begin{table*}
 	\caption{Correlation coefficients between parameters of the background model and the frequency shifts for KIC 8760414}           
 	\label{table:B14}
 	\centering
\begin{tabular}{c|c c c c c c c c }
	\hline\hline
	8760414 &shifts       &p  &height       &p   &noise       &p &$\tau_1$      &p        \\
	\hline
	height&   -0.54  &  0.22  &        &         &       &         &     &              \\
	noise&   -0.43  &  0.34  &  0.39  &  0.38   &       &         &     &              \\
	$\tau_1$ &  -0.64  &  0.12  & -0.11  &  0.82   & 0.07  &  0.88   &     &              \\
	$\tau_2$ &   0.39  &  0.38  & -0.54  &  0.22   &-0.86  &  0.01   &-0.14&    0.76      \\
\end{tabular}
 \end{table*}
 \begin{figure*}
 	\centering{
 		\includegraphics[angle=-90,width=0.70\textwidth]{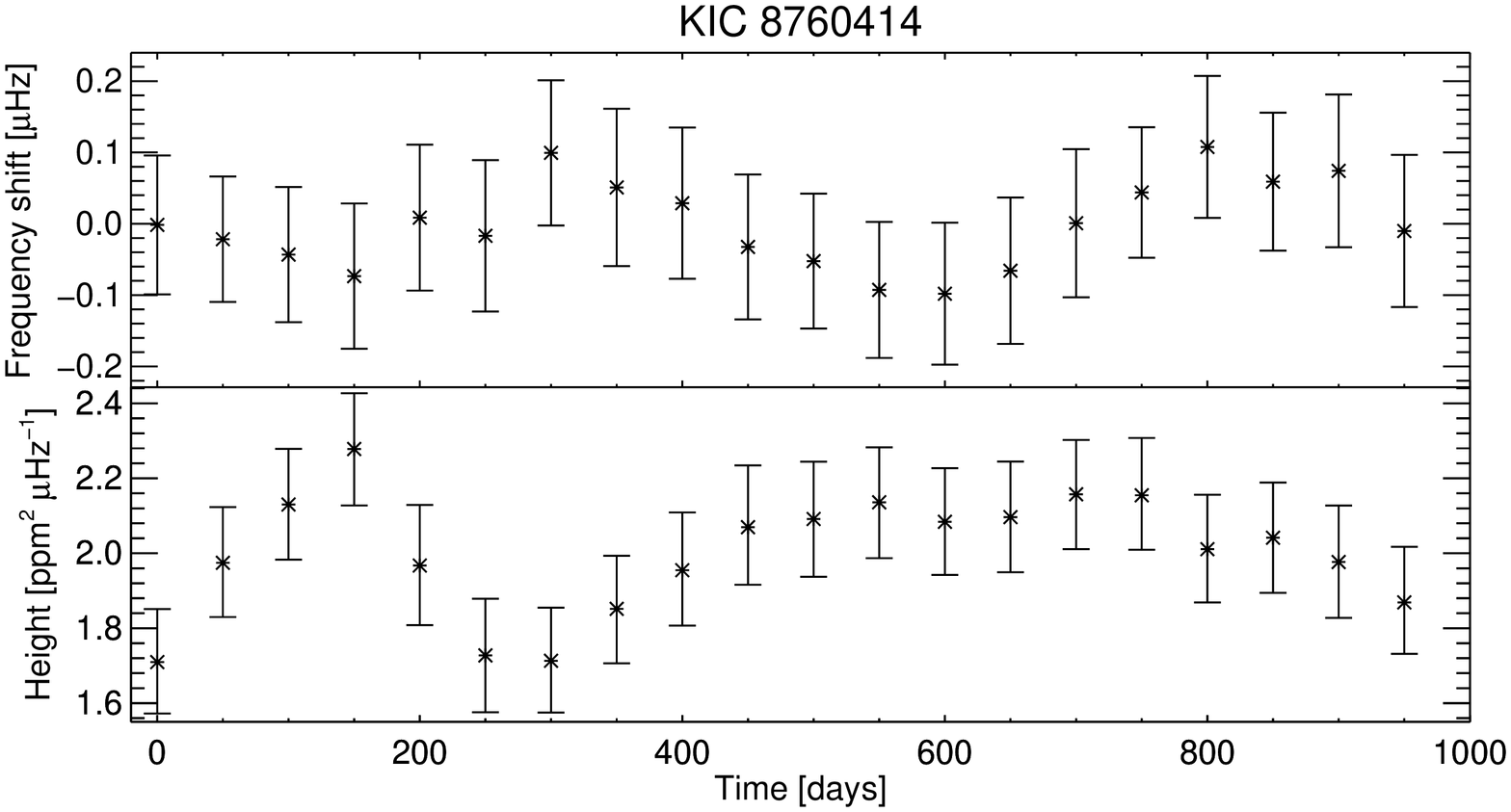}
 		\vspace{-1.em}
 		\caption{Frequency shifts (top half) and height of the p-mode envelope (bottom half) for KIC 8760414 as a function of time.}
 		\label{fig:B14}}
 \end{figure*} 
\clearpage
\pagebreak
 \begin{table*}
 	\caption{Correlation coefficients between parameters of the background model and the frequency shifts for KIC 9025370}           
 	\label{table:B15}
 	\centering
\begin{tabular}{c|c c c c c c c c }
	\hline\hline
	9025370 &shifts       &p  &height       &p   &noise       &p &$\tau_1$      &p        \\
	\hline
	 height&   -0.25    &   0.47 &         &        &        &        &        &            \\
	 noise&   -0.55     &   0.08 &  -0.32  &  0.34  &        &        &        &            \\
	 $\tau_1$  & -0.02  &   0.96 &   0.29  &  0.39  & -0.15  &  0.67  &        &            \\
	 $\tau_2$   & -0.13 &   0.71 &  -0.59  &  0.06  &  0.41  &  0.21  & -0.33  &  0.33      \\
\end{tabular}
 \end{table*}
 \begin{figure*}
 	\centering{
 		\includegraphics[angle=-90,width=0.70\textwidth]{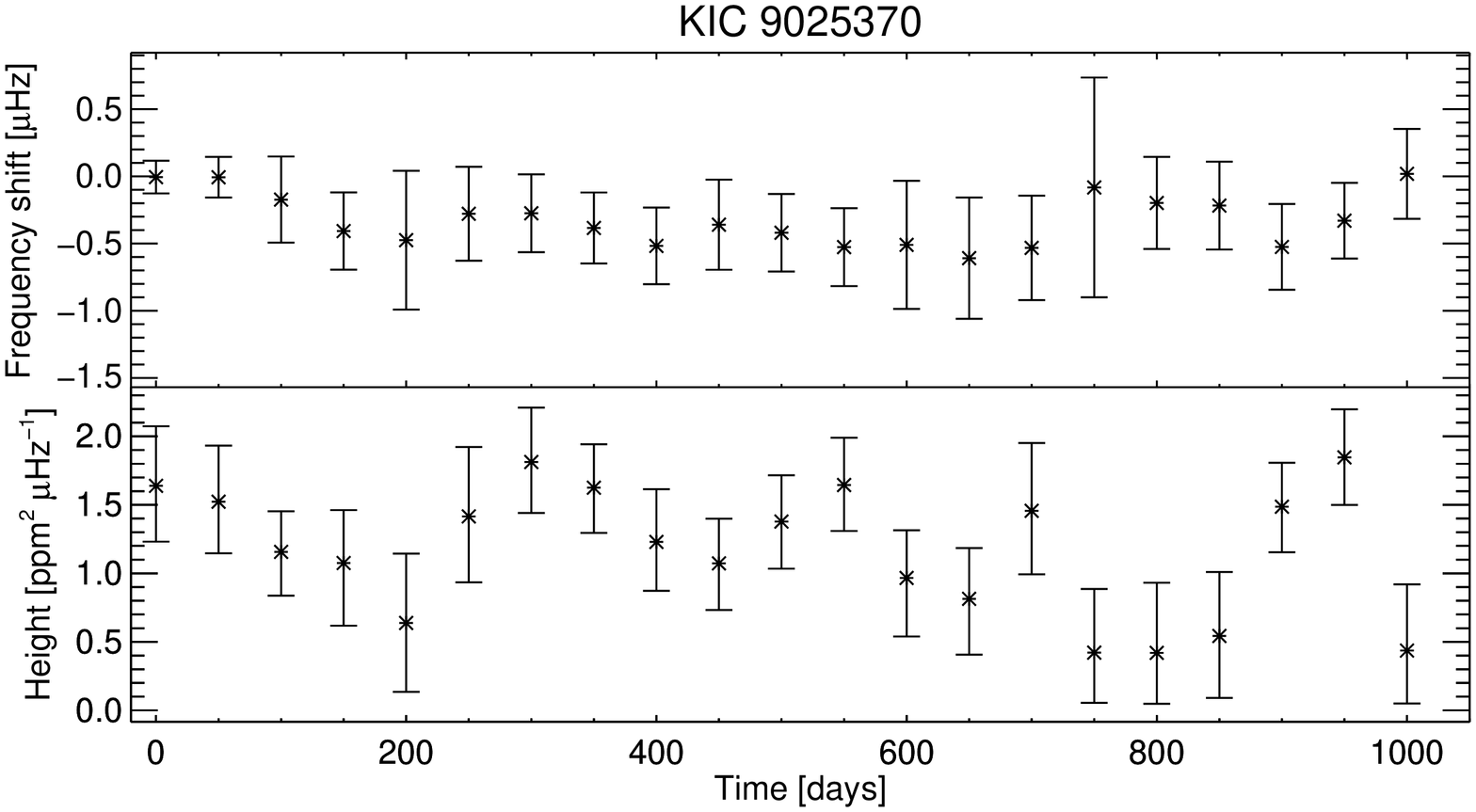}
 		\vspace{-1.em}
 		\caption{Frequency shifts (top half) and height of the p-mode envelope (bottom half) for KIC 9025370 as a function of time.}
 		\label{fig:B15}}
 \end{figure*} 

 \begin{table*}
	 \caption{Correlation coefficients between parameters of the background model and the frequency shifts for KIC 9955598}           
	 \label{table:B16}
 	\centering
\begin{tabular}{c|c c c c c c c c }
	\hline\hline
	9955598 &shifts       &p  &height       &p   &noise       &p &$\tau_1$      &p        \\
	\hline
	 height&   -0.75   &   0.05  &        &        &        &         &        &           \\
	 noise&    0.71    &   0.07  & -0.54  &  0.22  &        &         &        &           \\
	 $\tau_1$  &  0.00 &   1.00  & -0.14  &  0.76  & -0.43  &  0.34   &        &           \\
	 $\tau_2$  & -0.21 &   0.64  & -0.18  &  0.70  & -0.50  &  0.25   & 0.89   & 0.01      \\
\end{tabular}
 \end{table*}
 \begin{figure*}
 	\centering{
 		\includegraphics[angle=-90,width=0.70\textwidth]{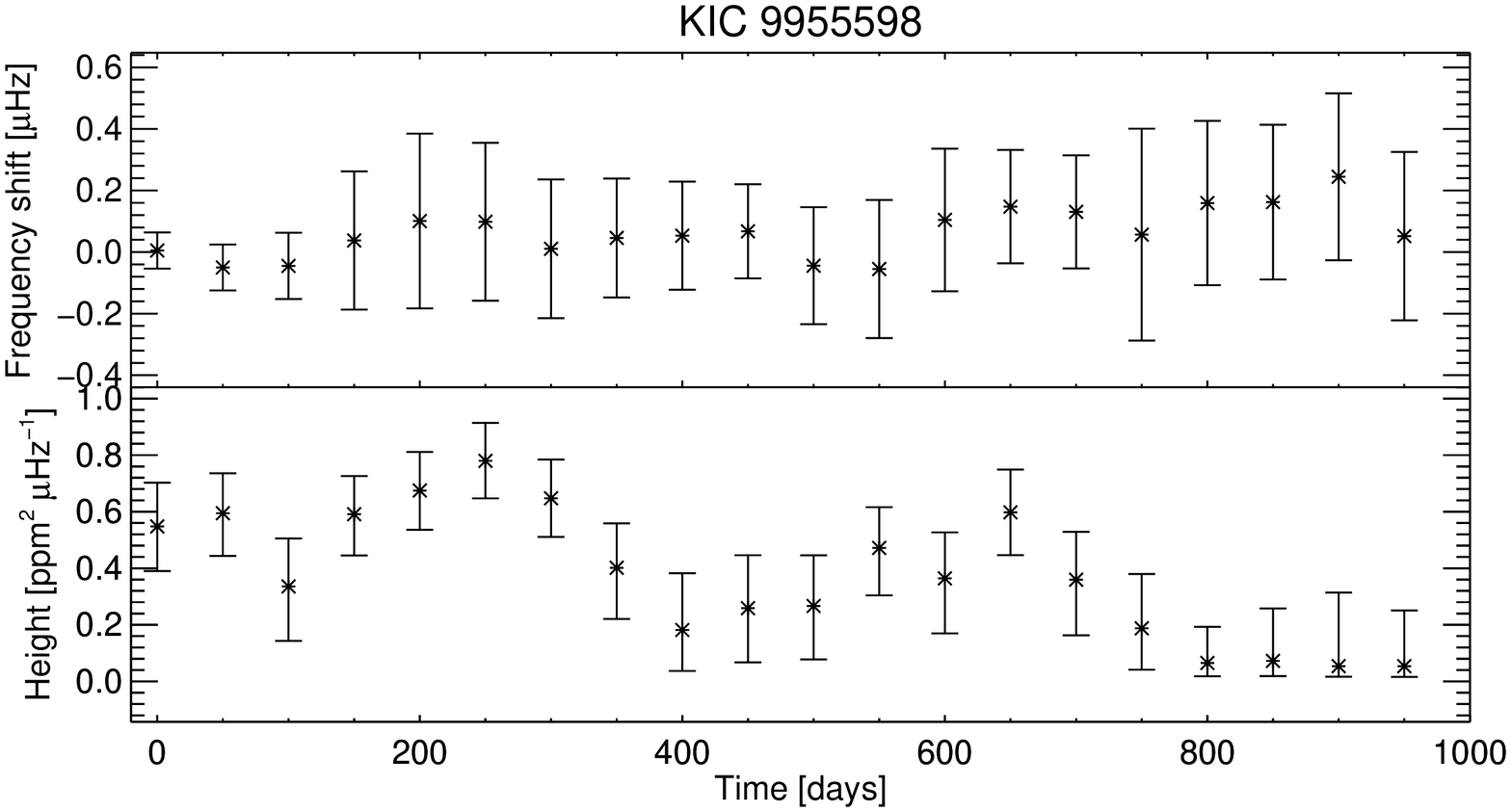}
 		\vspace{-1.em}
 		\caption{Frequency shifts (top half) and height of the p-mode envelope (bottom half) for KIC 9955598 as a function of time.}
 		\label{fig:B16}}
 \end{figure*} 
\clearpage
\pagebreak
 \begin{table*}
 	\caption{Correlation coefficients between parameters of the background model and the frequency shifts for KIC 10018963}           
 	\label{table:B17}
 	\centering
\begin{tabular}{c|c c c c c c c c }
	\hline\hline
	10018963 &shifts       &p  &height       &p   &noise       &p &$\tau_1$      &p        \\
	\hline
	height&    0.46   & 0.15 &         &        &       &         &        &            \\
	noise&    0.15   & 0.65 &  -0.26  &  0.43  &       &         &        &            \\
	$\tau_1$ &  -0.13   & 0.71 &   0.23  &  0.50  &  0.06 &   0.85  &        &            \\
	$\tau_2$ &  -0.09   & 0.79 &   0.17  &  0.61  & -0.31 &   0.36  & -0.08  &  0.81      \\
\end{tabular}
 \end{table*}
 \begin{figure*}
 	\centering{
 		\includegraphics[angle=-90,width=0.70\textwidth]{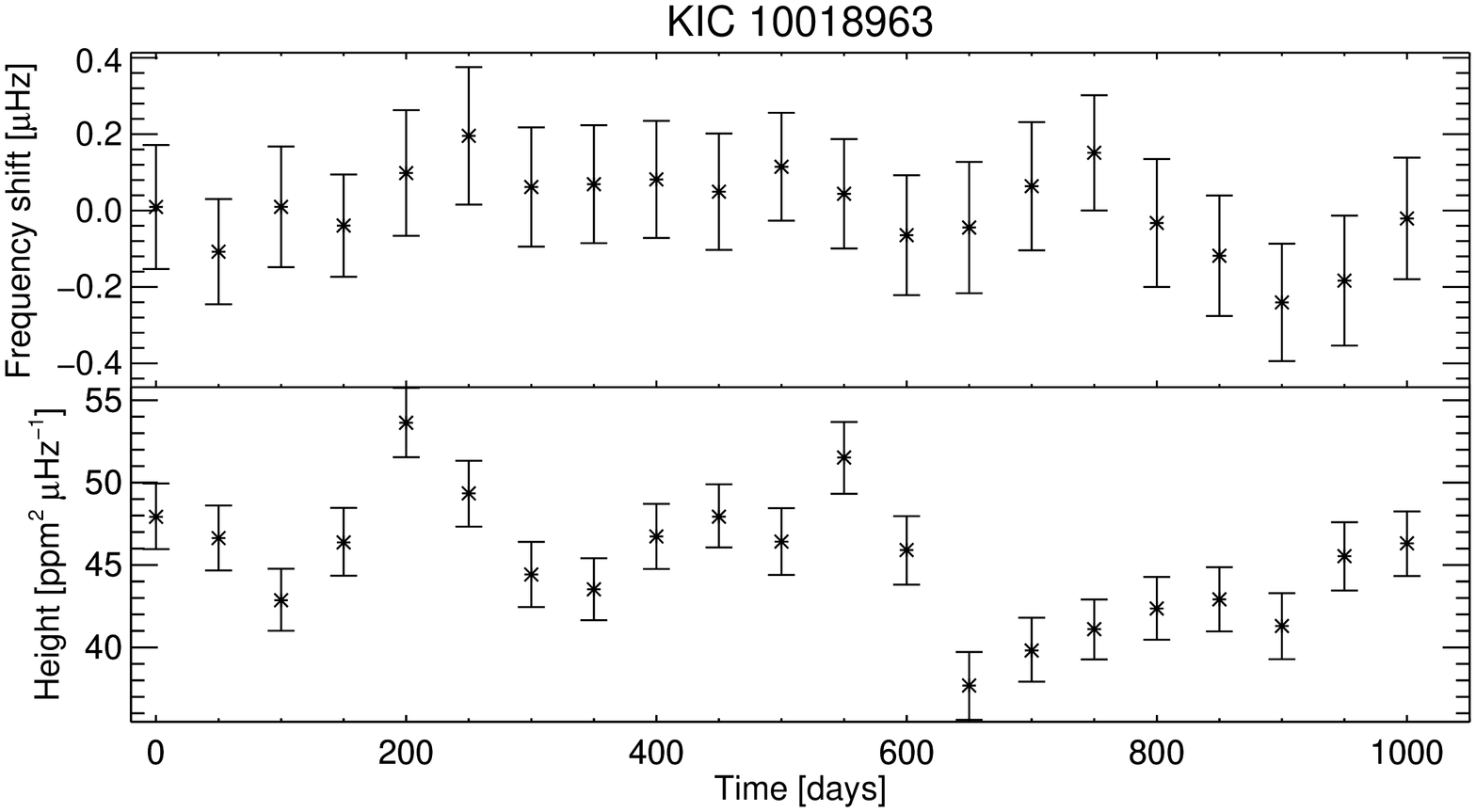}
 		\vspace{-1.em}
 		\caption{Frequency shifts (top half) and height of the p-mode envelope (bottom half) for KIC 10018963 as a function of time.}
 		\label{fig:B17}}
 \end{figure*} 
 
 \begin{table*}
 	\caption{Correlation coefficients between parameters of the background model and the frequency shifts for KIC 10516096}           
 	\label{table:B18}
 	\centering
\begin{tabular}{c|c c c c c c c c }
	\hline\hline
	10516096 &shifts       &p  &height       &p   &noise       &p &$\tau_1$      &p        \\
	\hline
	height&   -0.26  &  0.47 &        &        &         &       &          &           \\
	noise&    0.25  &  0.49 &   0.43 &   0.21 &         &       &          &           \\
	$\tau_1$ &  -0.75  &  0.01 &   0.38 &   0.28 &   0.05  &  0.88 &          &           \\
	$\tau_2$ &   0.14  &  0.70 &  -0.28 &   0.43 &  -0.56  &  0.09 &  -0.22   & 0.53      \\
\end{tabular}
 \end{table*}
 \begin{figure*}
 	\centering{
 		\includegraphics[angle=-90,width=0.70\textwidth]{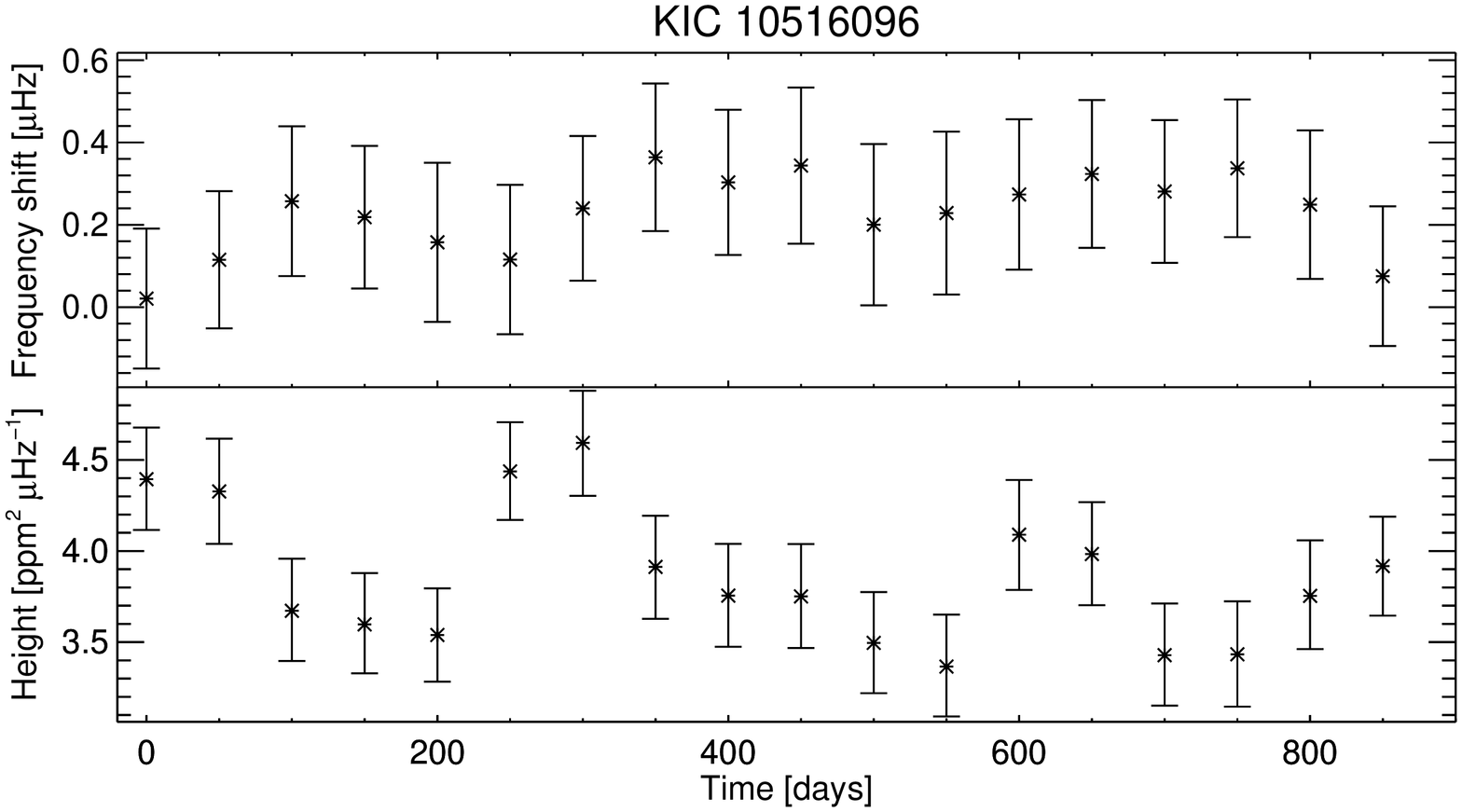}
 		\vspace{-1.em}
 		\caption{Frequency shifts (top half) and height of the p-mode envelope (bottom half) for KIC 10516096 as a function of time.}
 		\label{fig:B18}}
 \end{figure*} 
\clearpage
\pagebreak
 \begin{table*}
 	\caption{Correlation coefficients between parameters of the background model and the frequency shifts for KIC 10644253}           
 	\label{table:B19}
 	\centering
\begin{tabular}{c|c c c c c c c c }
	\hline\hline
	10644253 &shifts       &p  &height       &p   &noise       &p &$\tau_1$      &p        \\
	\hline
	height&   -0.93  &  $3\cdot 10^{-3}$ &         &        &       &         &       &            \\
	noise&    0.50  &  0.25 &  -0.75  &  0.05  &       &         &       &             \\
	$\tau_1$ &  -0.71  &  0.07 &   0.68  &  0.09  & -0.14 &   0.76  &       &             \\
	$\tau_2$ &   0.86  &  0.01 &  -0.82  &  0.02  &  0.43 &   0.34  & -0.75 &   0.05      \\
\end{tabular}
 \end{table*}
 \begin{figure*}
 	\centering{
 		\includegraphics[angle=-90,width=0.70\textwidth]{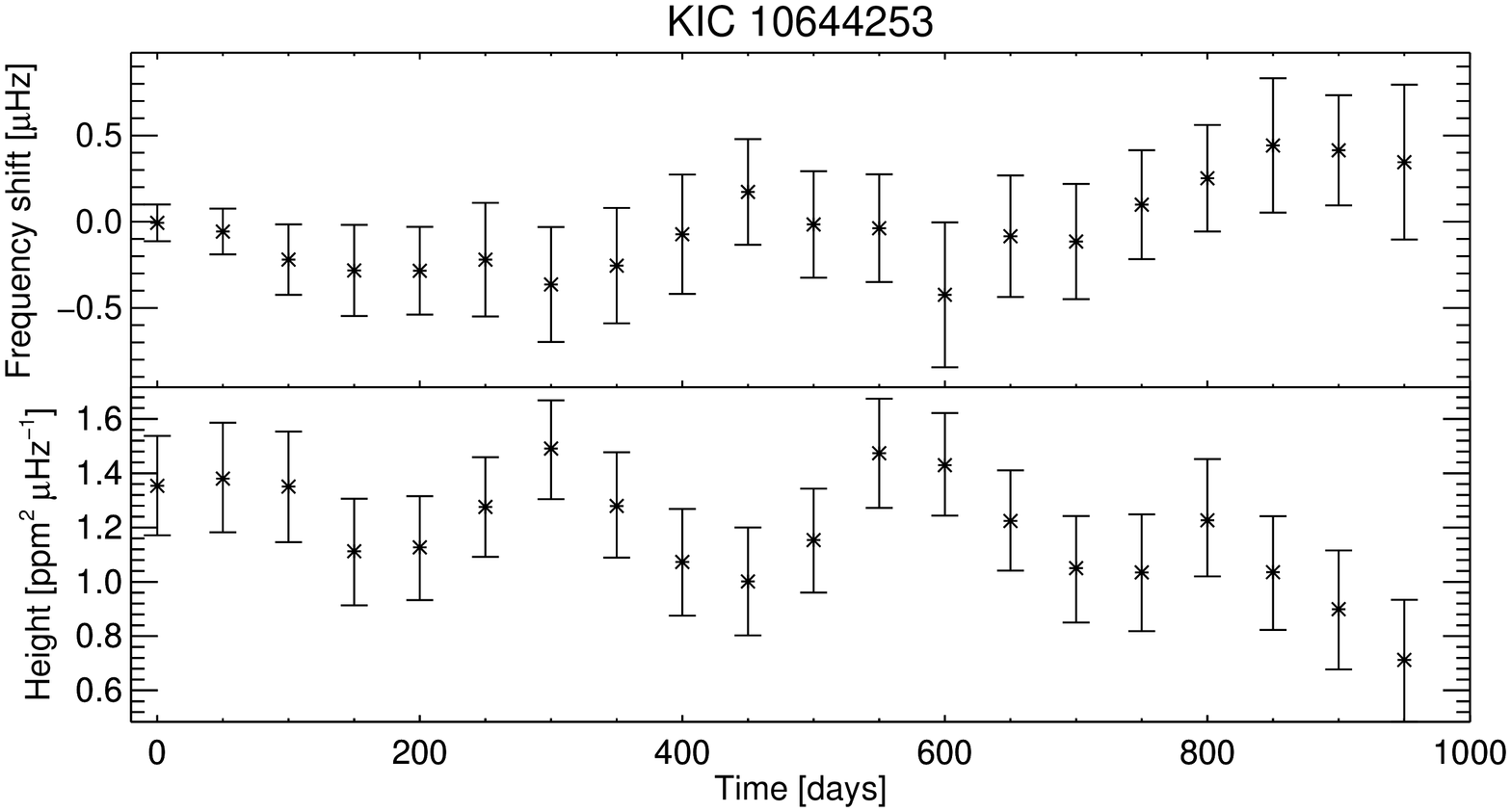}
 		\vspace{-1.em}
 		\caption{Frequency shifts (top half) and height of the p-mode envelope (bottom half) for KIC 10644253 as a function of time.}
 		\label{fig:B19}}
 \end{figure*} 

 \begin{table*}
	 \caption{Correlation coefficients between parameters of the background model and the frequency shifts for KIC 10963065}           
	 \label{table:B20}
 	\centering
\begin{tabular}{c|c c c c c c c c }
	\hline\hline
	10963065 &shifts       &p  &height       &p   &noise       &p &$\tau_1$      &p        \\
	\hline
	height&   -0.26   & 0.62  &        &       &         &        &       &             \\
	noise&    0.60   & 0.21  & -0.83  &  0.04 &         &        &       &             \\
	$\tau_1$ &   0.26   & 0.62  & -0.71  &  0.11 &   0.89  &  0.02  &       &             \\
	$\tau_2$ &   0.54   & 0.27  & -0.60  &  0.21 &   0.83  &  0.04  &  0.83 &   0.04      \\
\end{tabular}
 \end{table*}
 \begin{figure*}
 	\centering{
 		\includegraphics[angle=-90,width=0.70\textwidth]{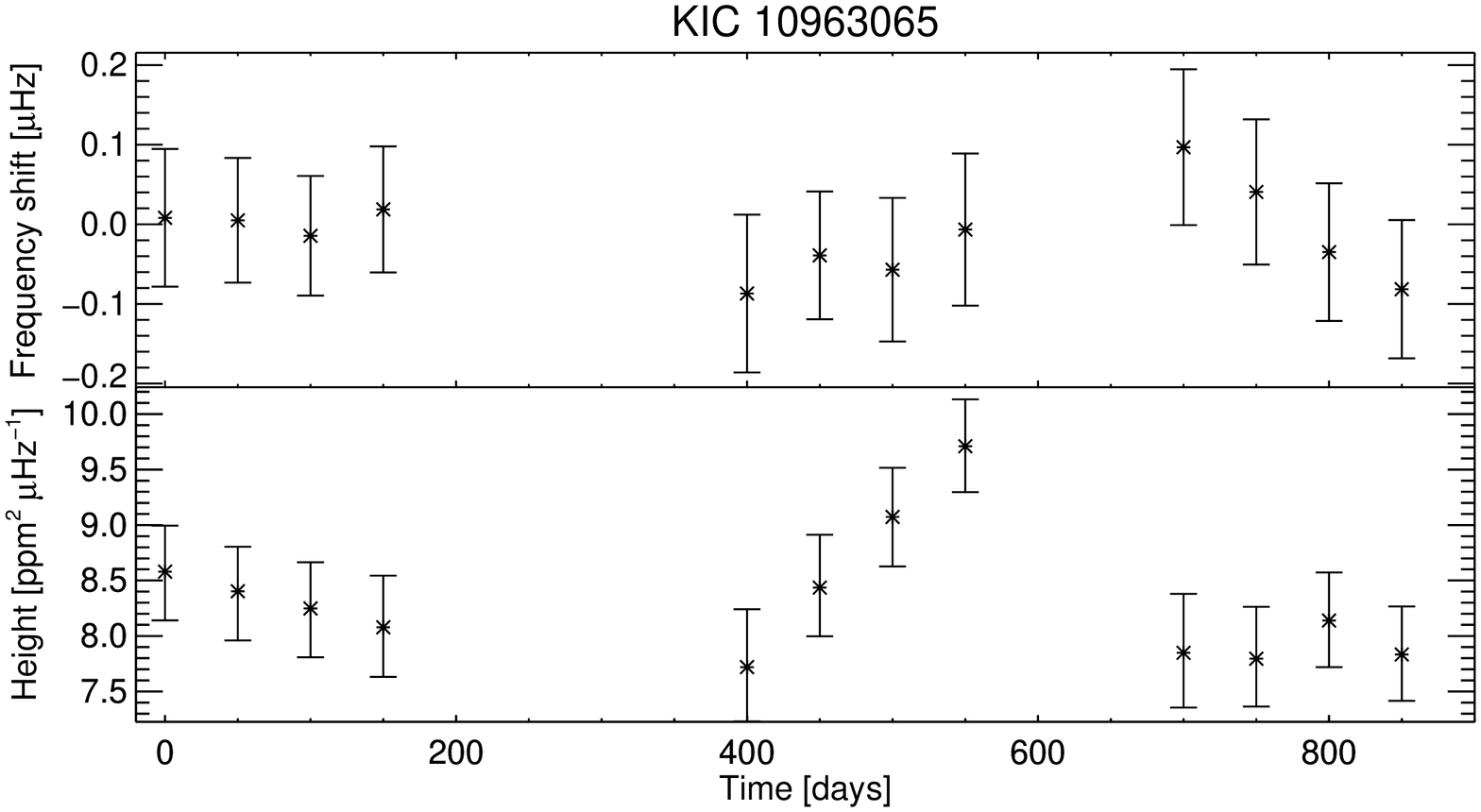}
 		\vspace{-1.em}
 		\caption{Frequency shifts (top half) and height of the p-mode envelope (bottom half) for KIC 10963065 as a function of time.}
 		\label{fig:B20}}
 \end{figure*} 
\clearpage
\pagebreak
 \begin{table*}
	\caption{Correlation coefficients between parameters of the background model and the frequency shifts for KIC 11244118}           
	\label{table:B21}
 	\centering
\begin{tabular}{c|c c c c c c c c }
	\hline\hline
	11244118 &shifts       &p  &height       &p   &noise       &p &$\tau_1$      &p        \\
	\hline
	height&   -0.61 &  0.05   &        &       &        &        &         &           \\
	noise&    0.25 &   0.47  & -0.37  &  0.26 &        &        &         &            \\
	$\tau_1$ &   0.25 &   0.47  & -0.12  &  0.73 &  -0.41 &   0.21 &         &            \\
	$\tau_2$ &   0.30 &   0.37  & -0.65  &  0.03 &   0.15 &   0.67 &   0.50  &  0.12      \\
\end{tabular}
\end{table*}
\begin{figure*}
	\centering{
		\includegraphics[angle=-90,width=0.70\textwidth]{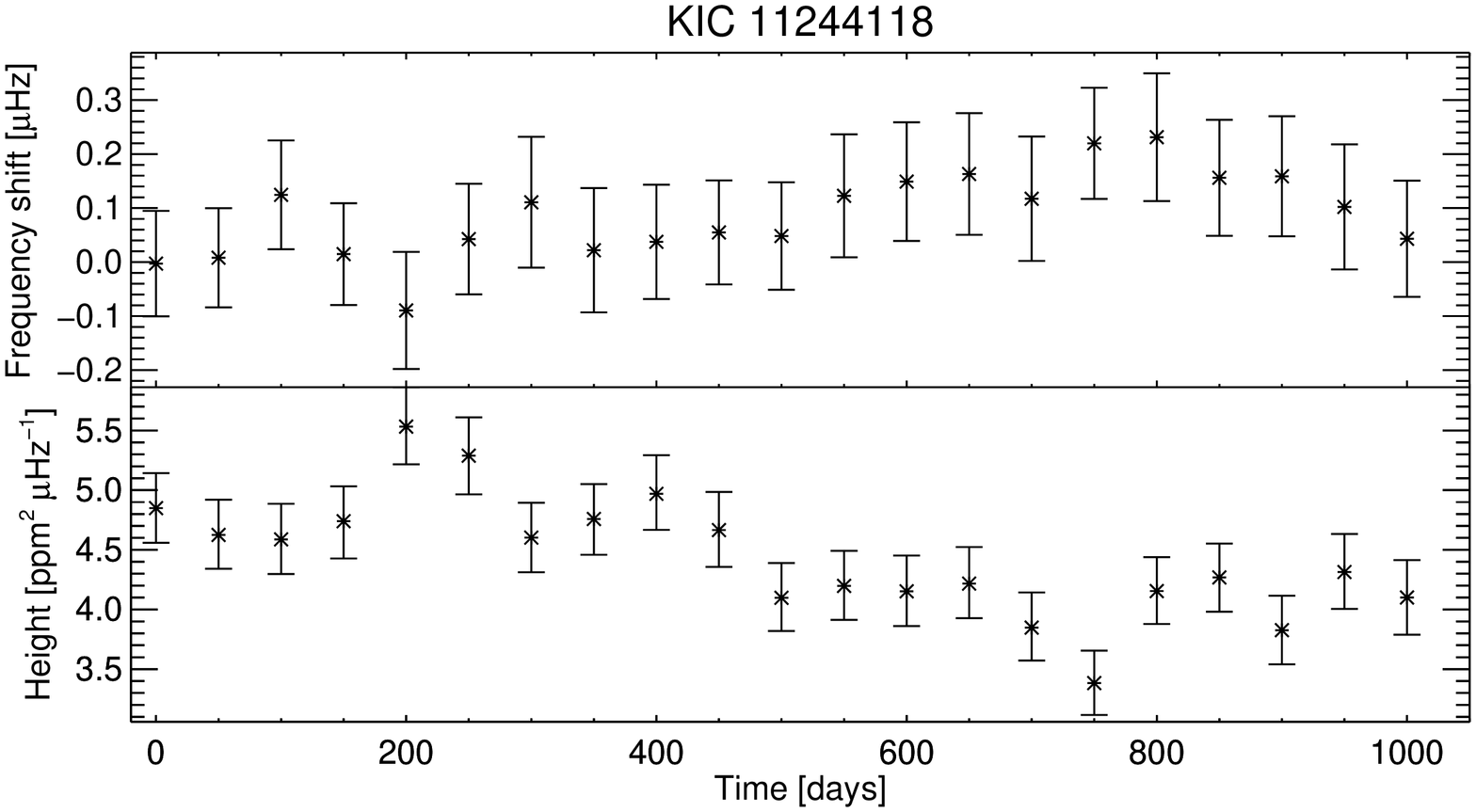}
		\vspace{-1.em}
		\caption{Frequency shifts (top half) and height of the p-mode envelope (bottom half) for KIC 11244118 as a function of time.}
		\label{fig:B21}}
\end{figure*} 
 
 \begin{table*}
 	\caption{Correlation coefficients between parameters of the background model and the frequency shifts for KIC 11295426}           
 	\label{table:B22}
 	\centering
\begin{tabular}{c|c c c c c c c c }
	\hline\hline
	11295426 &shifts       &p  &height       &p   &noise       &p &$\tau_1$      &p        \\
	\hline
	height&    0.36 &   0.43 &        &         &        &        &        &            \\
	noise&   -0.18 &   0.70 &   0.25 &   0.59  &        &        &        &            \\
	$\tau_1$ &  -0.14 &   0.76 &  -0.54 &   0.22  & -0.18  &  0.70  &        &            \\
	$\tau_2$ &   0.29 &   0.53 &  -0.64 &   0.12  & -0.79  &  0.04  &  0.46  &  0.29      \\
\end{tabular}
 \end{table*}
 \begin{figure*}
 	\centering{
 		\includegraphics[angle=-90,width=0.70\textwidth]{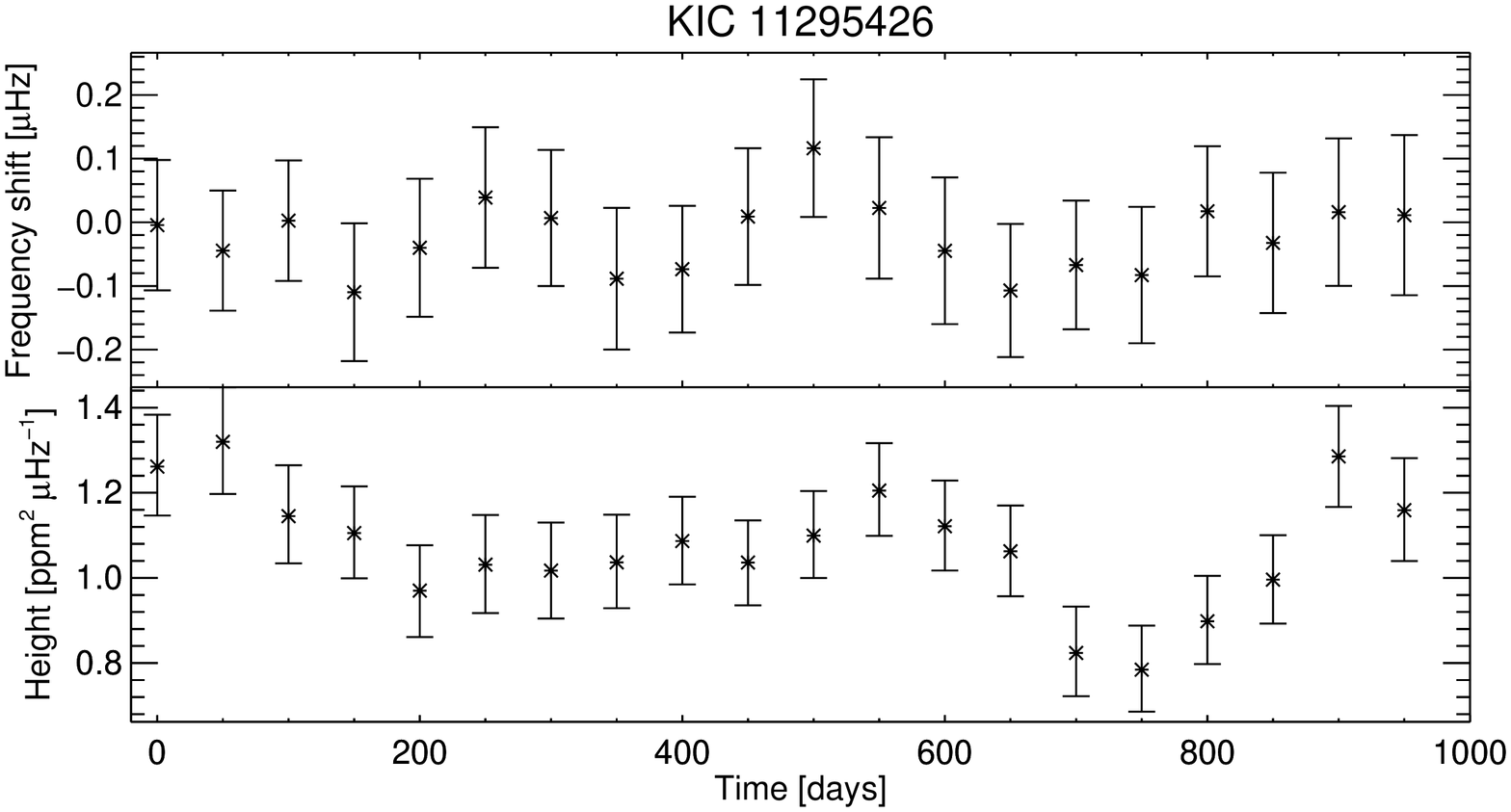}
 		\vspace{-1.em}
 		\caption{Frequency shifts (top half) and height of the p-mode envelope (bottom half) for KIC 11295426 as a function of time.}
 		\label{fig:B22}}
 \end{figure*} 
\clearpage
\pagebreak
 \begin{table*}
 	\caption{Correlation coefficients between parameters of the background model and the frequency shifts for KIC 12009504}           
 	\label{table:B23}
 	\centering
\begin{tabular}{c|c c c c c c c c }
	\hline\hline
	12009504 &shifts       &p  &height       &p   &noise       &p &$\tau_1$      &p        \\
	\hline
	height&   -0.13  &  0.71  &        &       &          &      &        &             \\
	noise&    0.17  &  0.61  &  0.63  &  0.04 &          &      &        &             \\
	$\tau_1$ &  -0.21  &  0.54  & -0.08  &  0.81 &   0.00   & 1.00 &        &             \\
	$\tau_2$ &  -0.12  &  0.73  & -0.45  &  0.17 &  -0.53   & 0.10 &  -0.15 &   0.65      \\
\end{tabular}
 \end{table*}
 \begin{figure*}
 	\centering{
 		\includegraphics[angle=-90,width=0.70\textwidth]{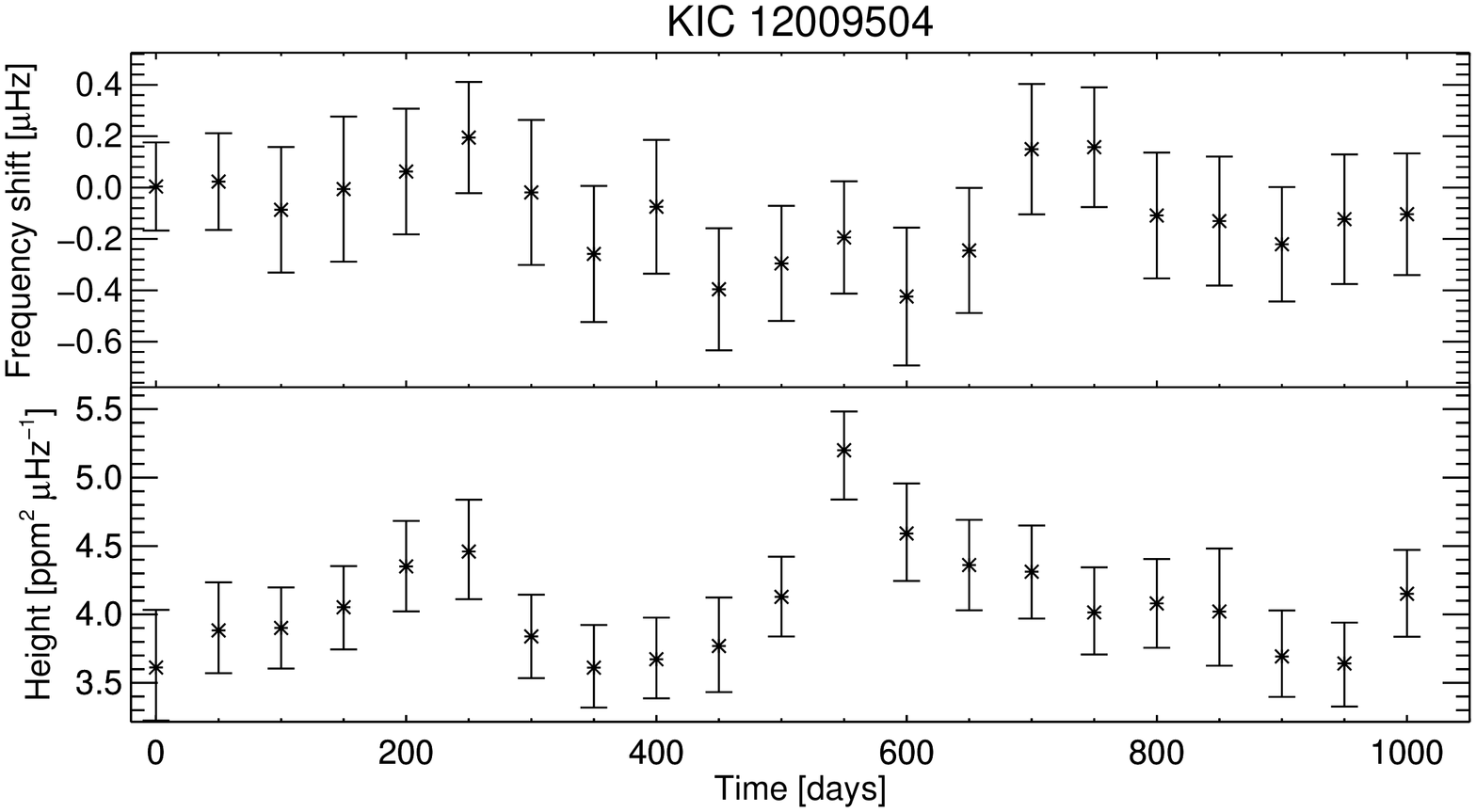}
 		\vspace{-1.em}
 		\caption{Frequency shifts (top half) and height of the p-mode envelope (bottom half) for KIC 12009504 as a function of time.}
 		\label{fig:B23}}
 \end{figure*} 

 \begin{table*}
	 \caption{Correlation coefficients between parameters of the background model and the frequency shifts for KIC~12258514}           
	 \label{table:B24}
 	\centering
\begin{tabular}{c|c c c c c c c c }
	\hline\hline
	12258514 &shifts       &p  &height       &p   &noise       &p &$\tau_1$      &p        \\
	\hline
	height &   -0.50  &  0.25  &        &       &        &         &        &            \\
	noise &    0.43  &  0.34  & -0.86  &  0.01 &        &         &        &            \\
	$\tau_1$&   -0.11  &  0.82  &  0.18  &  0.70 &  -0.14 &   0.76  &        &            \\
	$\tau_2$&    0.07  &  0.88  & -0.04  &  0.94 &  -0.29 &   0.53  & -0.61  &  0.15      \\
\end{tabular}
 \end{table*}
 \begin{figure*}
 \centering{
 	\includegraphics[angle=-90,width=0.70\textwidth]{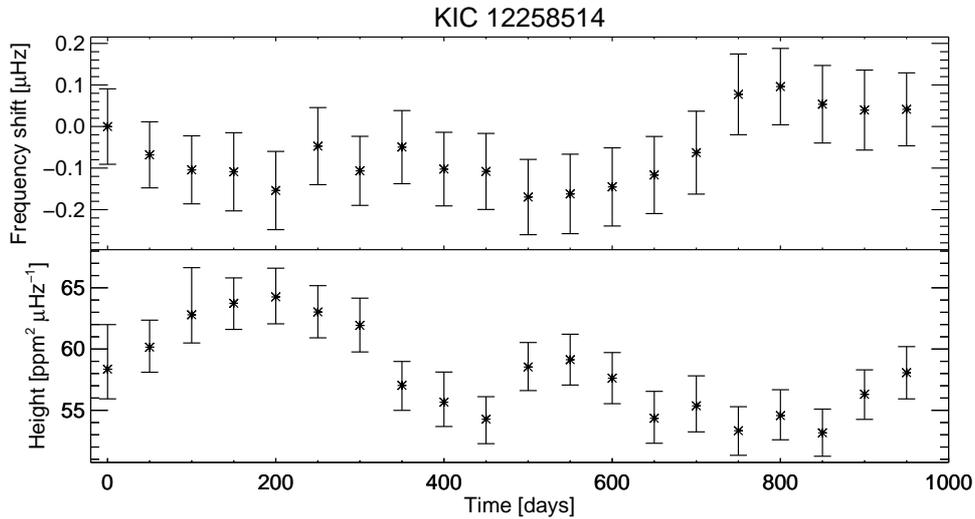}
 	\vspace{-1.em}
 	\caption{Frequency shifts (top half) and height of the p-mode envelope (bottom half) for KIC 12258514 as a function of time.}
 	\label{fig:B24}}
\end{figure*} 
\clearpage
\pagebreak

\section{Plot of frequency shifts as a function of height of the p-mode envelope for KIC~5184732}
\begin{figure*}[h!]
\begin{center}
\includegraphics[angle=-90,width=0.7\textwidth]{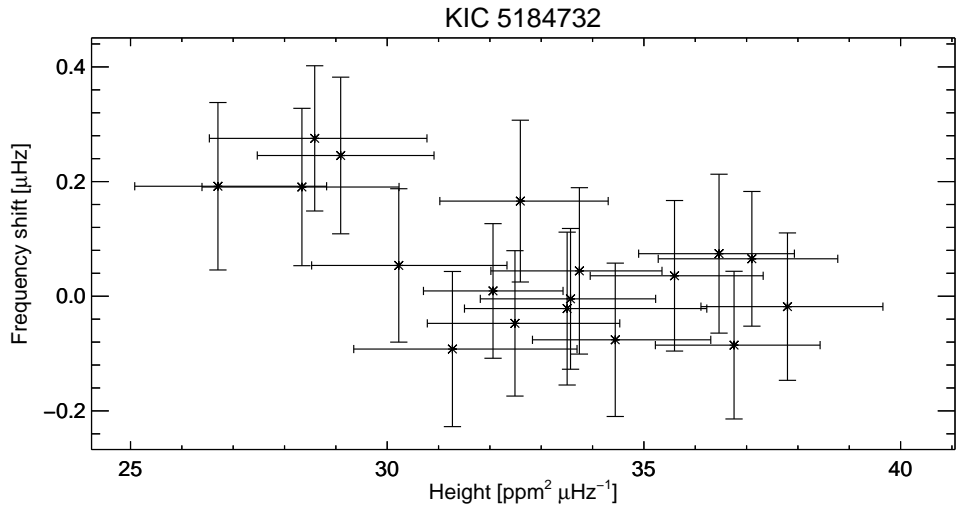}
\end{center}
\caption{Frequency shifts of KIC~5184732 as a function of height of the mode envelope. The correlation between the presented quantities is negative with a tendency towards no correlation for larger heights.}
\label{fig:C1}
\end{figure*}

\end{appendix}
\end{document}